\documentclass[aps,prb,longbibliography,twocolumn,groundaddress,floatfix]{revtex4-2}
\usepackage{amsmath,amssymb,bm,graphicx,color,gensymb,bbold,hyperref,keyval,url,latexsym}
\usepackage[dvipsnames,svgnames,table]{xcolor}
\usepackage[normalem]{ulem} 
\usepackage{multirow}
\usepackage{lipsum}
\usepackage{float}

\usepackage{hyperref}
\hypersetup{
    pdfstartview={FitH},
    colorlinks=true,    
    linkcolor=NavyBlue, 
    citecolor=Maroon,   
    filecolor=NavyBlue, 
    urlcolor=NavyBlue   
}

\begin{document}
    \title{Magnon interactions in the quantum paramagnetic phase of CoNb$_2$O$_6$}
    \author{C. A. Gallegos}
        \affiliation{Department of Physics, University of California, Irvine, California 92697, USA}
    \author{A. L. Chernyshev}
        \affiliation{Department of Physics, University of California, Irvine, California 92697, USA}
    \date{\today}

\begin{abstract}
In this work, we study effects of magnon interactions in the  excitation spectrum of CoNb$_2$O$_6$ in the quantum paramagnetic phase in transverse field, where the $1/S$ spin-wave theory exhibits unphysical divergences at the critical field. We propose a self-consistent Hartree-Fock approach that eliminates such unphysical singularities while preserving the integrity of the  singular threshold  phenomena of magnon decay and spectrum renormalization that are present in both theory and experiment. With the microscopic parameters adopted from previous studies, this method yields a close quantitative agreement with the available experimental data for CoNb$_2$O$_6$  in the relevant regime. Insights into the general structure of the spin-anisotropic model of CoNb$_2$O$_6$ and related zigzag chain materials are also provided and a discussion of the effects of  additional longitudinal field on the spectrum is given.
\end{abstract}

\maketitle

\section{Introduction}

Quantum magnets continue to generate an enormous interest  as a platform for realizing  unconventional ordered \cite{Penc2011Nematic, Starykh2015Review, Ross2011SpinIce, Rau2018OrderbyQDisorder, Rau2019ReviewPyrochlores, Jiang2023Nematic, Jiang2023J1J3} and exotic quantum disordered spin-liquid phases \cite{Balents2010QSL, Yan2011QSLKagome, Savary2016QSL, Knolle2019QSL}, which occur due to competing  interactions between their low-energy  spin degrees of freedom. The celebrated anisotropic-exchange Kitaev model, exhibiting a spin-liquid ground state and fractionalized excitations \cite{Kitaev2006, Nussinov2015ReviewCompass, Knolle2018KitaevReview}, has been particularly inspirational. However, the description of real materials consistently requires exchanges beyond the much-desired Kitaev one \cite{Jackeli2009JK, Chaloupka2010JK, Liu2018CobaltateKSL, Sano2018CobaltateKSL}, resulting in the models with several bond-dependent terms, which typically favor magnetically ordered, if unconventional, states \cite{Choi2012Na2IrO3, Biffin2014gammaLi2IrO3, Johnson2015RuCl3, Paddison2016YMGO, Winter2016KitaevMateriasl, Janssen2017RuCl3,  Zhu2017YMGO, Maksimov2019PRX, Maksimov2020RuCl3, Smit2020KHGamma}.

\begin{figure}[t]
\centering
\includegraphics[width=\linewidth]{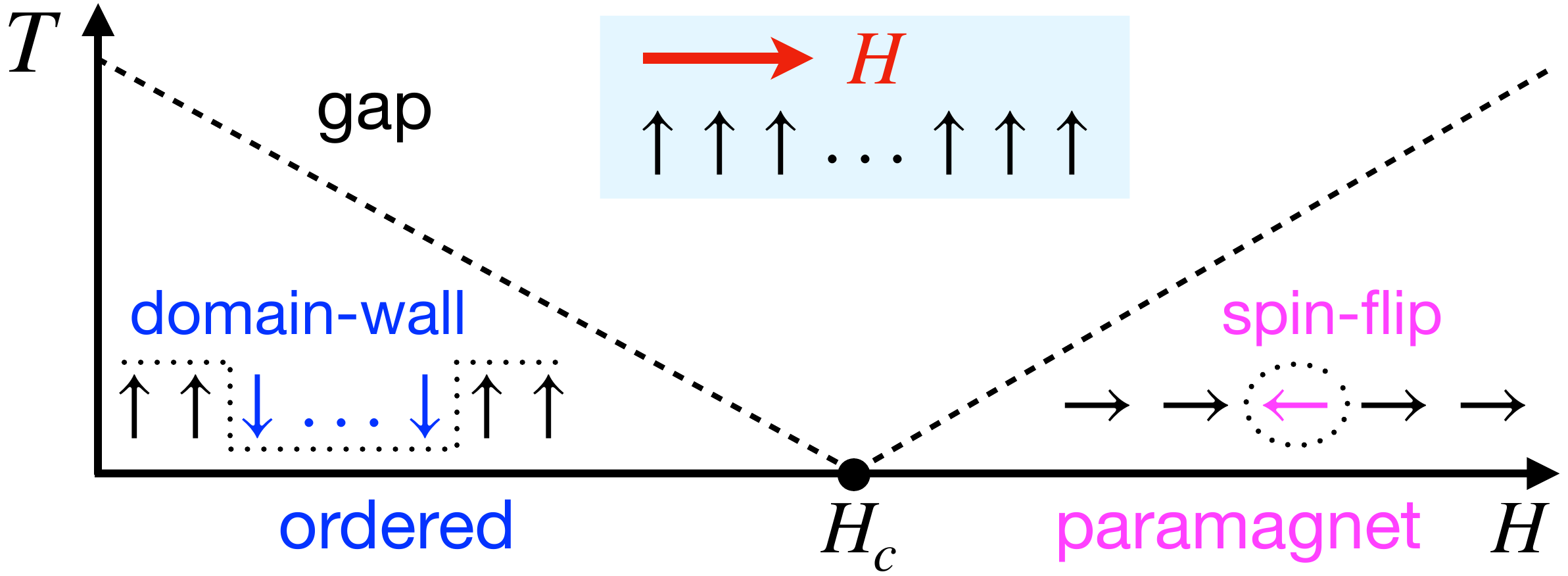} 
\vskip -0.2cm
\caption{The schematic $H$--$T$ phase diagram of the 1D Ising chain in a transverse magnetic field; $H_c$ is the critical field. Dashed lines indicate the spin excitation gap.}
\label{f:phase_d}
\vskip -0.5cm
\end{figure}

In the pursuit of the unusual physical outcomes of the bond-dependent exchanges, recent studies have been centered on the materials with strong spin-orbit coupling \cite{WKrempa2014ReviesSOC, Rau2016ReviewSOC, Schaffer2016SOCReview} and, specifically, on the transition-metal compounds with Co$^{2+}$ ions in an edge-sharing octahedral environment \cite{Songvilay2020Cobaltates, Liu2020CobaltateKSL, Zhong2020Cobaltate, Yang2022Cobaltate, Maksimov2022Cobaltate, Halloran2023Cobaltate, Yao2023Cobaltate}. Cobalt niobate, CoNb$_2$O$_6$, is such an anisotropic-exchange magnet, with spins forming quasi-one-dimensional ferromagnetic zigzag chains. This material is one of the closest realizations of the Ising model, which exhibits a paradigmatic quantum phase transition in transverse field \cite{Pfeuty1970}. The field-induced transition is from the ordered phase with the domain-wall-like excitations to the fluctuating paramagnetic phase, in which excitations are magnon-like spin flips; see Fig.~\ref{f:phase_d}.

CoNb$_2$O$_6$ has generated further excitement by providing experimental evidence of the bound states in its ordered phase, of the emergent E$_8$ symmetry near the critical field, and of the spectacular realization of the magnon decay effect in its paramagnetic phase \cite{coldea2010, Cabrera2014, Robinson2014}. More recently, all of these phenomena have received a consistent explanation within the microscopic spin model, which included important bond-dependent off-diagonal exchange interactions allowed by the crystal symmetry \cite{fava2020, Woodland2023Parameters, Woodland2023_DWsolitons}. 

Specifically, in the paramagnetic phase of CoNb$_2$O$_6$, these off-diagonal exchanges naturally yield the so-called cubic anharmonic term that couples  single-magnon excitations to the two-magnon continuum, leading to magnon decays, the scenario confirmed by the time-dependent density-matrix renormalization group (tDMRG)  calculations \cite{fava2020}. The magnon decay effect is well-documented in the isotropic and diagonal-exchange models, in which the noncollinear states are required for the anharmonic term to occur \cite{Zhitomirsky2013, Maksimov2016XXZTriangular}. Conversely, in the presence of the off-diagonal exchanges, the anharmonic term should be generally unavoidable even in the collinear states \cite{Chernyshev2016DampedKagome,Winter2017Nature, Maksimov2020RuCl3,Maksimov2022PRB}, which is the case of the fluctuating, nominally field polarized paramagnetic phase of CoNb$_2$O$_6$ for $H\!>\!H_c$.

Therefore, it is expected that the analytical insights into  magnon interactions and decays can shed further light on the important aspects of the excitation spectrum in the paramagnetic phase of CoNb$_2$O$_6$ and other anisotropic-exchange magnets. However, quantum fluctuations shift $H_c$ from its classical value, leaving a wide field range inaccessible to the standard spin-wave theory (SWT). Moreover, the $1/S$-expansion in anisotropic-exchange models, needed to account for magnon decays,  is contaminated by the unphysical divergences at the critical field. Different methods have been proposed to overcome similar problems in other systems, with auxiliary fields allowing for the shifts of the phase boundaries \cite{Coletta2013MASWT, Coletta2014MASWT} and self-consistent methods regularizing unphysical divergences \cite{Consoli2020JK1Sexpansion, Maksimov2022PRB, Rau2023GoldstoneSC}.

In this work, we propose a method that allows us to explore the paramagnetic phase of CoNb$_2$O$_6$ in the field range inaccessible to the standard SWT. It naturally regularizes the unphysical $1/S$-divergences, while preserving the integrity of the physical threshold singularities and affiliated decay phenomena. This method combines a self-consistent Hartree-Fock approach \cite{Loly1971,Zhitomirsky2022SCfccAFM} with the perturbative treatment of the cubic anharmonicities. Our results for the dynamical structure factor using microscopic parameters suggested in Ref.~\cite{Woodland2023Parameters} agree closely with the inelastic neutron scattering data \cite{Robinson2014, fava2020} that were previously reproduced by the tDMRG \cite{fava2020}.  Furthermore, we investigate the effects of the additional longitudinal fields in the spectrum of CoNb$_2$O$_6$ in the paramagnetic phase and make predictions of magnon band gaps and associated Van Hove singularities. We also provide insights into the general structure of the spin-anisotropic model for this and related materials.

The paper is organized as follows. In Section~\ref{Sec:Model}, we introduce the spin Hamiltonian for CoNb$_2$O$_6$, relate it to the other anisotropic models, and discuss phenomenological constraints. In Section~\ref{Sec:SWE}, we present the standard $1/S$ spin-wave expansion, demonstrate the unphysical divergences in it, and describe the self-consistent method that regularizes them. In Section~\ref{Sec:Results}, we compare our results with the experimental data. In Section~\ref{Sec:LongFields}, we present our predictions of the effects of additional longitudinal fields in the spectrum of CoNb$_2$O$_6$. We conclude by summarizing our results in Section~\ref{Sec:Conclusions}. Appendixes provide technical details.

\vspace{-0.2cm}
\section{Model}
\label{Sec:Model}
\vskip -0.1cm

In this Section, we introduce the anisotropic-exchange model that should describe magnetic properties of CoNb$_2$O$_6$ and related quasi-1D materials. Following  the prior analysis \cite{Woodland2023Parameters,fava2020}, we use the space group of the crystal structure,   provide a connection of this model to the broader class of anisotropic-exchange models, and  discuss phenomenological constraints on the spin Hamiltonian given in Refs.~\cite{Woodland2023Parameters,fava2020}.

\vspace{-0.3cm}
\subsection{Crystal structure}
\vskip -0.2cm

\begin{figure}[t]
\centering
\includegraphics[width=\linewidth]{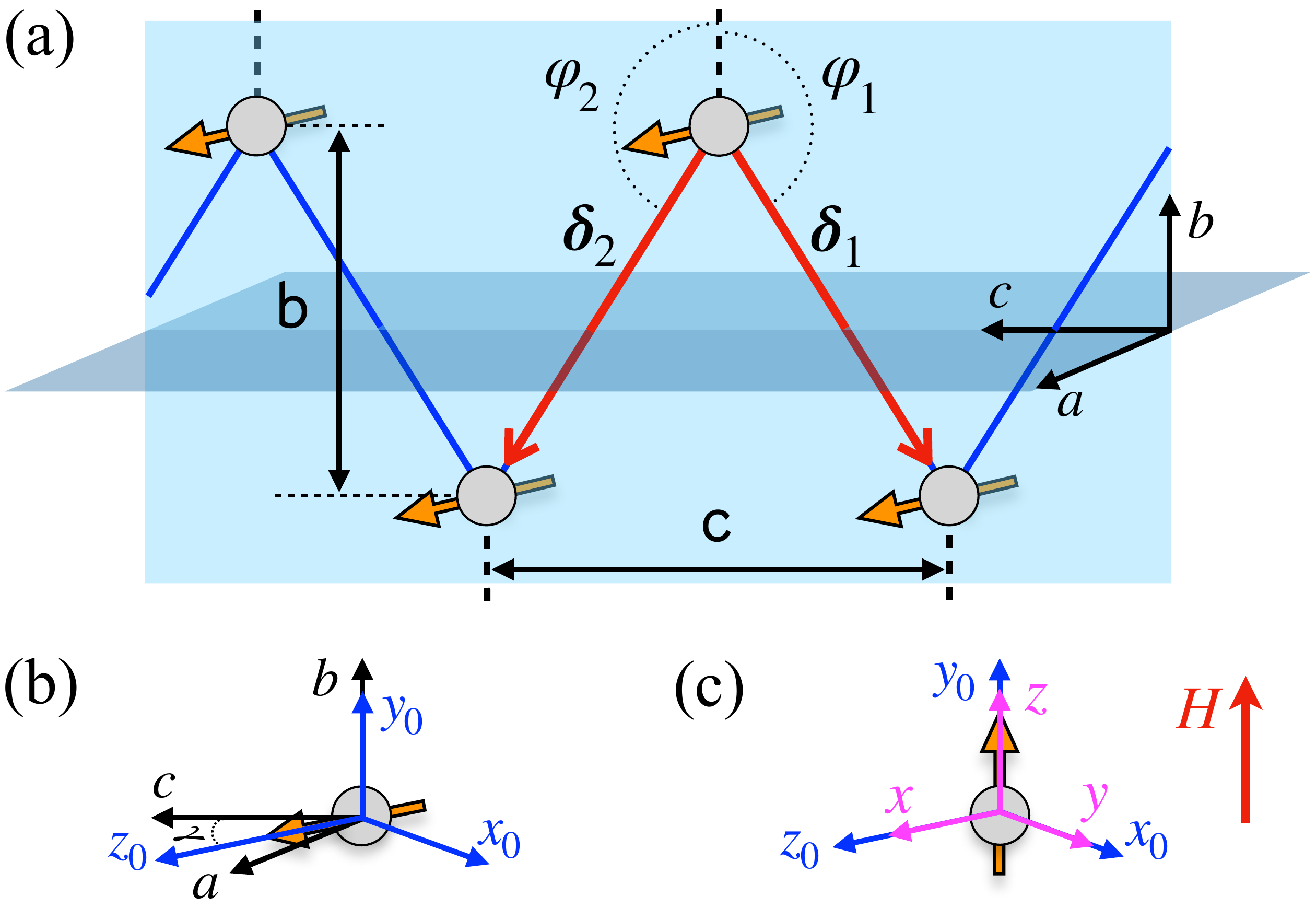} 
\vskip -0.2cm
\caption{(a) The schematic representation of a segment of the zigzag spin chain, with the crystallographic $\{a,b,c\}$ axes, bond-dependent angles $\varphi_\alpha$, nearest-neighbor vectors $\bm{\delta}_\alpha$, lattice constant ${\sf c}$, width of the chain ${\sf b}$, glide $ac$ plane, and $bc$ plane of the zigzag structure. Dashed lines in  $b$ direction are the imaginary missing bonds of the hypothetical honeycomb lattice, see text. (b) The laboratory reference frame $\{x_0,y_0,z_0\}$ with $z_0$ axis along the zero-field Ising direction of spins and angle $\gamma$ in the $ac$ plane. (c) The ``local''  reference frame $\{x,y,z\}$ in the paramagnetic phase with $z$ axis along the transverse-field-induced  spin orientation for $H\!>\!H_c$.}
\label{f:chain}
\vskip -0.4cm
\end{figure}

The crystal structure of CoNb$_2$O$_6$ is orthorhombic, space group $Pbcn$. The combination of the crystal-field and spin-orbit coupling splits the $j\!=\!3/2$ multiplet of the Co$^{2+}$, leading to an effective spin-$1/2$ ground state on each magnetic site \cite{HEID1995}. The magnetic 
Co$^{2+}$ ions are arranged in 1D zigzag chains oriented along the crystallographic $c$ axis with the staggered displacement along the $b$ axis; see Fig.~\ref{f:chain}(a). In the basal $ab$ plane,  Co$^{2+}$ spins form a  weakly-coupled deformed triangular lattice; see Ref.~\cite{Cabrera2014} for details.  The schematic representation of the isolated spin-$1/2$ zigzag chain is shown in Fig.~\ref{f:chain}(a) together with the crystallographic $\{a, b, c\}$ axes.

At low temperatures, Co$^{2+}$ moments in each chain order ferromagnetically, pointing along the Ising easy-axis, which lies in the $ac$ plane at an angle $\gamma\!\approx\!30^{\circ}$ to the $c$ axis \cite{HEID1995}. Therefore, it is natural to introduce another reference frame, referred to as the laboratory frame $\{x_0, y_0, z_0\}$, obtained by a rotation of the crystallographic frame  about the $b$ axis by the angle $\gamma$, so that $z_0$ is aligned with the Ising direction; see Fig.~\ref{f:chain}(b).

\subsection{Symmetries and the nearest-neighbor model}
\label{Sec:Symmetries_NNmodel}

Here, we consider the 1D zigzag spin-chain model with interactions only between the nearest-neighbor sites. Given the translational invariance, the most general nearest-neighbor spin Hamiltonian   of such a chain is
\begin{align}
\mathcal{\hat{H}}_1 &\!=\! \sum_{\langle ij \rangle} \textbf{S}^{\text{T}}_{i} 
\hat{\boldsymbol{J}}_{\alpha} \textbf{S}_{j},
\label{eq:spin_hamiltonian}
\end{align}
where $\textbf{S}_i^{\text{T}}\!=\!(S_i^{x},S_i^{y},S_i^{z})$, $\langle ij \rangle$ denotes summation over the nearest-neighbor bonds, $\alpha\!=\!1,2$ numerates the two distinct bonds of the zigzag structure with the nearest-neighbor vectors $\boldsymbol{\delta}_{1(2)}$  depicted in Fig.~\ref{f:chain}(a),  and $\hat{\boldsymbol{J}}_{1(2)}$ being their respective $3\!\times\!3$ exchange matrices.  At this stage, the two exchange matrices in the model \eqref{eq:spin_hamiltonian} have eighteen independent parameters in total. 

The number of independent parameters in the model \eqref{eq:spin_hamiltonian} is reduced by the space group symmetry of the lattice. The effect of these symmetries on the form of the exchange matrices  $\hat{\boldsymbol{J}}_{\alpha}$ have been thoroughly discussed in Ref.~\cite{fava2020}. Here, we provide a complementary derivation.

CoNb$_2$O$_6$ has two space-group symmetries, the bond-center inversion of the nearest-neighbor bond and the glide symmetry. The inversion with respect to the bond center transposes individual exchange matrices $\hat{\boldsymbol{J}}_{\alpha}$, but must leave them invariant, permitting only symmetric off-diagonal terms. This reduces the number of independent parameters in the model \eqref{eq:spin_hamiltonian} to six per bond. 

The glide symmetry consists of the spatial reflection in the $ac$ plane, followed by a translation by half of the unit cell ${\sf c}_0\!=\!{\sf c}/2$; see Fig.~\ref{f:chain}(a). The spatial reflection flips the sign of the spin components that are parallel to the $ac$ plane, leaving the $b$-component intact. The half-translation completes the space-group operation, but swaps $\hat{\boldsymbol{J}}_{1}$ and $\hat{\boldsymbol{J}}_{2}$, yielding the exchange matrices for the two bonds in the crystallographic $\{a,b,c\}$ reference frame given by
\begin{equation}
\hat{\boldsymbol{J}}_\alpha \!=\! 
\begin{pmatrix}
{J}_{aa} & (-1)^\alpha{J}_{ab} & {J}_{ac}\\
(-1)^\alpha{J}_{ab} & {J}_{bb} & (-1)^\alpha{J}_{bc}\\
{J}_{ac} & (-1)^\alpha{J}_{bc} & {J}_{cc}\\
\end{pmatrix}.\label{eq:J1stagerred}
\end{equation}
Thus, the nearest-neighbor model \eqref{eq:spin_hamiltonian} has only six independent parameters, $\{J_{aa}, J_{bb}, J_{cc}, J_{ab}, J_{ac}, J_{bc}\}$.

An important aspect of the exchange matrices in \eqref{eq:J1stagerred} is the presence of the two off-diagonal staggered terms that alternate between the zigzag bonds. Such a bond-dependence suggests a broad relation of the model for CoNb$_2$O$_6$ with the other well-known forms of the anisotropic-exchange models,  discussed next.

\subsection{Alternative parametrizations}

Given the bond-dependence of the exchange matrices in \eqref{eq:J1stagerred}, it is tempting to establish a connection between the zigzag chain model and the much-studied bond-dependent models on the honeycomb and other lattices. To make this parallel more explicit geometrically, one can perceive the 1D zigzag structure as an element of a hypothetical honeycomb lattice, which is missing bonds in one direction \cite{Agrapidis2018KH1D, Yang2020KG1D, Yang2020KHG1D}. For CoNb$_2$O$_6$, one can introduce the imaginary missing bonds in the $b$ direction; see Fig.~\ref{f:chain}(a). As is noted below, the mutual 2D arrangement of the chains in the $bc$ plane of CoNb$_2$O$_6$ {\it does not} correspond to the honeycomb lattice, but an important symmetry that is needed to make such a  construct possible is present. However, there are a few nuances in the discussed connection that are worth highlighting.

First, the angles of the nearest-neighbor vectors $\boldsymbol{\delta}_\alpha$ with the imaginary missing bonds shown in Fig.~\ref{f:chain}(a), $\varphi_{1,2}\!=\!\mp 127{\degree}$, are close but not equal to those in an ideal honeycomb lattice. Second, unlike in the honeycomb lattice, the physical bonds of the zigzag chain are not the $C_2$-symmetry axes, or, alternatively, the zigzag chain has only one of the three glide planes of the honeycomb lattice. While the imaginary bonds {\it are} the true $C_2$-symmetry axes, the $\pi$-rotation in them is equivalent to a combination of the glide and bond-center inversion symmetries discussed above, providing no further restrictions on the parameters of the model in Eq.~\eqref{eq:J1stagerred}.

\begin{figure}[t]
\centering
\includegraphics[width=\linewidth]{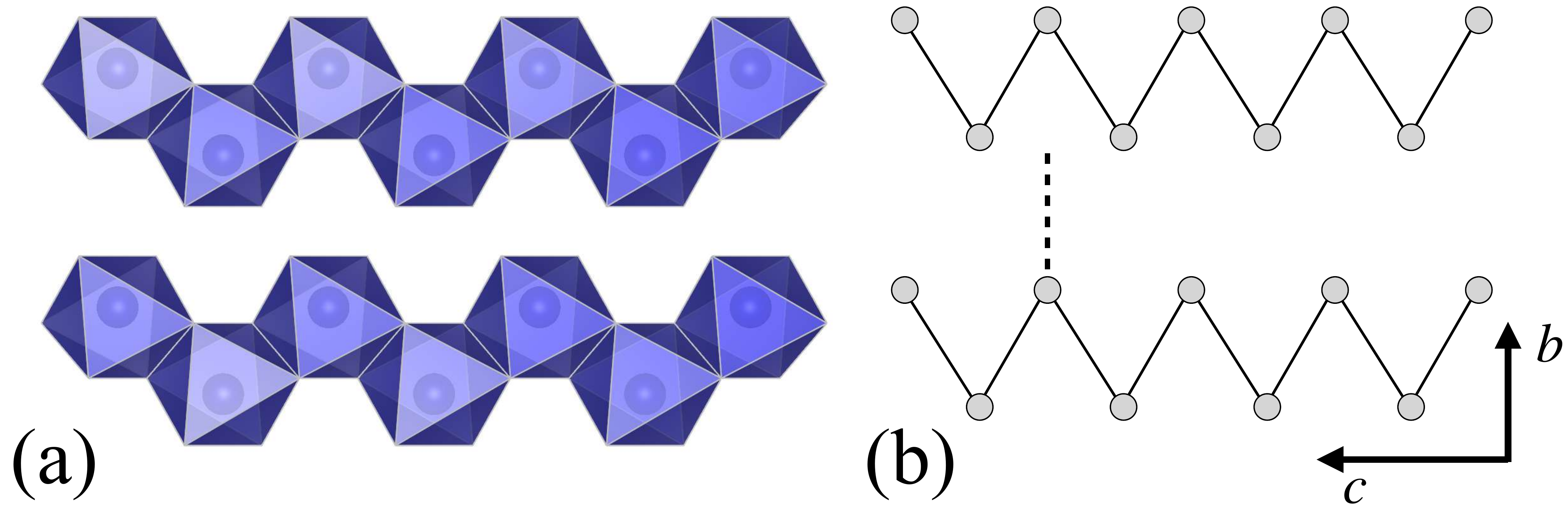} 
\vskip -0.2cm
\caption{The crystal structure of CoNb$_2$O$_6$ as seen in the $bc$ plane, (a) with and (b) without the oxygen octahedral environment. The Nb ions are not shown.}
\label{f:2chain}
\vskip -0.4cm
\end{figure}

Curiously, the true 2D arrangement of the chains in the $bc$ plane of CoNb$_2$O$_6$ is that of a distorted centered rectangular lattice, see Fig.~\ref{f:2chain}, which  has  the $C_2$-symmetry axis for the imaginary bonds.

With these insights, the model \eqref{eq:J1stagerred} can be straightforwardly cast into the ``ice-like" form \cite{Ross2011SpinIce}. Within this parametrization, one remains in the same crystallographic reference frame $\{a,b,c\}$, but the diagonal elements in (\ref{eq:J1stagerred}) are rewritten as
\begin{equation}
J_{bb}+J_{cc}\!=\!2J, \ J_{bb}-J_{cc}\!=\!4J_{\pm\pm}\cos{\varphi_\alpha}, \ J_{aa}\!=\!\Delta J, 
\label{eq:J1diag}
\end{equation}
where $\Delta$ is the $XXZ$ anisotropy parameter, with $a$ being anisotropy axis, and $\varphi_\alpha$ are the bond angles in Fig.~\ref{f:chain}(a). The off-diagonal terms can be rewritten as $J_{ac}\!=\!J_{z\pm}\cos{\varphi_\alpha}$ and 
\begin{eqnarray}
(-1)^\alpha \! J_{ab} \! = \! -\lambda_{z} J_{z\pm} \! \sin{\varphi_\alpha},\,  (-1)^\alpha \! J_{bc} \! = \! 2 \lambda_{\pm} J_{\pm\pm} \! \sin{\varphi_\alpha},
\label{eq:J1offdiag}
\end{eqnarray}
thus encoding the staggered nature of the bond-dependent terms in that of the bond angles, $\varphi_1\!=\!-\varphi_2$; see Fig.~\ref{f:chain}(a). This converts the bond-dependent exchange matrix in (\ref{eq:J1stagerred}) to
\begin{equation}
\hat{\boldsymbol{J}}_{\alpha} \!=\! 
\begin{pmatrix}
\Delta J & -\lambda_{z} J_{z\pm}s_\alpha & J_{z\pm}c_\alpha \\
-\lambda_{z}J_{z\pm}s_\alpha & J+2J_{\pm\pm}c_\alpha & 2\lambda_{\pm}J_{\pm\pm}s_\alpha \\
J_{z\pm}c_\alpha & 2\lambda_{\pm} J_{\pm\pm}s_\alpha & J-2J_{\pm\pm}c_\alpha\\
\end{pmatrix}
,\label{eq:J1ice}
\end{equation}
where the notations $c_\alpha \!\equiv\! \cos {\varphi}_\alpha$ and $s_\alpha\! \equiv\! \sin {\varphi}_\alpha$ are used for brevity. The form \eqref{eq:J1ice} provides an alternative parametrization to the exchange matrix, translating the set of six independent variables in \eqref{eq:J1stagerred} to $\{J, \Delta, J_{\pm\pm}, J_{z\pm}, \lambda_{\pm}, \lambda_{z}\}$, see Appendix~\ref{A:Parametrizations}.

This form in \eqref{eq:J1ice} is similar to the anisotropic-exchange matrices for the triangular and honeycomb lattices within the same ``ice-like" parametrization, up to a cyclic permutation of the axes \cite{Maksimov2019PRX, Maksimov2022PRB}, but it is enriched by the two additional independent terms, which are introduced as the multiplicative factors, $\lambda_{\pm}$ and $\lambda_{z}$. The presence of these extra terms is due to the lower symmetry of the zigzag chains discussed above. An obvious utility of the form (\ref{eq:J1ice}) is that one can straightforwardly characterize the deviation of the zigzag model from the more symmetric honeycomb-lattice case by using the actual parameters proposed for  CoNb$_2$O$_6$ and examining the differences of $\lambda_{\pm}$ and $\lambda_{z}$ from unity. 

We also note that recently, an attempt to introduce the bond-dependent exchanges of the Kitaev honeycomb-lattice model to describe CoNb$_2$O$_6$ was made in the form of the twisted Kitaev-chain Hamiltonian; see Ref.~\cite{Morris2021TwistedKitaev}. This  Hamiltonian corresponds to an interpolation between the Ising chain and the 1D Kitaev model \cite{You2014_1DCompass}. However, a limited number of independent exchange terms in this model restricts its ability to describe quantitatively generic anisotropic-exchange zigzag chain materials and, specifically, CoNb$_2$O$_6$ \cite{Woodland2023Parameters}. Naturally, a more complete description should be  achievable within this Kitaev-like parametrization, but it would require an extended version of the model with all  symmetry-allowed terms present in the exchange matrices,  corresponding to a generalized Kitaev-Heisenberg chain \cite{Agrapidis2018KH1D, Yang2020KG1D, Yang2020KHG1D}.

Next, we discuss  phenomenological constraints on the model (\ref{eq:J1stagerred}) for CoNb$_2$O$_6$.

\subsection{Phenomenological constraint}

In principle, having taken advantage of all of the lattice symmetries, the full six-parameter nearest-neighbor model (\ref{eq:spin_hamiltonian}) with the exchange matrices from (\ref{eq:J1stagerred}) and minimal additional further neighbor and three-dimensional terms should be used to provide the best fit of the experimental data in order to determine the actual values of these parameters for a specific material.

However,  in the most comprehensive studies of CoNb$_2$O$_6$ in Refs.~\cite{fava2020, Woodland2023Parameters}, the number of independent terms in the nearest-neighbor exchange matrices (\ref{eq:J1stagerred}) has been reduced  to four by utilizing a phenomenological constraint  {\it before} the parameter fitting procedure. 

\vspace{-0.35cm}
\subsubsection{Ising axis direction}
\label{Sec:Ising_Axis}
\vskip -0.25cm

Since the zero-field magnetic order in CoNb$_2$O$_6$ has a preferred direction, it is natural to rotate the crystallographic reference frame to align one of the axes ($z_0$) with the observed  Ising axis of the spins. It is done by a rotation of the  $\{a,b,c\}$ axes by $\gamma$ about the $b$ axis; see Fig.~\ref{f:chain}(b). Then, the exchange matrices (\ref{eq:J1stagerred}) are transformed to this  laboratory reference frame $\{x_0,y_0,z_0\}$ via $\hat{{\textbf{J}}}_{\alpha}\!=\!\hat{\textbf{R}}_\gamma^{\phantomsection} \hat{{\boldsymbol{J}}}_{\alpha} \hat{\textbf{R}}_\gamma^{\text{T}}$, where $\hat{\textbf{R}}_\gamma$ is the rotation matrix and
\begin{equation}
\hat{\textbf{J}}_{\alpha} \!=\! 
\begin{pmatrix}
J_{x_0x_0} & (-1)^{\alpha}J_{x_0y_0} & J_{x_0z_0}\\
(-1)^{\alpha}J_{x_0y_0} & J_{y_0y_0} & (-1)^{\alpha}J_{y_0z_0}\\
J_{x_0z_0} & (-1)^{\alpha}J_{y_0z_0} & {J}_{z_0z_0}\\
\end{pmatrix},
\label{eq:J1stagerredRotated}
\end{equation}
with the relations of the $\{x_0,y_0,z_0\}$ exchanges to the ones in the $\{a,b,c\}$  frame given in Appendix~\ref{A:Parametrizations}. One can notice that  matrices in (\ref{eq:J1stagerredRotated}) retain the  structure of (\ref{eq:J1stagerred}).

Since the  Ising  direction is minimizing the energy of the zero-field spin configuration, it provides an implicit phenomenological constraint on the matrix elements of $\hat{\textbf{J}}_{\alpha}$ in (\ref{eq:J1stagerredRotated}), such that the spins in the ground state of the model should stay aligned along the $z_0$ axis.

The essence of the  approach proposed in Refs.~\cite{fava2020, Woodland2023Parameters} is to impose such a  constraint explicitly by eliminating {\it all individual} terms in (\ref{eq:J1stagerredRotated}) that generate  an unphysical tilt of spins away from  the $z_0$ axis. One of such offending terms is  $J_{x_0z_0}$. Since it  creates an $x_0$-tilt of spins in the $x_0z_0$ plane already in the classical limit of the model, it is rendered zero in this approach. Curiously, the $J_{y_0z_0}$-term does not provide a  $y_0$-tilt because of its staggered nature stemming from the glide symmetry of the lattice. 

Less obviously, in the quantum case, the $x_0$-tilt is also generated by the combination of the two staggered terms, $J_{x_0y_0}$ and $J_{y_0z_0}$, as we demonstrate  below. The $J_{y_0z_0}$-term was found crucial for the CoNb$_2$O$_6$ phenomenology as the key microscopic source of the  domain-wall dispersion observed in the ordered phase \cite{fava2020, Woodland2023Parameters}. Then, it follows that the only way to eliminate the unphysical $x_0$-tilt completely is to vanish $J_{x_0y_0}$-term, yielding the four-parameter exchange matrix advocated in  Refs.~\cite{fava2020, Woodland2023Parameters},
\begin{equation}
\hat{\textbf{J}}_{\alpha} \!=\! 
\begin{pmatrix}
J_{x_0x_0} & 0 & 0\\
0 & J_{y_0y_0} & (-1)^{\alpha}J_{y_0z_0}\\
0 & (-1)^{\alpha}J_{y_0z_0} & {J}_{z_0z_0}\\
\end{pmatrix}.\label{eq:J1}
\end{equation}
While we will adopt this form of the exchange matrix in the main part of the present work below, the following note is in order. Although the approach of Refs.~\cite{fava2020, Woodland2023Parameters} is simple, seemingly unambiguous, and potentially generic,  it is not without a caveat. 

One may suspect that such an approach is overconstraining, because a single phenomenological constraint is used to eliminate  {\it two} symmetry-allowed terms  from the exchange matrix.  Instead, both offending terms,  $J_{x_0z_0}$ and $J_{x_0y_0}$, may be allowed to be non-zero, but exactly compensating each others' spin tilting  and leaving the physical Ising direction intact. Of course, the technical implementation of such an indirect constraint as a part of the parameter-fitting procedure is more challenging, so the approach  of making $J_{x_0z_0}\!=\!J_{x_0y_0}\!=\!0$ can be taken as a mild assumption in the search of a minimal model.

To vindicate the assumption  of Refs.~\cite{fava2020, Woodland2023Parameters} in the case of CoNb$_2$O$_6$ further, we note that the  compensating tilts from $J_{x_0z_0}$ and $J_{x_0y_0}(J_{y_0z_0})$ terms  appear in different orders of the quasiclassical theory. $J_{x_0z_0}$ creates a tilt already in the classical limit of the model, while the tilt due to  $J_{x_0y_0}$ is a purely quantum effect. Given this hierarchy and using the fact that the off-diagonal exchanges in CoNb$_2$O$_6$ are secondary to the main Ising term, below we provide a perturbative consideration of the effects of the ``residual'' $J_{x_0z_0}$ and $J_{x_0y_0}$ terms.

\vspace{-0.35cm}
\subsubsection{Perturbative consideration}
\label{Sec:Perturbative_gamma}
\vskip -0.25cm

Coming back momentarily to the exchange matrix in the crystallographic reference frame \eqref{eq:J1stagerred}, a straightforward  minimization of the {\it classical} energy of the model yields the tilt angle of spins away from the $c$ axis as
\begin{equation}
\tan 2\widetilde{\gamma} = \frac{2J_{ac}}{J_{cc}-J_{aa}}=-\frac{2J_{z\pm}\cos \varphi_{\alpha}}{(\Delta-1)J+2J_{\pm\pm}\cos \varphi_{\alpha}},
\label{eq:tan2gamma}
\end{equation}
given here for both parametrizations of the exchange matrix in \eqref{eq:J1stagerred} and \eqref{eq:J1ice}. The latter illustrates one of the broader perspectives provided by the form in \eqref{eq:J1ice} as the $J_{z\pm}$ term is known to produce such an out-of-the-plane tilt in the previously discussed models \cite{Iaconis2018SpinIceTriangularLattice,Maksimov2019PRX, Maksimov2022PRB}. 

Then, within the classical approximation, one would equate the tilt angle $\widetilde{\gamma}$ to its experimentally observed value $\gamma$, thus using the preferred direction of the magnetic order in CoNb$_2$O$_6$ shown in Fig.~\ref{f:chain}(b) as a phenomenological constraint that provides a relation between exchanges given by Eq.~(\ref{eq:tan2gamma}).  As a result, the number of independent terms in the nearest-neighbor exchange matrix would be reduced to five, setting  $J_{x_0z_0}\!\equiv\!0$ in (\ref{eq:J1stagerredRotated}).

However,  quantum fluctuations can 
renormalize the tilt angle of the ordered magnetic moment, producing deviations from the classical result \eqref{eq:tan2gamma}.  In other words, if one would calculate the angle between the Ising $z_0$ and $c$ axes in the quantum case with $\hat{{\textbf{J}}}_{\alpha}$ in (\ref{eq:J1stagerredRotated}) and $J_{x_0z_0}\!=\!0$, it would generally deviate from $\gamma$.

This quantum renormalization can be accessed perturbatively by considering virtual spin-flip processes that are generated by the staggered terms $J_{x_0y_0}$ and $J_{y_0z_0}$. Using the real-space perturbation theory \cite{Bergman2007RSPT, Chernyshev2014RSPTKagome, Zhitomirsky2015RSPT} for the $S\!=\!1/2$ model in (\ref{eq:J1stagerredRotated}) with only the main Ising  and staggered terms, we  derive the tilt angle in the second order of the theory as given by
\begin{equation}
\delta\gamma  \approx -\frac{J_{x_0y_0}J_{y_0z_0}}{J_{z_0z_0}^2},
\label{eq:delta_gamma}
\end{equation}
which is supported by our DMRG calculations \cite{ITensor}, with the details for both deferred to Appendix~\ref{A:RSPT}. 

We can further assert that the higher-order corrections to the tilt angle also require {\it both} staggered terms, because they have to cancel their symmetry-related staggered form as in (\ref{eq:delta_gamma}). Moreover, the higher-order corrections to  (\ref{eq:delta_gamma})  need to carry odd powers of each of the staggered terms because they generate different number of spin flips, as can also be verified numerically; see Appendix~\ref{A:RSPT}.

Thus, from the quasiclassical perspective, the choice of $J_{x_0y_0}\!=\!0$ made in Refs.~\cite{fava2020, Woodland2023Parameters} automatically renders {\it all} quantum corrections to the classical Ising axis angle $\widetilde{\gamma}$ equal to zero, leaving it equal to the experimental value $\gamma$ by construction. For the parametrizations  of the  exchange matrices in \eqref{eq:J1stagerred} and \eqref{eq:J1ice}, the choice $J_{x_0y_0}\!=\!0$  provides another relation between the components of the exchange matrices that reads
\begin{equation}
\tan\gamma = \frac{J_{ab}}{J_{bc}} = -\frac{\lambda_z J_{z\pm}}{2\lambda_{\pm}J_{\pm\pm}}.
\end{equation}
Altogether,  for the nearest-neighbor model written in the laboratory reference frame $\{x_0,y_0,z_0\}$, this results in the four-parameter exchange matrix given in \eqref{eq:J1}.

Lastly, one can use the perturbative consideration for the tilt angle in (\ref{eq:delta_gamma}) together with the numerically precise DMRG calculations in order to quantify the potential values of the ``residual,'' mutually compensating $J_{x_0z_0}$ and $J_{x_0y_0}$ terms in the quantum $S\!=\!1/2$ model of CoNb$_2$O$_6$, if these terms are allowed to deviate from zero. A straightforward derivation gives the relation between such terms that would leave the Ising axis direction intact, 
\begin{equation}
J_{x_0z_0}=J_{x_0y_0}\,\left(\frac{J_{y_0z_0}}{J_{z_0z_0}}\right),
\label{eq:Jxz_vs_Jxy}
\end{equation}
which explicates the different order of their corresponding effects on the spin orientation in the quasiclassical expansion. We note that this result is obtained for a simplified model as is Eq.~(\ref{eq:delta_gamma}).

\begin{figure}[t]
\centering
\includegraphics[width=\linewidth]{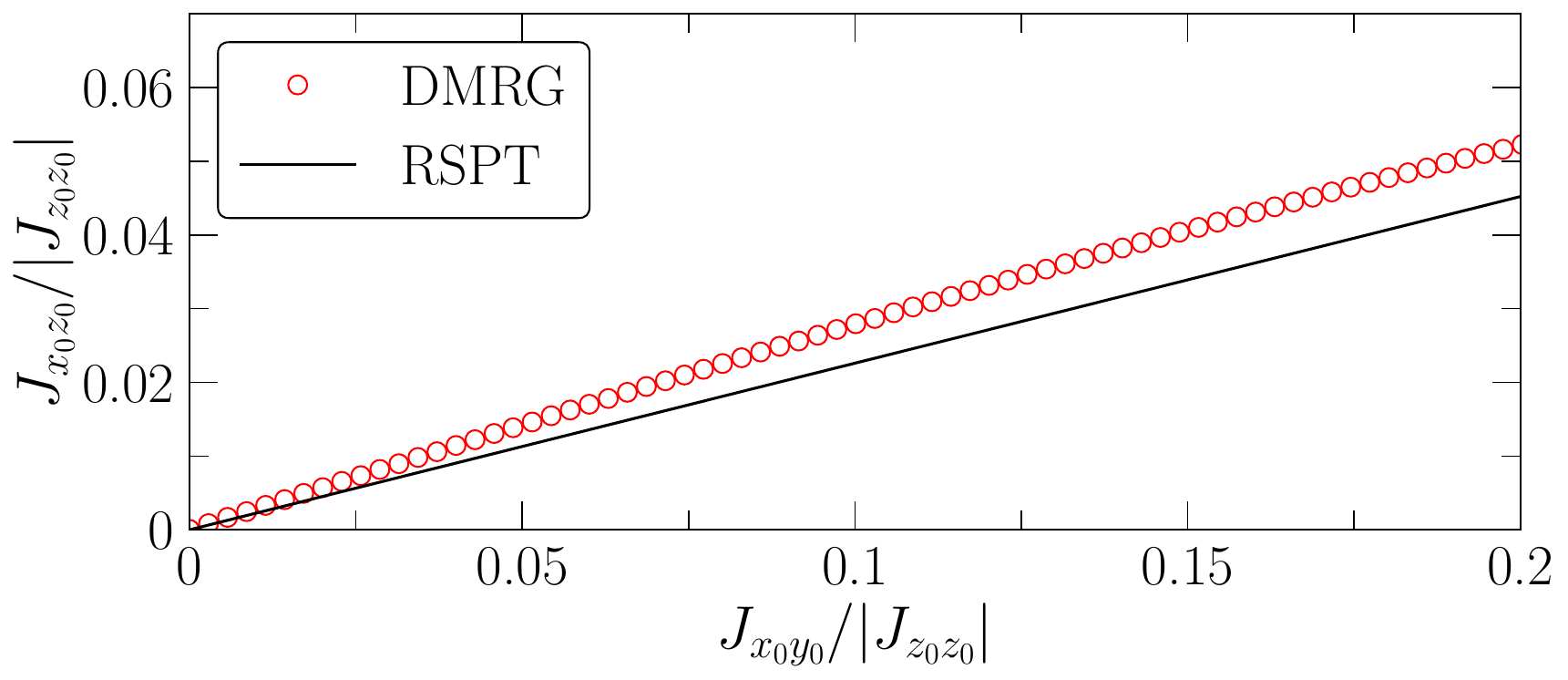} 
\vskip -0.2cm
\caption{$J_{x_0z_0}$ vs $J_{x_0y_0}$ for the  CoNb$_2$O$_6$ parameters in Sec.~\ref{Sec:best_fit}. Line is the real-space perturbation theory (RSPT) result (\ref{eq:Jxz_vs_Jxy}) for the simplified model with only the $J_{z_0z_0}$, $J_{x_0y_0}$, and $J_{y_0z_0}$ terms; symbols are DMRG results for the full model, see Appendix~\ref{A:RSPT} for details.}
\label{f:Jzx_vs_Jxy}
\vskip -0.4cm
\end{figure}

In Fig.~\ref{f:Jzx_vs_Jxy}, we show this dependence for the choice of $J_{y_0z_0}$ and $J_{z_0z_0}$ values that  correspond to CoNb$_2$O$_6$; see  Sec.~\ref{Sec:best_fit} below. It is plotted together with the DMRG results for the full model using the best-fit parameters discussed in the next Section. According to Ref.~\cite{fava2020}, the $J_{x_0y_0}$ term in CoNb$_2$O$_6$ should be small as it produces the momentum space periodicity of the domain-wall excitations in the ordered phase that is different from the observed one. As is shown in  Fig.~\ref{f:Jzx_vs_Jxy}, this should render the limit on the residual $J_{x_0z_0}$ to nearly zero, thus providing a further partial exoneration to the approach of Refs.~\cite{fava2020, Woodland2023Parameters}. 

\vspace{-0.3cm}
\subsection{Best-fit parameters}
\label{Sec:best_fit}
\vskip -0.2cm

An extensive comparison of the experimental data with the numerical modeling carried out in Ref.~\cite{Woodland2023Parameters} has resulted in the set of the best-fit parameters for CoNb$_2$O$_6$. Translating them  to the notations  of our Eq.~\eqref{eq:J1}, the nearest-neighbor exchanges are
\begin{align}
J_{x_0x_0} \!&=\! -0.57(2)\ \text{meV},\ J_{y_0y_0}\!=\! -0.67(2)\ \text{meV},\nonumber\\
J_{z_0z_0}\!&=\! -2.48(2)\ \text{meV},\ J_{y_0z_0}\!=\!-0.56(1)\ \text{meV}.
\label{eq:values_JNNlab}
\end{align}
We note that in  Refs.~\cite{fava2020, Woodland2023Parameters} a different parametrization has been used for the  two diagonal exchanges: $J_{x_0x_0}\!=\!J_{z_0z_0}(\lambda_s\!+\!\lambda_a)$ and $J_{y_0y_0}\!=\!J_{z_0z_0}(\lambda_s\!-\!\lambda_a)$. 

For completeness, we also translate these numerical values to that of the exchange parameters  in the original crystallographic $\{a,b,c\}$ frame in Eq.~\eqref{eq:J1stagerred} and to the ``ice-like" parametrization of Eq.~\eqref{eq:J1ice}; see Appendix~\ref{A:Parametrizations}.  For the ``ice-like" form \eqref{eq:J1ice}, the best-fit parameters $\{J, \Delta, J_{\pm\pm}, J_{z\pm}, \lambda_{\pm}, \lambda_{z}\}$ are given by
\begin{align}
J \!&=\! -1.34(1)\ \text{meV},\ \Delta\!=\! 0.78(1),\ J_{\pm\pm}\!=\! -0.55(1)\ \text{meV},\nonumber\\
J_{z\pm}\!&=\! 1.37(2)\ \text{meV},\ \lambda_{z}\!=\!0.26(1),\ \lambda_{\pm}\!=\!0.55(1),
\label{eq:values_Jice}
\end{align}
where the last two  parameters,  $\lambda_z$ and $\lambda_\pm$, quantify the substantial degree to which the zigzag chain differs from the hypothetical honeycomb lattice, where $\lambda_z\!=\!\lambda_\pm\!=\!1$.  Clearly, the diagonal $XXZ$ exchanges are ferromagnetic, and the values of $J_{\pm\pm}$ and $J_{z\pm}$ underscore the pronounced anisotropic nature of CoNb$_2$O$_6$. Interestingly, the value of $\Delta\!<\!1$ implies that in this parametrization, CoNb$_2$O$_6$ can be regarded as an easy-plane anisotropic-exchange magnet, thus establishing a  connection to the other members of the cobaltate family \cite{Elliot2021, Maksimov2022Cobaltate, Halloran2023Cobaltate}. Moreover, the dominant $J_{z\pm}$ and smaller $J_{\pm\pm}$ anisotropies are reminiscent of the other transition-metal anisotropic-exchange materials, such as $\alpha$-RuCl$_3$ \cite{Maksimov2020RuCl3}.

As is stated in Ref.~\cite{Woodland2023Parameters}, the nearest-neighbor model \eqref{eq:spin_hamiltonian} with the exchange matrix in \eqref{eq:J1}, needs to be supplemented by the next-nearest-neighbor $XXZ$ term 
\begin{equation}
\hat{\mathcal{H}}_2 \!= \!\sum_{i}
\Big\{J_{2}\left(S_i^{x_0}S_{i+2}^{x_0}+S_i^{y_0}S_{i+2}^{y_0}\right)\!+\!J_{2z_0}S_i^{z_0}S_{i+2}^{z_0}\Big\},
\label{eq:J2}
\end{equation}
providing a consistent quantitative agreement with the experimental data for the excitation spectrum of CoNb$_2$O$_6$ in different regions of its phase diagram and for the fields applied in the transverse and longitudinal directions. The best parameter choice for the next-nearest-neighbor part of the  Hamiltonian \eqref{eq:J2} is
\begin{align}
J_{2} \!&=\! 0.077(3)\ \text{meV},\ J_{2z_0}\!=\! 0.19(1)\ \text{meV},\nonumber\\
g_{x_0}\!&=\!3.29(6),\ g_{y_0}\!=\!3.32(2),\  g_{z_0}\!=\!6.90(5),
\label{eq:values_J2g}
\end{align}
where we have also listed the principal moments of the g-tensor. Below, we will use the best-fit parameter values in \eqref{eq:values_JNNlab} and \eqref{eq:values_J2g} to compare our results with the  neutron scattering data in the paramagnetic phase of CoNb$_2$O$_6$ \cite{Robinson2014,fava2020}. We will also follow prior works~\cite{Woodland2023Parameters} in a simplifying assumption that the g-tensor is diagonal in the laboratory frame.

We also point out that one may need to include small interchain couplings if considering the full 3D spectrum, or use an effective longitudinal field to account for their confining effect in the ordered phase \cite{Cabrera2014, fava2020, Woodland2023Parameters, Woodland2023_DWsolitons}. However, in this work, we focus on the spectrum in the field-induced paramagnetic phase of CoNb$_2$O$_6$, in which spins are aligned in the transverse ($b\!=\!y_0$) direction, making the effect of the interchain couplings negligible.

\vspace{-0.1cm}
\section{Self-consistent spin-wave theory}
\label{Sec:SWE}
\vskip -0.1cm

In this Section, we briefly outline the steps of the  $1/S$ spin-wave expansion as applied to the paramagnetic phase of CoNb$_2$O$_6$, demonstrate the problem of the unphysical divergences in it, and describe the self-consistent Hartree-Fock method that regularizes them.

\vspace{-0.2cm}
\subsection{Critical field and Hamiltonian in local axes}
\label{Sec:evenodd}
\vskip -0.1cm

The paramagnetic phase in CoNb$_2$O$_6$ is induced by a transverse magnetic field. To model it,
the six-parameter spin Hamiltonian  $\hat{\mathcal{H}}_1+\hat{\mathcal{H}}_2$ from  Eqs.~(\ref{eq:spin_hamiltonian}), (\ref{eq:J1}), and (\ref{eq:J2}) has to be augmented by the transverse-field term
\begin{equation}
\hat{\mathcal{H}}_\perp=g_{y_0}\mu_B B\sum_i S_i^{y_0},
\label{eq:H_perp}
\end{equation} 
with the field $B$ along the high-symmetry $y_0 (b)$ axis. Using the classical energy consideration detailed in Appendix~\ref{A:SWExpansion} with exchanges and relevant g-tensor component from Eqs.~(\ref{eq:values_JNNlab}) and (\ref{eq:values_J2g}) gives the classical critical field for the transition to the paramagnetic phase
\begin{equation}
H_c=2S (J_{y_0y_0}\!-\!J_{z_0z_0}\!+\!J_2\!-\!J_{2z_0}),\ B_c\approx 8.8(1)\ \mathrm{T},
\label{eq:critical_field}
\end{equation} 
where the field in the energy units, $H\!=\!g_{y_0}\mu_B B$, is introduced. For the sake of the future discussion, we note that this critical field is considerably larger than the one found by the DMRG in the 1D model, $B_c^{\text{1D}}\!\approx \!4.5\ \mathrm{T}$, see Section~\ref{Sec:Results}, suggesting strong renormalization due to quantum effects.

For the spin-wave expansion in the paramagnetic phase, we perform a rotation from the laboratory $\{x_0,y_0,z_0\}$ to the local reference frame $\{x,y,z\}$, depicted in Fig.~\ref{f:chain}(c),  aligning the local quantization axis $z$ with the direction of the field. This leads to the cyclic permutation of the spin components
\begin{equation}
\left(S_i^{x_0}, S_i^{y_0}, S_i^{z_0} \right)_{\text{lab}} = \left(S_i^y, S_i^z, S_i^x \right)_{\text{loc}}.
\label{eq:localframe}
\end{equation}
After the transformation \eqref{eq:localframe}, it is convenient to divide  the Hamiltonian into two parts, referred to as the even and the odd, in order to separate even and odd powers of the bosonic operators in the subsequent spin bosonization. Using the Hamiltonian in Eqs.~(\ref{eq:spin_hamiltonian}), (\ref{eq:J1}),  (\ref{eq:J2}), and (\ref{eq:H_perp}), the even term reads   
\begin{align}
\hat{\mathcal{H}}_{\text{even}}& \!=\! \sum_{i}
\Big\{J_{z_0z_0}S_i^xS_{i+1}^x+J_{x_0x_0} S_i^yS_{i+1}^y +J_{y_0y_0} S_i^zS_{i+1}^z 
\nonumber \\ 
& \hspace{-.5cm} + J_{2z_0}S_i^xS_{i+2}^x + J_{2} \left(S_i^yS_{i+2}^y + S_i^zS_{i+2}^z\right)\! -\! H S_i^{z}\Big\},
\label{eq:Heven}
\end{align}
while the odd part is given by
\begin{eqnarray}
\hat{\mathcal{H}}_{\text{odd}} =J_{y_0z_0}\sum_{i}e^{i{\bf Q}{\bf r}_i}\left(S_i^x S_{i+1}^z + S_i^zS_{i+1}^x\right).
\label{eq:Hodd}
\end{eqnarray} 
In the latter,  the factor $e^{i{\bf Q}{\bf r}_i}\!=\!(-1)^i$ replicates the staggered structure of the $J_{y_0z_0}$-term, where ${\bf Q}\!=\!2\pi \hat{\textbf{{\sf c}}}/{\sf c}$ is the reciprocal lattice vector of the {\it zigzag} chain, ${\sf c}$ is its lattice constant, and $\hat{\textbf{{\sf c}}}$ is the unit vector along the $c$ axis in Fig.~\ref{f:chain}(a). As is discussed below, the relevant unit cell is smaller, with the  lattice constant ${\sf c}_0\!=\!{\sf c}/2$ and the reciprocal lattice vector  ${\bf G}\!=\!2{\bf Q}$.

Clearly, in the absence of the odd part \eqref{eq:Hodd}, there would be no memory of the zigzag structure left in the spin model, as the even part \eqref{eq:Heven} describes a ``simple'' Ising-like chain with the  transverse field term and second-neighbor exchanges, but no bond-dependent terms. Since \eqref{eq:Heven} yields the linear spin-wave theory, one can anticipate that it will have only a single bosonic branch, with no zone-folding  in the reciprocal space from the zigzag structure of CoNb$_2$O$_6$. 

On the other hand, the odd part of the model \eqref{eq:Hodd} arises precisely from such bond-dependent terms. However, in the paramagnetic phase, it contributes only to the nonlinear, anharmonic coupling of the spin flips, bringing about an important  ${\bf Q}$-shift of the two-magnon continuum that couples to the single-magnon branch. This  feature of the spin model of the zigzag chains in CoNb$_2$O$_6$ has been recognized and thoroughly discussed in Ref.~\cite{fava2020} as crucial for explaining puzzling kinematics of the observed magnon decays.

We also note that a similar structure of the theory was recently discussed in Ref.~\cite{Maksimov2022PRB} in the context of the easy-plane honeycomb-lattice model with bond-dependent exchanges, underscoring the  connection of the present consideration to a broader class of models and materials with spin-orbit-generated anisotropic exchanges.  

\vspace{-0.2cm}
\subsection{Linear spin-wave theory}
\label{Sec:LSWT}
\vskip -0.1cm

The harmonic, or linear spin-wave theory (LSWT) order of the $1/S$-expansion about the classical ground state is obtained via the standard Holstein-Primakoff (HP) bosonization of  spin operators in the local reference frame: $S_i^z\! = \!S-n_i$ and, to the lowest order,  $S_i^+ \!\approx \!\sqrt{2S} a_i$. 

In the field-polarized paramagnetic phase of CoNb$_2$O$_6$ considered here, it is only the even part of the Hamiltonian \eqref{eq:Heven} that contributes to LSWT. As is discussed above, this part of the Hamiltonian is  invariant to the translations by ${\sf c}_0\!=\!{\sf c}/2$, that is, half of the primitive lattice vector of the zigzag chain. In other words, spin states on all sites of the chain are equivalent and only one bosonic species needs to be introduced. Using the HP bosonization in \eqref{eq:Heven} and the standard  Fourier transformation 
\begin{equation}
a_i = \frac{1}{\sqrt{N}} \sum_{\bf k} e^{i{\bf k} {\bf r}_i }a_{\bf k},
\label{eq:fourier}
\end{equation}
where $N$ is the number of lattice sites in the chain, we obtain the LSWT Hamiltonian
\begin{equation}
\mathcal{\hat{H}}^{(2)} = \sum_{\bf k} \Big\{ A_{\bf k} a^\dagger_{\bf k} a^{\phantom\dag} _{\bf k}-\frac{B_{\bf k}}{2}\left(a^\dagger_{{\bf k} }a^\dagger_{-{\bf k}}+\text{H.c.} \right) \Big\},
\label{eq:H2k}
\end{equation}
where $A_{\bf k}$ and $B_{\bf k}$  are 
\begin{align}
A_{\bf k} =  H - 2S(J_{y_0y_0}+J_2) + SJ_{{\bf k}+},\ B_{\bf k}  = SJ_{{\bf k}-},
\label{eq:AkBk}
\end{align}
with 
\begin{align}
J_{{\bf k} \pm} = \left(J_{x_0x_0}\pm J_{z_0z_0}\right)\gamma_{\bf k}^{(1)} +\left(J_{2}\pm J_{2z_0}\right)\gamma_{\bf k}^{(2)},   
\label{eq:Jkpm}
\end{align}
and the  nearest- and next-nearest-neighbor hopping amplitudes $\gamma_{\bf k}^{(n)} \!=\! \cos (n k_c {\sf c}_0)$, where $n\!=\!1(2)$,  $k_c\!=\!{\bf k} \hat{\textbf{{\sf c}}}$ is a projection of the momentum ${\bf k}$ on the chain direction, and ${\sf c}_0\!=\!{\sf c}/2$, with ${\sf c}$ being the lattice constant of the zigzag chain, as before; see  Fig.~\ref{f:chain}(a) and Appendix~\ref{A:SWExpansion} for more details and Ref.~\cite{Cabrera2014} for similar expressions.

The LSWT Hamiltonian \eqref{eq:H2k} is diagonalized by a textbook Bogolyubov transformation, $a_{\bf k}\!=\!u_{\bf k} b_{\bf k}^{\phantom \dag}\!+v_{\bf k} b_{-{\bf k}}^\dag$, with  $u_{\bf k}^2+v_{\bf k}^2 \!= \!A_{\bf k}/\varepsilon_{\bf k}$,  $2u_{\bf k}v_{\bf k} \!=\! B_{\bf k}/\varepsilon_{\bf k}$, and magnon energy 
\begin{equation}
\varepsilon_{\bf k}=\sqrt{A_{\bf k}^2-B_{\bf k}^2}.
\label{eq:omega_onemagn}
\end{equation}
The excitation gap of the LSWT spectrum (\ref{eq:omega_onemagn}) at the $\boldsymbol{\Gamma}$ point $({\bf k}\!=\!0)$ is given by
\begin{equation}
\Delta_{0}=\sqrt{(H-H_c)(H+2S(J_{x_0x_0}-J_{y_0y_0}))},
\label{eq:EkGamma}
\end{equation}
vanishing at the critical field $H_c$ in (\ref{eq:critical_field}), as expected. Given the low spin-symmetry of the model, the spectrum in (\ref{eq:omega_onemagn}) has a relativistic form near $\boldsymbol{\Gamma}$,  with  $\varepsilon_{\bf k}\!\propto\!|{\bf k}|$ at $H\!=\!H_c$, the behavior that will be important for the unphysical divergences discussed below.

\begin{figure}[t]
\centering
\includegraphics[width=\linewidth]{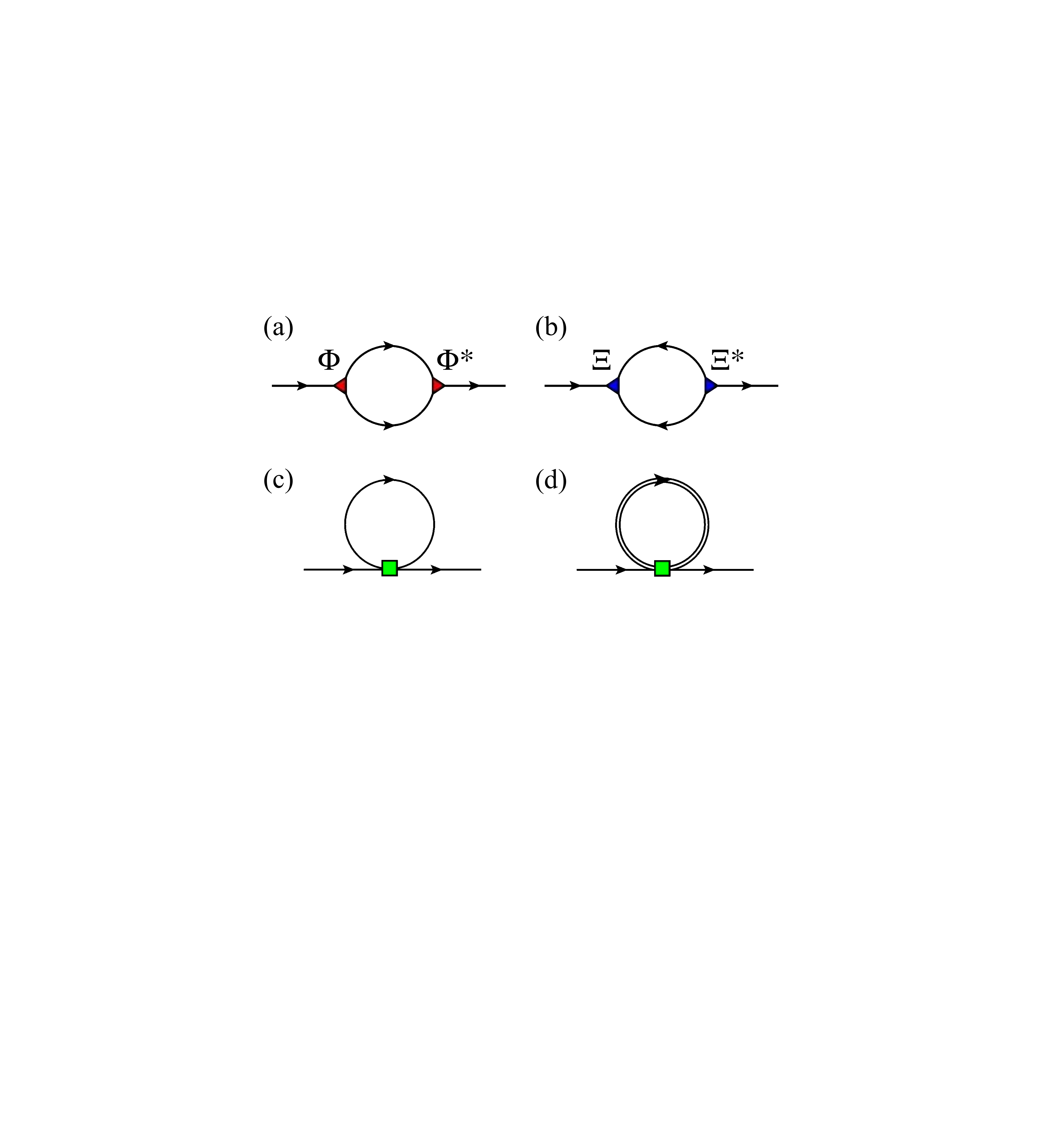}
\vskip -0.2cm
\caption{(a) Decay, (b) source, (c) Hartree-Fock, and (d) self-consistent Hartree-Fock self-energies.}
\label{f:Diagrams}
\vskip -0.4cm
\end{figure}

\vspace{-0.2cm}
\subsection{Non-linear spin-wave theory and divergences}
\label{Sec:NLSWT}
\vskip -0.1cm

The $1/S$-expansion of the Hamiltonian in \eqref{eq:Heven} and \eqref{eq:Hodd} beyond the LSWT yields two anharmonic terms, cubic and quartic,  describing three- and four-magnon interaction, respectively.  The cubic anharmonicity  comes from the odd part of the model \eqref{eq:Hodd} and  carries an important umklapp-like ${\bf Q}$-shift of the momentum in the one-to-two-magnon coupling, similar to the  other models with the staggered structure of the cubic terms studied in the past; see Refs.~\cite{ZhCh99,Mourigal2010,Maksimov2022PRB}. The quartic term is from the even part of the model \eqref{eq:Heven}, in which higher $1/S$-terms of the HP bosonization of spins are kept.  

Diagrammatically,  these interactions result in a loop expansion, with the lowest-order diagrams shown in  Figs.~\ref{f:Diagrams}(a)-(c). In a strict $1/S$ sense, their contributions  to the magnon excitation spectrum are of the same order. Three more  diagrams with the same number of loops, corresponding to the anomalous self-energies, are not shown as they yield corrections of the higher $1/S$-order \cite{ChZh_triPRB09}.  

Deferring some essential but technical details concerning three-magnon vertex symmetrization  to Appendix~\ref{A:SWExpansion}, the two self-energies in Figs.~\ref{f:Diagrams}(a) and \ref{f:Diagrams}(b)  are the decay and the source ones, respectively,  
\begin{align}
\Sigma^{(3)}({\bf k}, \omega)= \Sigma^{(d)}({\bf k},\omega)+\Sigma^{(s)}({\bf k},\omega)\,.
\label{eq:self_energy}
\end{align}
While both come from the same cubic anharmonicities, it is the decay diagram that is relevant to the description of some of the most dramatic modifications that may occur in the magnon excitation spectra, such as the anomalous broadening due to quasiparticle breakdown \cite{Maksimov2020RuCl3,Winter2017Nature,Zhitomirsky2013},  strong renormalization due to avoided decays \cite{Mourigal2013DSF,Coldea_tri_avoided20}, and threshold singularities \cite{pitaevskii1959,Zhitomirsky2013}. These effects occur when the  single-particle and two-particle spectra overlap, with the lower dimensions of the spin system \cite{Verresen_avoided_19}, symmetry of the spin model \cite{Winter2017Nature}, and favorable kinematics \cite{Chernyshev2016DampedKagome,Zhitomirsky2013,ChernyshevKagome15} all playing a significant role in the resultant magnitude of these effects.

All these phenomena manifest themselves quite spectacularly in the CoNb$_2$O$_6$ excitation spectrum in the transverse-field-induced polarized phase  \cite{Robinson2014,fava2020}, owing to the 1D nature of the zigzag chains, low spin-symmetry leading to a direct one-to-two-magnon coupling  (\ref{eq:Hodd}), and  a favorable overlap with the ${\bf Q}$-shifted two-magnon continuum, also allowing for the field-variation of it,  the features thoroughly discussed in Ref.~\cite{fava2020}.

Therefore,   analytical insights by the $1/S$ nonlinear SWT (NLSWT) into the magnon interactions can be expected to shed further light on the important aspects of the decays, level repulsion, and singularities in the excitation spectra of CoNb$_2$O$_6$,  related zigzag chain materials, and other anisotropic-exchange magnets. However, this expectation is undermined by the unphysical divergences in NLSWT at the critical field, characteristic to  anisotropic models \cite{Maksimov2022PRB, Consoli2020JK1Sexpansion}.

The problem can be seen in the strict $1/S$-expansion for the magnon spectrum, in which corrections to the LSWT energy  (\ref{eq:omega_onemagn}) are given by the on-shell ($\omega\!=\!\varepsilon_{\bf k}$) self-energies from Figs.~\ref{f:Diagrams}(a)-(c)
\begin{eqnarray}
&&\tilde{\varepsilon}_{\bf k}=\varepsilon_{\bf k}+\delta \varepsilon_{\bf k}^{(3)}+\delta \varepsilon_{\bf k}^{(4)}, \quad \Gamma_{\bf k}=-\text{Im} \big[\Sigma^{(3)}({\bf k}, \varepsilon_{\bf k})\big],\quad
\nonumber\\
&& 
\delta \varepsilon_{\bf k}^{(3)}=\text{Re} \big[\Sigma^{(3)}({\bf k}, \varepsilon_{\bf k})\big], \quad \delta \varepsilon_{\bf k}^{(4)}=\Sigma^{HF}({\bf k}), 
\label{eq:NLSWT_Ek}
\end{eqnarray}
where $\tilde{\varepsilon}_{\bf k}$ is the renormalized spectrum, $\Gamma_{\bf k}$ is the decay-induced  broadening, $\Sigma^{(3)}({\bf k}, \omega)$ from  (\ref{eq:self_energy}) is discussed above, and the $\omega$-independent Hartree-Fock self-energy $\Sigma^{HF}({\bf k})$ is shown Fig.~\ref{f:Diagrams}(c), see Appendix~\ref{A:NLSWT}.

Our Figure~\ref{f:1S_Spectrum} shows the NLSWT result of such a  $1/S$-renormalization of the magnon spectrum (\ref{eq:NLSWT_Ek}), calculated using the best-fit model for CoNb$_2$O$_6$ from (\ref{eq:values_JNNlab}) and (\ref{eq:values_J2g}) and  for the field just above the classical  value of the critical one in (\ref{eq:critical_field}), $H\!=\!1.01H_c$, all for $S\!=\!1/2$. An artificial broadening of $10^{-3}$~meV was used in calculating $\Sigma^{(3)}({\bf k}, \varepsilon_{\bf k})$ \eqref{eq:self_energy}. Also shown are the LSWT single-magnon branch (\ref{eq:omega_onemagn}) together with the ${\bf Q}$-shifted two-magnon LSWT continuum, black dashed line and  the shaded area, respectively.

As is expected, significant singular modifications of the spectrum close to the decay threshold boundaries, which correspond to the crossing of the single-magnon branch with the edges of the two-magnon continuum, are present in the NLSWT spectrum in Fig.~\ref{f:1S_Spectrum}. The  decay-induced scattering rate  $\Gamma_{\bf k}$, divergent at the same thresholds ${\bf k}^*$ and ${\bf G}-{\bf k}^*$, with ${\bf G}\!=\!2{\bf Q}$, is also shown. This is in accord with the similarly stark modifications of the magnon spectra in a variety of other models \cite{ChZh_triPRB09,Mourigal2010}. Not only do they demonstrate the anomalous broadening and strong  repulsion of the single-magnon spectrum from the two-magnon continuum within the limited capacity of the na\"{i}ve perturbation theory, but they also signify a breakdown of the $1/S$-expansion in the vicinity of the decay thresholds and call for a more consistent treatment of these effects, going beyond the $1/S$-approximation to regularize the associated divergences \cite{ChZh_triPRB09,Zhitomirsky2013, Winter2017Nature, Chernyshev2016DampedKagome}. We offer further analysis of these {\it physical} threshold singularities in Sec.~\ref{Sec:Results}.

\begin{figure}[t]
\centering
\includegraphics[width=\linewidth]{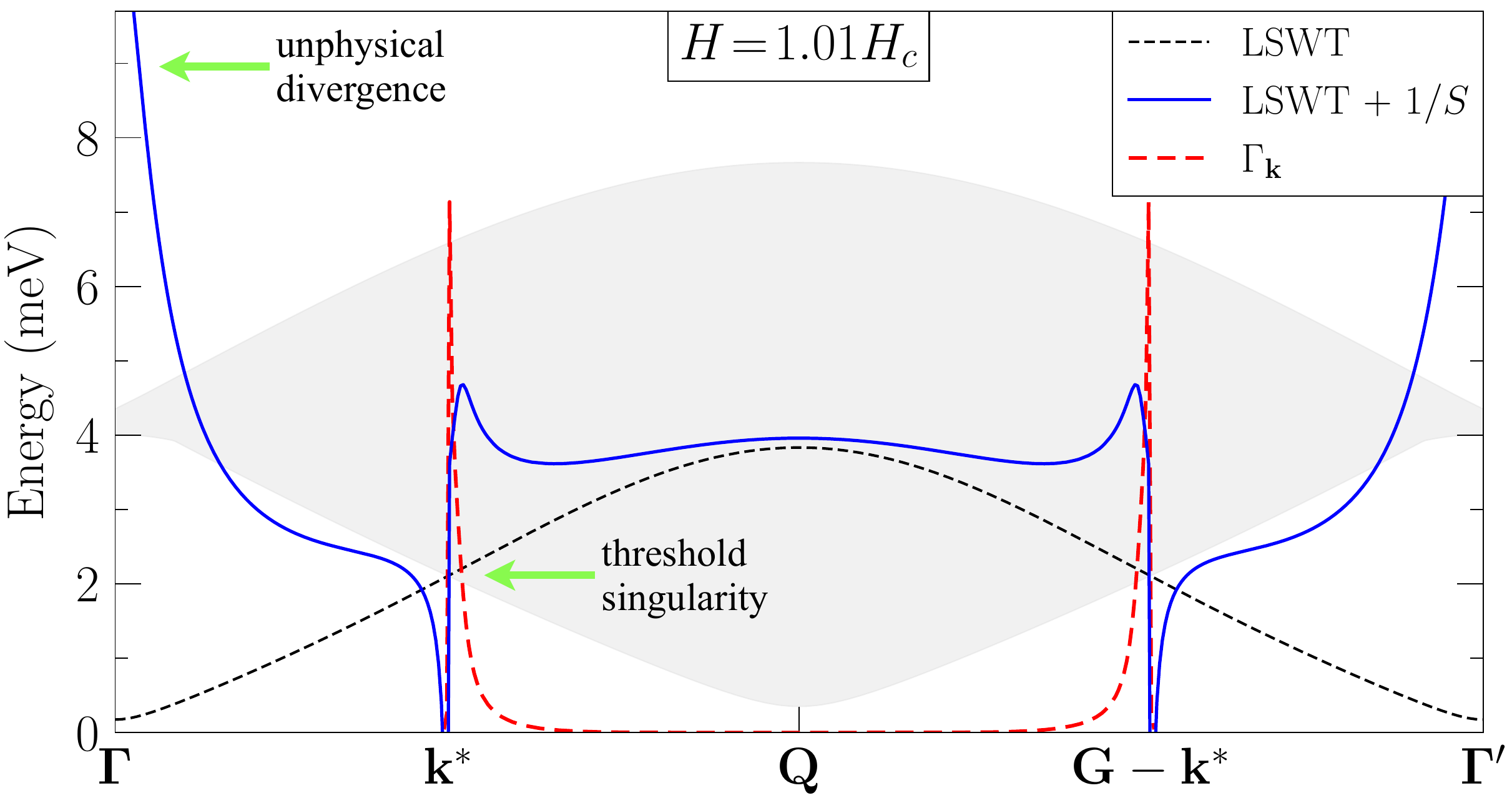} 
\vskip -0.2cm
\caption{Magnon spectrum in the LSWT (\ref{eq:omega_onemagn}) and NLSWT (\ref{eq:NLSWT_Ek}) $1/S$-approximations, two magnon continuum (shaded area), and $1/S$ decay rate $\Gamma_{\bf k}$ for $H\!=\!1.01H_c$, best-fit parameters for CoNb$_2$O$_6$, and $S\!=\!1/2$. Arrows indicate the unphysical divergences and physical threshold singularities, see the text.}
\label{f:1S_Spectrum}
\vskip -0.5cm
\end{figure}

However, this discussion of the physical aspects of the interacting magnon spectrum is completely undermined, as the results in Fig.~\ref{f:1S_Spectrum} are dominated instead by the nearly divergent behavior of the spectrum near the $\boldsymbol{\Gamma}$ point, which is far away from the decay thresholds. Therefore, it should not be affected by the anharmonicities and should correspond to a minimum of the magnon mode in the ferromagnetically-dominated model.

This behavior is due to the $1/S$-expansion, which can also be seen as an expansion in  $1/\varepsilon_{\bf k}$. Because of the closing of the excitation gap in (\ref{eq:EkGamma}), the $1/S$-corrections to the  magnon energy in (\ref{eq:NLSWT_Ek})  diverge as $1/|{\bf k}|$  at $H\!\rightarrow\!H_c$. 

While clearly unphysical, this failure of the NLSWT in the proximity of the field-induced transition with vanishing excitation gap is not unexpected, as it is characteristic of the  models with the lower spin symmetry, such as anisotropic-exchange ones  \cite{Maksimov2022PRB, Consoli2020JK1Sexpansion}. 

In order to analyze the technical anatomy of this failure, it is instructive to consider the Hartree-Fock self-energy in Fig.~\ref{f:Diagrams}(c). To derive it  from the quartic terms in (\ref{eq:Heven}), one decouples the four-boson combinations from the $1/S$-expansion down to the two-boson ones using the real-space HF averages  $\{\langle a^\dag_i a_i\rangle,\langle a^\dag_i a_j\rangle,\langle a^\dag_i a_j^\dag\rangle,\dots\}$, which can be straightforwardly evaluated from the Bogolyubov parameters of the LSWT; see Appendix~\ref{A:SWExpansion} for the explicit expressions and technical steps \cite{Mourigal2010, ChZh_triPRB09}.  

As a result, the quartic terms are reduced to the LSWT form of  Eq.~(\ref{eq:H2k}), only with the $1/S$-corrections $\delta A_{\bf k}$ and $\delta B_{\bf k}$ instead of the $A_{\bf k}$ and $B_{\bf k}$ terms in (\ref{eq:AkBk}). Then, the $1/S$-expansion yields
\begin{equation}
    \delta \varepsilon_{\bf k}^{(4)}= \Sigma^{HF}({\bf k})=\frac{A_{\bf k} \delta A_{\bf k} - B_{\bf k} \delta B_{\bf k}}{\varepsilon_{\bf k}},
\label{eq:dEk4}
\end{equation}
which is explicitly divergent  in $1/\varepsilon_{\bf k}$  at $H_c$ due to  the gapless  mode at the $\boldsymbol{\Gamma}$ point. One should note that in the highly-symmetric spin-isotropic models such an expansion is benign, because $\delta A_{\bf k}$ and $\delta B_{\bf k}$ follow the same ${\bf k}$-dependence as the LSWT $A_{\bf k}$ and $B_{\bf k}$ terms, canceling the divergence for the vanishing $\varepsilon_{\bf k}$ \cite{Zhitomirsky2022SCfccAFM,Gochev95,Yang97}.

It is important to observe that the strong divergence in Eq.~\eqref{eq:dEk4} originates from the strict use of the $1/S$-approximation that can be straightforwardly avoided by replacing $A_{\bf k}\!\rightarrow\! A_{\bf k}+\delta A_{\bf k}$ and $B_{\bf k} \!\rightarrow\!B_{\bf k}+\delta B_{\bf k}$ in the renormalized spectrum, an approach used in a variety of models \cite{Maksimov2022PRB, Chubukov_1994}.  In our case, the solution is more subtle, first because of the cubic terms, but also because of the 1D character of the problem, which leads to the logarithmically divergent real-space HF averages for the gapless spectrum. Nevertheless, such an approach hints at the self-consistent regularization scheme, which does not only remove the singularity at $H\!\rightarrow\!H_c$, but also allows us to access the field range that is inaccessible to the standard spin-wave theory.  This method is discussed next.

\vspace{-0.2cm}
\subsection{Self-consistent Hartree-Fock method} 
\label{Sec:SCHF}
\vskip -0.1cm

For the  CoNb$_2$O$_6$ model discussed in this work, there is a clear hierarchy of the exchange terms, with the dominant Ising exchange $J_{z_0z_0}$; see Eq.~(\ref{eq:values_JNNlab}). Moreover, since the staggered $J_{y_0z_0}$ term enters only via the higher-order anharmonic coupling, one can expect that its contribution to the magnon spectrum away from the threshold singularities is perturbatively small, $\Sigma^{(3)}({\bf k}, \omega)\!\sim\!{\cal O}(J_{y_0z_0}^2/J_{z_0z_0})$, while the role of the problematic correction from the quartic terms is $\Sigma^{HF}({\bf k})\!\sim\!{\cal O}(J_{z_0z_0})$. This suggests the following two-step regularization procedure.

At the first stage, we neglect the contribution of the cubic terms and perform a self-consistent calculation of the renormalized eigenvalues $\bar{\varepsilon}_{\bf k}$ and eigenstates $\bar{u}_{\bf k}$ and $\bar{v}_{\bf k}$ of the SWT using an iterative procedure with the quartic-term contribution, referred to as the self-consistent Hartree-Fock (SCHF) method. It goes beyond the standard SWT by combining different orders in $1/S$ \cite{Rau2023GoldstoneSC, Loly1971,Zhitomirsky2022SCfccAFM}, as is depicted in Fig.~\ref{f:Diagrams}(d), which emphasizes the self-consistency in the inner line of the HF self-energy. The self-consistency loop is depicted below 
\begin{align}
\begin{array}{ccc}
\{\mbox{HFs}\}&\Longrightarrow&\{\delta \bar{A}_{\bf k},\delta \bar{B}_{\bf k}\}\\
\Uparrow&&\Downarrow\\
\{\bar{\varepsilon}_{\bf k},\bar{u}_{\bf k},\bar{v}_{\bf k}\}&\Longleftarrow&\{\bar{A}_{\bf k},\bar{B}_{\bf k}\}
\end{array}
\label{eq:SCHF_diagram}
\end{align}
The set of the real-space HF averages,  denoted as $\{\mbox{HFs}\}$, is used to obtain the quartic-term contributions to the harmonic theory, $\delta \bar{A}_{\bf k}$ and $\delta \bar{B}_{\bf k}$, as is described in Appendix~\ref{A:NLSWT}. They result in the modified, but an LSWT-like eigenvalue problem of the same form as in Eq.~(\ref{eq:H2k}) with $\bar{A}_{\bf k}\!=\! A_{\bf k}+\delta \bar{A}_{\bf k}$ and $\bar{B}_{\bf k}\!=\! B_{\bf k}+\delta \bar{B}_{\bf k}$, which, in turn, leads to the new set of the energies  $\bar{\varepsilon}_{\bf k}$ and Bogolyubov parameters $\bar{u}_{\bf k}$ and $\bar{v}_{\bf k}$, with the latter used as an input for the HF averages. The cycle is continued until numerical convergence in the HF averages is reached.

With the additional important technical details and step-by-step implementation described in Appendix~\ref{A:SCHF}, the following point should be made. Since the described approach regularizes the energy gap in the magnon spectrum, the gap remains finite at the nominal LSWT critical field $H_c$. Because of that, the SCHF method also allows us to extend our study to the field values below  $H_c$, the feat which is  unattainable by the standard $1/S$ SWT approaches. For that, we perform the SCHF calculations at a field $H\!>\!H_c$, and then use the outcome for the converged HF averages at a higher field as an input for the next  SCHF calculation for a continuously decreasing field. The stability and consistency of this procedure is verified by varying the initial field,  the step in the field decrease, and other iterative parameters; see Appendix~\ref{A:SCHF}.

Finally, the magnon energy spectrum is obtained by reinstating the cubic terms \eqref{eq:self_energy}. Importantly, all Bogolyubov coefficients  that enter decay and source vertices as well as the magnon energies in the loops of the diagrams in Figs.~\ref{f:Diagrams}(a) and \ref{f:Diagrams}(b) (see Appendix~\ref{A:NLSWT}) are replaced with their regularized values obtained within the SCHF method described above. Then, the regularized on-shell ($\omega\!=\!\bar{\varepsilon}_{\bf k}$) magnon energy is  given by 
\begin{align}
& \tilde{\varepsilon}_{\bf k}=\bar{\varepsilon}_{\bf k}+\delta \bar{\varepsilon}_{\bf k}^{(3)}, 
\label{eq:SCSWT_Ek}\\
& \delta \bar{\varepsilon}_{\bf k}^{(3)}=\text{Re} \big[\Sigma^{(3)}({\bf k}, \bar{\varepsilon}_{\bf k})\big], \quad \bar{\Gamma}_{\bf k}=-\text{Im} \big[\Sigma^{(3)}({\bf k}, \bar{\varepsilon}_{\bf k})\big].
\nonumber
\end{align}
The success of the described regularization procedure in removing the unphysical singularity at the critical field and in describing the physical spectrum of CoNb$_2$O$_6$  is demonstrated in the next Section. We also note that in the results for the dynamical structure factor discussed below, the full $\omega$-dependence of the cubic self-energies in \eqref{eq:self_energy} is used. In the following, we refer to the method outlined here as the SCHF$+\Sigma^{(3)}$.

\section{Results}
\label{Sec:Results}

In this Section we present the outcome of the self-consistent method advocated above, demonstrating its power in regularizing the unphysical divergences and ability to extend the theory beyond the restrictive classical boundaries. We also compare our results for the  dynamical structure factor in the field-induced paramagnetic phase with the inelastic neutron scattering data of Refs.~\cite{Robinson2014, fava2020}.

\subsection{Regularization of the unphysical divergences}

\begin{figure}[t!]
\centering
\includegraphics[width=\linewidth]{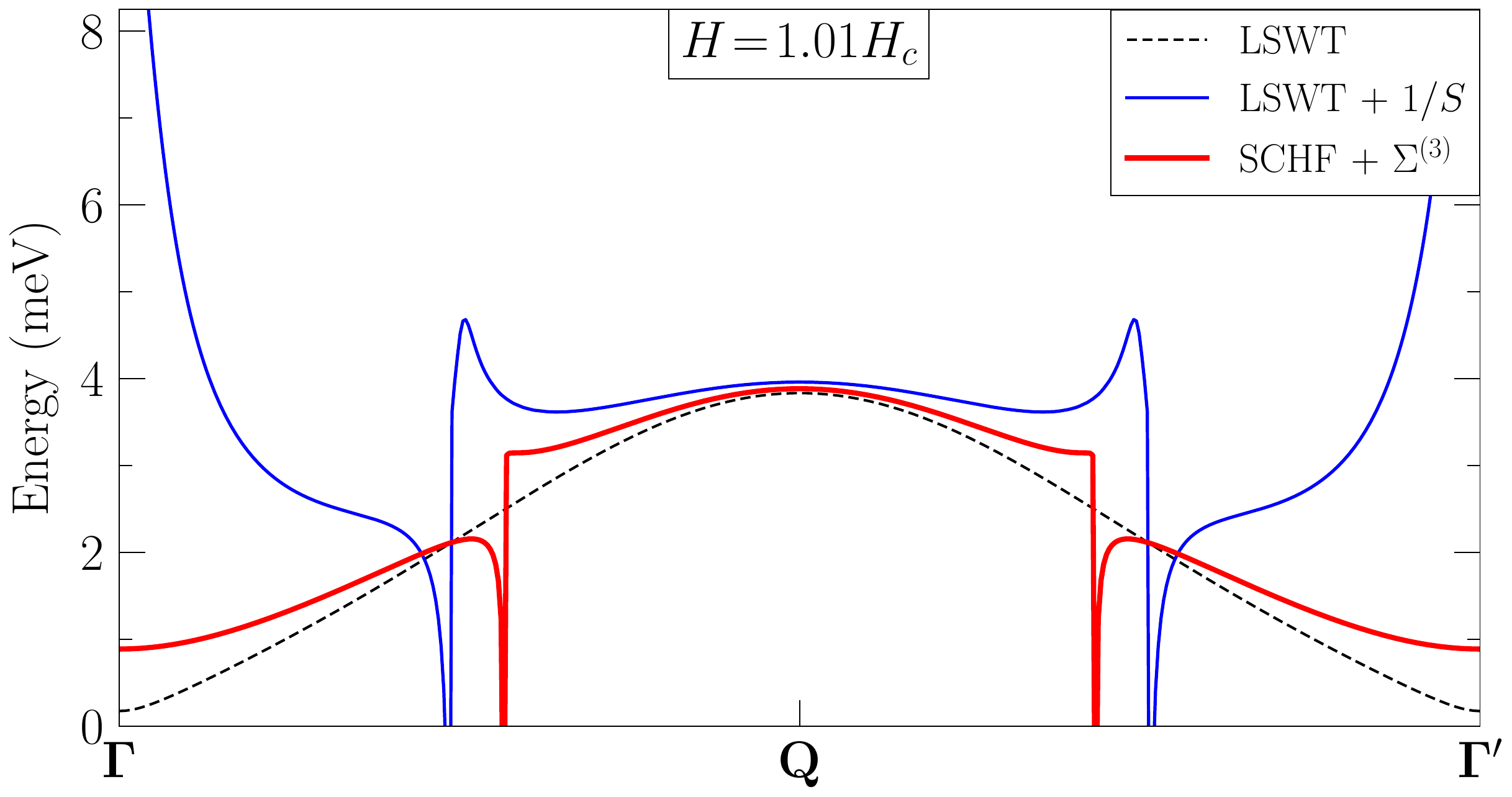}
\vskip -0.2cm
\caption{Same as in Fig.~\ref{f:1S_Spectrum}.  LSWT \eqref{eq:omega_onemagn}, NLSWT \eqref{eq:NLSWT_Ek}, and SCHF$+\Sigma^{(3)}$ \eqref{eq:SCSWT_Ek} magnon spectra for $H=1.01H_c$.}
\vskip -0.5cm
\label{f:SCHF_101Hc}
\end{figure}

The success of our method \eqref{eq:SCSWT_Ek} in regularizing the unphysical divergences discussed in Sec.~\ref{Sec:NLSWT} is demonstrated in Fig.~\ref{f:SCHF_101Hc}, where the  magnon energy spectrum  by the SCHF$+\Sigma^{(3)}$ for $H\!=\!1.01H_c$ and the best-fit model of CoNb$_2$O$_6$  is shown together with the LSWT \eqref{eq:omega_onemagn} and NLSWT \eqref{eq:NLSWT_Ek} results  from Fig.~\ref{f:1S_Spectrum}.  In the regularized spectrum   in Fig.~\ref{f:SCHF_101Hc}, the offending divergent behavior of the NLSWT near the ${\bm \Gamma}$ point is  gone, and one is able to focus on the physical effects of magnon interaction in the decay-related phenomena.

The second achievement is the following. The 1D critical field calculated by DMRG for the CoNb$_2$O$_6$ model, $B_c^{\text{1D}}\! \approx \! 4.52(1)\ \mathrm{T}$, is close to the experimentally estimated one, $B_{c,\text{exp}}^{1D}\approx\! 5 \ \mathrm{T}$~\cite{coldea2010}, both much smaller than the classical critical field, $B_c^{\text{cl}}\! \approx \! 8.8\ \mathrm{T}$, obtained for the best-fit parameters in Eq.~\eqref{eq:critical_field}. For $B\!< \!B_c^{\text{cl}}$, the paramagnetic phase is not a minimum of the classical energy and cannot be studied by means of the $1/S$-expansion, because the LSWT Hamiltonian \eqref{eq:H2k} is not positive-definite, with its spectrum \eqref{eq:omega_onemagn} becoming imaginary in some regions of ${\bf k}$.  

\begin{figure}[t]
\centering
\includegraphics[width=\linewidth]{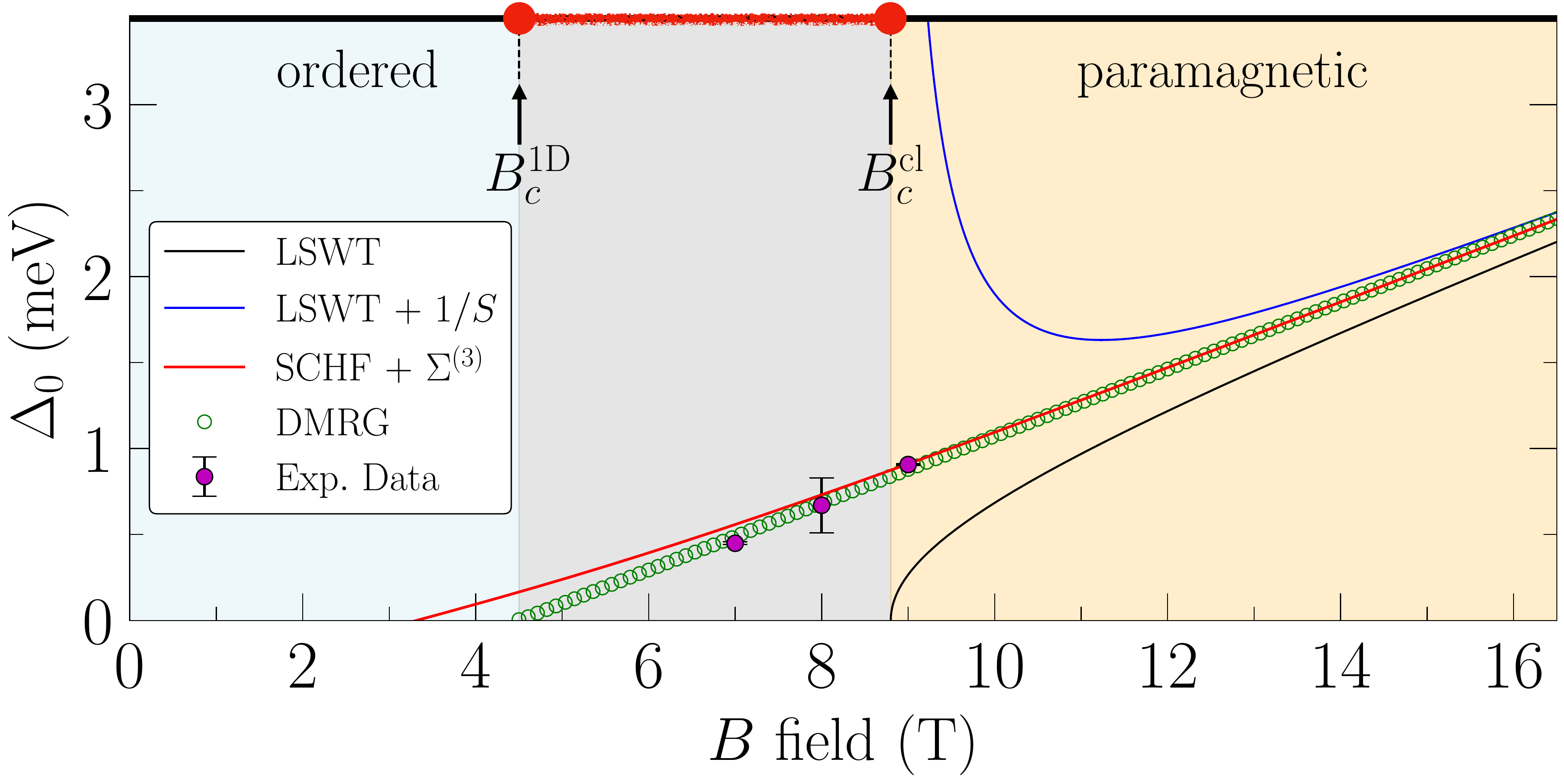}
\vskip -0.2cm
\caption{Magnon gap $\Delta_{0}$ vs $B$ by LSWT \eqref{eq:omega_onemagn}, NLSWT \eqref{eq:NLSWT_Ek}, SCHF$+\Sigma^{(3)}$ \eqref{eq:SCSWT_Ek}, and DMRG for the best-fit model of CoNb$_2$O$_6$, compared to experimental gaps (solid points) \cite{footnote}. The DMRG and classical critical fields, $B_c^{1D}$ and $B_c^{\text{cl}}$, and the region inaccessible by standard SWT (gray shaded area) are highlighted, see text.}
\label{f:SC_Gap}
\vskip -.5cm
\end{figure}

Our Fig.~\ref{f:SC_Gap} shows the magnon excitation gap $\Delta_0$ at the ${\bm \Gamma}$ point  as a function of the field for the best-fit model of CoNb$_2$O$_6$ obtained by different methods. The vanishing of this gap corresponds to a phase transition from the paramagnetic to the ordered phase at $T\!=\!0$.  The red horizontal line on top of the figure and the gray shaded area emphasize the difference between the  classical and experimental results for it. According to the LSWT,  the gap vanishes at the classical critical field \eqref{eq:critical_field}, and it diverges in the NLSWT $1/S$-approximation. The SCHF$+\Sigma^{(3)}$ method regularizes this divergence at $B_c^{\text{cl}}$ and allows us to extend the study of the magnon  spectrum into the field region below the classical boundary to the paramagnetic phase. These results also shows an excellent agreement with the experimental data for the gap at 7, 8, and 9 $\mathrm{T}$ \cite{footnote}, highlighting the quantitative accuracy of our approach.

Last but not the least,  Fig.~\ref{f:SC_Gap}  shows the results of the DMRG simulations for the gap in the same model, which agree  closely with both  experimental data and  results of the self-consistent theory, except for the close proximity of the critical field. With the details of the DMRG calculations deferred to Appendix~\ref{A:Results}, it should be noted that the DMRG critical field for the single-chain 1D model is $B_c^{\text{1D}}\! = \! 4.52(1)\ \mathrm{T}$, also below the experimental value. This is because the 3D interchain terms  play important role near the transition; see Refs.~\cite{Woodland2023Parameters, Woodland2023_DWsolitons}. 

Deviation of the SCHF$+\Sigma^{(3)}$  from the DMRG results is of a different, but related nature. While remarkably successful otherwise, the self-consistent method is not entirely consistent for the gap approaching zero. The SCHF method naturally prevents the gap from closing, because the HF averages would diverge logarithmically for the gapless spectrum due to the 1D character of the model. The finite critical field in the SCHF$+\Sigma^{(3)}$ method is due to the non-self-consistent perturbative treatment of the cubic term $\Sigma^{(3)}$, and the value of the critical field it yields at approximately $3.4\ \mathrm{T}$ is not physically meaningful. 

Notwithstanding these minor limitations and concerns, one should not lose the sight offered by Fig.~\ref{f:SC_Gap}, which demonstrates the ability of our approach to provide quantitatively meaningful description of the magnetic excitations for a wide field range in the paramagnetic phase of CoNb$_2$O$_6$. This is  despite strong quantum fluctuations, anisotropic exchanges, and low dimensionality of the problem, the factors that make the standard SWT fail.  Not only does the SCHF$+\Sigma^{(3)}$ method regularize the unphysical divergences, but it also preserves the physical features of the threshold singularities, which are  discussed next.

\vspace{-0.2cm}
\subsection{Decay threshold singularities}
\label{Sec:Decay_thresholds}
\vskip -0.1cm

While the main results and comparison with the experimental data for the dynamical structure factor will be discussed in the next section, we would like to briefly recall the origin and the nature of the decay-related phenomena in the magnon spectra; see also Refs.~\cite{ZhCh99,Zhitomirsky2013,ChZh_triPRB09}. 

\begin{figure}[t]
\centering
\includegraphics[width=\linewidth]{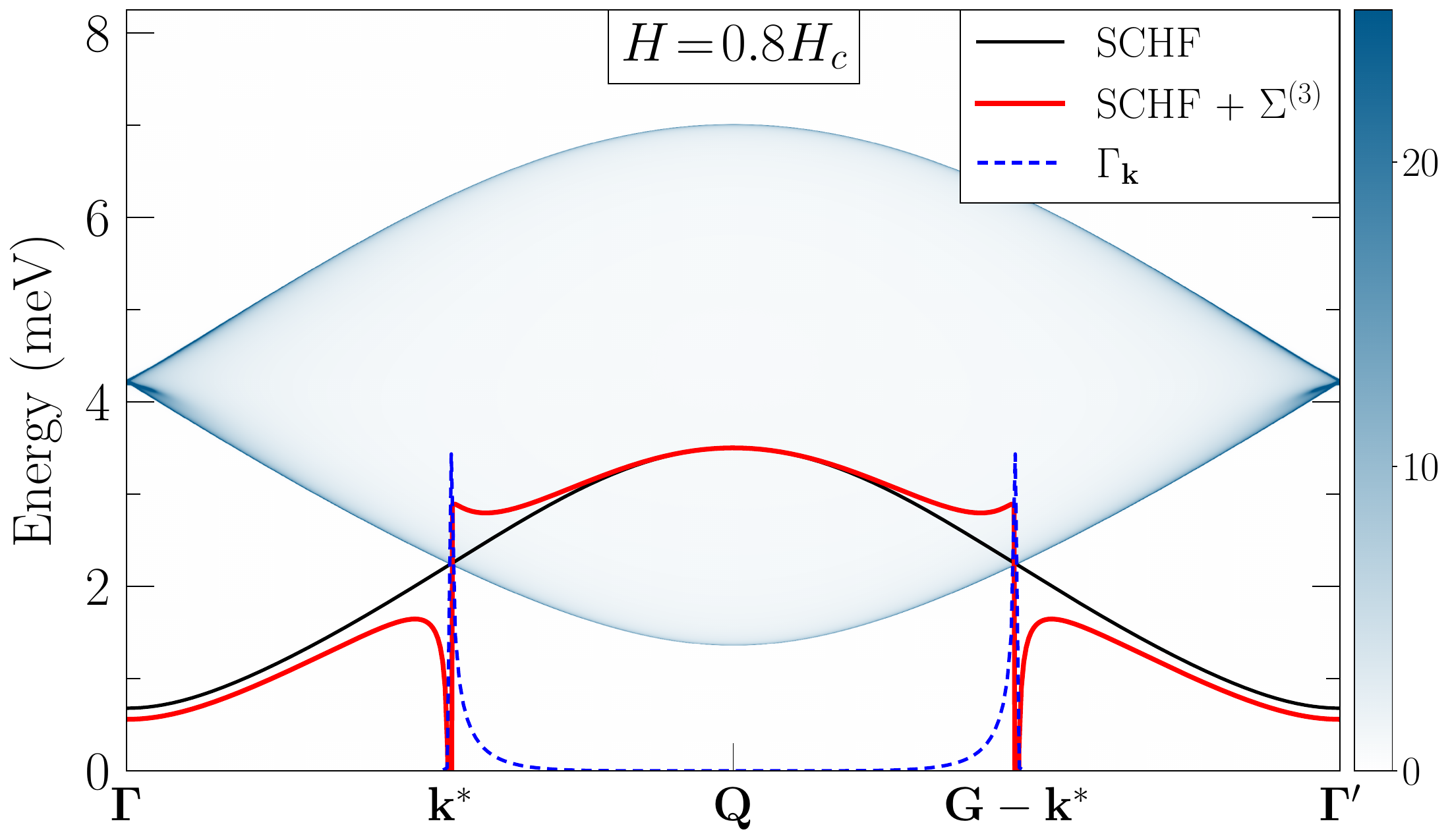}
\vskip -0.3cm
\caption{Magnon spectrum by the SCHF \eqref{eq:SCSWT_Ek} with and without the cubic self-energy $\Sigma^{(3)}({\bf k}, \bar{\varepsilon}_{\bf k})$ for the best-fit model of CoNb$_2$O$_6$ and $B\!=\!7$~T ($H\!=\!0.8H_c$). The intensity plot is the two-magnon DoS calculated using the SCHF energies $\bar{\varepsilon}_{\bf q}$.}
\vskip -0.5cm
\label{f:SCHF_08Hc}
\end{figure}

In Fig.~\ref{f:SCHF_08Hc} we show the magnon spectrum $\bar{\varepsilon}_{\bf k}$ obtained by SCHF method discussed in Sec.~\ref{Sec:SCHF}, together with $\tilde{\varepsilon}_{\bf k}$ from Eq.~\eqref{eq:SCSWT_Ek}  that includes contribution of  the on-shell cubic self-energy $\Sigma^{(3)}({\bf k}, \bar{\varepsilon}_{\bf k})$, for the best-fit model of CoNb$_2$O$_6$ and $H\!=\!0.8H_c$ ($B\!\approx\!7$~T), well below the classical critical field $H_c$. The two-magnon density of states (DoS), $D^{(2)}({\bf k}, \omega)\!=\!\frac{1}{N}\sum_{\bf q} \delta(\omega-\bar{\varepsilon}_{{\bf q}}-\bar{\varepsilon}_{{\bf k}-{\bf q}+{\bf Q}})$, for the SCHF energies $\bar{\varepsilon}_{\bf q}$, is shown as an intensity plot. 

As one can see in Fig.~\ref{f:SCHF_08Hc}, the contribution of the cubic term $\Sigma^{(3)}({\bf k}, \bar{\varepsilon}_{\bf k})$ to the magnon spectrum away from the crossings with the two-magnon continuum is indeed small, as is anticipated in the discussion of the SCHF$+\Sigma^{(3)}$ approach in Sec.~\ref{Sec:SCHF}. In fact,  the cubic term vanishes entirely at ${\bf k} \!= \!{\bf Q}$, owing to the staggered structure of the corresponding spin-exchange terms (\ref{eq:Hodd}), which, in turn, is translated into the antisymmetric structure of the cubic vertices, see Appendix~\ref{A:NLSWT}. 

The divergent behavior exhibited by the on-shell SCHF$+\Sigma^{(3)}$ spectrum near the crossing with the  two-magnon continuum at ${\bf k}^*$ and equivalent points, referred to as the decay threshold boundaries~\cite{Zhitomirsky2013},  is due to a resonance-like coupling of the single-magnon branch with the  two-magnon continuum, which is provided by the cubic terms. Since the lowest two-magnon energy must necessarily correspond to a minimum of $E_{{\bf k}, {\bf q}}\!=\!\bar{\varepsilon}_{\bf q}+\bar{\varepsilon}_{{\bf k}-{\bf q}+{\bf Q}}$ at any given ${\bf k}$, it follows that the corresponding two-magnon DoS must be singular at that minimum in 1D, as one can observe in Fig.~\ref{f:SCHF_08Hc}; see also Appendix~\ref{A:Kinematics}. It also follows that one can expand  the denominator of the decay part of the on-shell self-energy, $\bar{\varepsilon}_{\bf k}-E_{{\bf k}, {\bf q}}$,  near such a minimum in the proximity of the threshold ${\bf k}^*$ for small $\Delta {\bf k} \!=\!{\bf k}- {\bf k}^*$. Because the decay vertex has no symmetry constraints at a generic ${\bf k}^*$ and must generally be finite, one can obtain the asymptotic behavior for the real and imaginary parts of the on-shell self-energy on the two sides of the threshold
\begin{align}
&\text{Re} \big[\Sigma^{(3)}({\bf k}, \bar{\varepsilon}_{\bf k})\big] \propto \begin{cases}
-1/\sqrt{|\Delta {\bf k}|}, & \text{for}\ \Delta {\bf k} <0, \\
\Lambda+\gamma \Delta {\bf k}, & \text{for}\ \Delta {\bf k} >0, \end{cases} \nonumber
\\
& \phantom{\text{Re} \big[\Sigma^{(3)}({\bf k}, \bar{\varepsilon}\,}\bar{\Gamma}_{\bf k}\propto \begin{cases}
0, & \text{for}\ \Delta {\bf k} <0, \\
1/\sqrt{\Delta {\bf k}}, & \text{for}\ \Delta {\bf k} >0, \ \ \ \ \ 
\end{cases}
\label{Sigma3_Threshold}
\end{align}
where $\Lambda\!>\!0$ is the cut-off parameter and $\gamma$ is a constant. The inverse square-root singularities in (\ref{Sigma3_Threshold}) are from the 1D Van Hove singularity at the edge of the two-magnon continuum that gets imprinted on the single-magnon spectrum via the anharmonic coupling. These asymptotic results explain the behavior observed in Fig.~\ref{f:SCHF_08Hc} and should be contrasted with a typically weaker singularities in the higher dimensions and for the more symmetric models~\cite{Zhitomirsky2013, ChZh_triPRB09}.

Note that, given the relative simplicity of the magnon dispersion in the paramagnetic phase of CoNb$_2$O$_6$, the two-magnon energy, $\bar{\varepsilon}_{\bf q}+\bar{\varepsilon}_{{\bf k}-{\bf q}+{\bf Q}}$, can be well-approximated as an energy of two particles with the nearest-neighbor hopping, $\bar{\varepsilon}_{\bf q}\!\approx\!{\sf E_0}+{\sf J}_1\gamma_{\bf q}^{(1)}$, total momentum ${\bf k}$, and the ${\bf Q}$ shift, straightforwardly yielding the bow-tie form of the continuum with zero width at the ${\bm \Gamma}$ point, see Fig.~\ref{f:SCHF_08Hc}. In that case, the minimum of the  two-magnon energy  corresponds to the energy of two  magnons with equivalent momenta, $E^{\rm min}_{{\bf k}, {\bf q}^*}\!=\!2\bar{\varepsilon}_{{\bf q}^*}$ with ${\bf q}^*\!=\!({\bf k}-{\bf Q})/2$,  for any ${\bf k}$. This latter condition also holds for most ${\bf k}$ for the true form of $\bar{\varepsilon}_{\bf q}$, see Appendix~\ref{A:Kinematics}.

However, because of the further-neighbor exchanges  and a relativistic form of the magnon dispersion, there is more structure in the two-magnon continuum in the vicinity of the ${\bm \Gamma}$ point, providing the bow-tie region with a finite width and a richer set of the Van Hove singularities visible in  Fig.~\ref{f:SCHF_08Hc}. Since they are far away from the physical decay thresholds, we refrain from discussing them here and delegate a more detailed analysis of the two-magnon kinematics to Appendix~\ref{A:Kinematics}.

As is discussed above, the perturbative consideration of the decay diagram within the on-shell approach  offered by Fig.~\ref{f:SCHF_08Hc}, Eq.~(\ref{eq:SCSWT_Ek}), and Eq.~(\ref{Sigma3_Threshold}) signifies a breakdown of the perturbation theory in the vicinity of the decay thresholds and calls for a more consistent treatment of these effects to regularize the associated divergences. Qualitatively, upon a self-consistent treatment of the higher-order contributions, which  is typically difficult to implement, these singularities are expected to lead to the anomalous broadening, renormalization of the magnon spectrum, and the so-called termination points~\cite{pitaevskii1959,ChZh_triPRB09}. 

One of the less-sophisticated regularizations, which, nevertheless, avoids the divergences of the on-shell approach, is based on a straightforward use of the explicit $\omega$-dependence in the cubic self-energy in (\ref{eq:self_energy}). It corresponds to the one-loop approximation  for the single-magnon Green's function $G({\bf k}, \omega)$ and its spectral function $A({\bf k}, \omega)$, which is able to yield  the quantitatively faithful description of the quasiparticle-like and incoherent parts of the single-magnon spectrum~\cite{ChZh_triPRB09,Verresen_avoided_19,Mourigal2013DSF}. Since the spectral function is directly related to the dynamical structure factor $S({\bf k}, \omega)$, measured in the inelastic neutron-scattering experiments, we will use this approach as is discussed in the next Section.

\subsection{Dynamical structure factor} 
\label{Sec:DSF}

The general form of the dynamical structure factor (DSF) for the neutron scattering is \cite{Woodland2023Parameters}
\begin{equation}
{\cal S}({\bf k},\omega) = \sum_{\alpha, \beta} g_\alpha g_\beta \left(\delta_{\alpha,\beta}-\frac{k_\alpha k_\beta}{k^2} \right){\cal S}^{\alpha\beta}({\bf k},\omega),
\label{eq:dsf}
\end{equation}
with  the momentum and energy transfer ${\bf k}$ and $\omega$, axes of the reference frame $\alpha$ and  $\beta$,  g-tensor components $g_\alpha$, and the spin-spin dynamical correlation function
\begin{equation}
\mathcal{S}^{\alpha \beta}({\bf k}, \omega)=\frac{1}{\pi}\,\text{Im}\int_{-\infty}^{\infty}dt \, e^{i\omega t}\,i\big\langle \mathcal{T} S_{\bf k}^\alpha(t) S_{-{\bf k}}^\beta (0) \big\rangle\,.
\label{eq:dcf}
\end{equation}
For the  field-induced fluctuating paramagnetic state of CoNb$_2$O$_6$, with the choice of the local $\{x,y,z\}$ or laboratory $\{x_0,y_0,z_0\}$  axes in Fig.~\ref{f:chain}(c), only diagonal components of (\ref{eq:dcf}) are considered \cite{Woodland2023Parameters}. Two of them are in the $ac$ plane normal to the field, one  along the Ising axis, ${\cal S}^{xx}$ $({\cal S}^{z_0z_0})$, and one perpendicular to it,  ${\cal S}^{yy}$ $({\cal S}^{x_0x_0})$ \cite{Cabrera2014}. One more component is along the field, ${\cal S}^{zz}$ $({\cal S}^{y_0y_0})$.

In the studies of the CoNb$_2$O$_6$ spectrum in the polarized phase that are discussed in Refs.~\cite{Cabrera2014,fava2020,Woodland2023Parameters}, the momentum transfer is not aligned exclusively along the chain direction $c$, see Fig.~\ref{f:chain}(a), but  has other components. In our consideration, which is  focused on the single spin-chain model, the additional momentum component along the $b$ axis is important because it is able to detect the zigzag structure of the chain. 

Assuming the momentum transfer in the $bc$ plane and keeping only diagonal components in the DSF, the general expression in Eq.~\eqref{eq:dsf} is simplified  to
\begin{align}
{\cal S}({\bf k}, \omega)&=g_{z_0}^2\left(1-\tilde{\lambda}\,\cos^2\gamma\right){\cal S}^{xx}({\bf k}, \omega) \label{eqA:dsf} \\
&+g_{x_0}^2\left(1-\tilde{\lambda}\,\sin^2\gamma \right){\cal S}^{yy}({\bf k}, \omega)+g_{y_0}^2\tilde{\lambda}\,{\cal S}^{zz}({\bf k}, \omega)\nonumber,
\end{align}
where we use the shorthand notation $\tilde{\lambda}\!=\!k_c^2/(k_c^2+k_b^2)$ and the angle $\gamma$ is between the Ising  and $c$ axes, as before, with the transfer momentum ${\bf k}$ in the local frame
\begin{equation}
{\bf k} = k_b \hat{\sf b}+k_c \hat{\sf c} = k_b\hat{z}+k_c(\cos\gamma\ \hat{x}-\sin \gamma\  \hat{y}),
\label{eq:transfer_k}
\end{equation}
see Fig.~\ref{f:chain}. As is discussed in Ref.~\cite{fava2020}, the non-zero $b$-component of the  momentum in (\ref{eq:transfer_k}) is responsible for the secondary ${\bf Q}$-shifted signal in the structure factor from the doubling of the unit cell \cite{Cabrera2014,fava2020, Woodland2023Parameters}. The general form of the diagonal components of the dynamical correlation function is given by
\begin{align}
{\cal S}^{\alpha\alpha}({\bf k},\omega) = & \cos^2 (k_b{\sf b})\,\widetilde{{\cal S}}^{\alpha\alpha}({\bf k},\omega)\nonumber \\
+ & \sin^2 (k_b{\sf b})\,\widetilde{{\cal S}}^{\alpha\alpha}({\bf k}+{\bf Q},\omega),
\label{eq:Sshadow}
\end{align}
where ${\sf b}$ is the width of the zigzag chain in the $b$ direction, see  Fig.~\ref{f:chain}(a) and Ref.~\cite{Cabrera2014}, and $\widetilde{{\cal S}}^{\alpha\alpha}({\bf k},\omega)$ is the correlation function that depends only on the momentum in the $c$ direction, ${\bf k} \!= \!k_c \hat{\textbf{{\sf c}}}$. We note that in Eq.~\eqref{eq:Sshadow} the main signal is associated with the first term and the secondary, ``shadow'' signal, with the  ${\bf Q}$-shifted one.

The $\widetilde{{\cal S}}^{xx}$ and $\widetilde{{\cal S}}^{yy}$ components of the structure factor are the transverse ones and  can be straightforwardly related to the single-magnon spectral function \cite{Cabrera2014,Mourigal2013DSF} as 
\begin{equation}
\widetilde{{\cal S}}^{\alpha\alpha}({\bf k},\omega) = \mathcal{F}^{\alpha \alpha}({\bf k})A({\bf k},\omega),
\label{eq:Sab}
\end{equation}
where the  kinematic formfactors 
\begin{equation}
\mathcal{F}^{xx}_{\bf k} = \frac{S}{2}\,(\bar{u}_{\bf k}+\bar{v}_{\bf k})^2, \ \ \ \  \mathcal{F}^{yy}_{\bf k} = \frac{S}{2}\, (\bar{u}_{\bf k}-\bar{v}_{\bf k})^2,
\end{equation}
produce the ${\bf k}$-dependent modulation of the  single-magnon spectral peaks throughout the Brillouin zone. 

The DSF can also be expected to exhibit  significant decay-related features, such as incoherent parts of the single-magnon spectrum and strong renormalizations, due to  the cubic self-energy~\eqref{eq:self_energy} in the single-magnon  spectral function $A({\bf k},\omega)\!=\!-\frac{1}{\pi}{\rm Im}[G({\bf k},\omega)]$, where   
\begin{eqnarray}
G({\bf k},\omega)=\frac{1}{\omega-\bar{\varepsilon}_{\bf k}-\Sigma^{(3)}({\bf k},\omega)+i0^{+}}\, , 
\label{eq:GkwAkw}
\end{eqnarray}
is the  Green's function   in the SCHF$+\Sigma^{(3)}$ approach.

The DSF  component along the field, $\widetilde{{\cal S}}^{zz}$, corresponds to the longitudinal fluctuations, which account for the direct two-magnon continuum contribution to it,
\begin{align}
\widetilde{{\cal S}}^{zz}({\bf k},\omega) &= \sum_{\bf q} \mathcal{F}^{zz}_{{\bf q},{\bf k}} \, \delta(\omega-\bar{\varepsilon}_{\bf q}-\bar{\varepsilon}_{{\bf k}-{\bf q}}), \nonumber \\
\mathcal{F}^{zz}_{{\bf q},{\bf k}} &= \frac{1}{2}(\bar{u}_{\bf q} \bar{v}_{{\bf k}-{\bf q}} + \bar{v}_{\bf q} \bar{u}_{{\bf k}-{\bf q}})^2.
\label{eq:Szz}
\end{align}
Note that in contrast to the two-magnon continuum in the anharmonic coupling, this continuum is {\it not} umklapp-shifted by the  momentum ${\bf Q}$.

\begin{figure*}[t!]
\centering
\includegraphics[width=\linewidth]{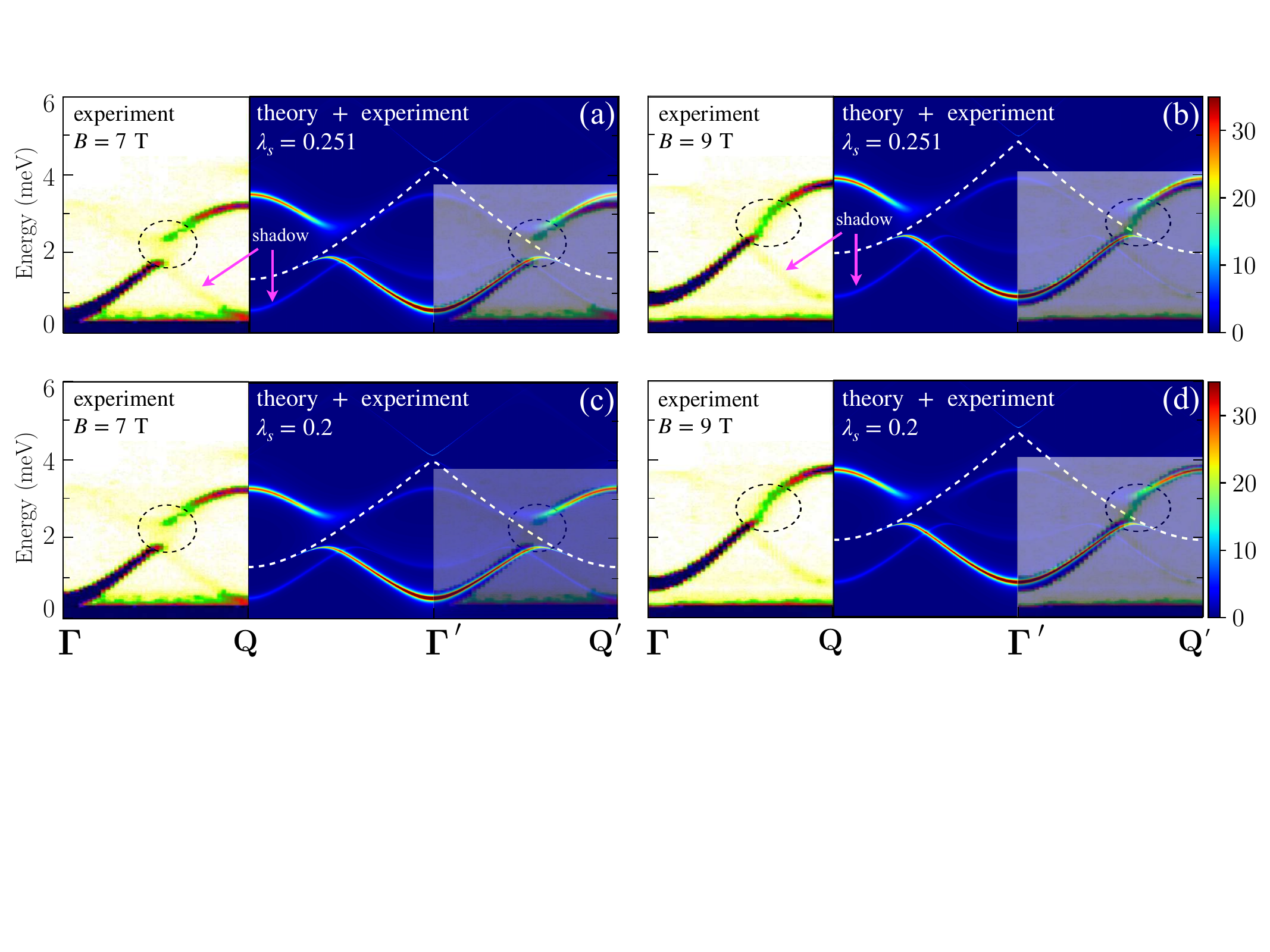} 
\vskip -0.1cm
\caption{Intensity plots of the dynamical structure factor ${\cal S}(\textbf{k},\omega)$ in the paramagnetic phase of CoNb$_2$O$_6$ for the field $7\ \mathrm{T}$ [(a) and (c)] and $9\ \mathrm{T}$ [(b) and (d)], best-fit parameters [(a) and (b)] and adjusted $\lambda_s$ [(c) and (d)]. In each plot, experimental data adapted from Ref.~\cite{fava2020} are shown in the left panel, theoretical results are in the middle panel, and the two are overlaid in the right panel. Artificial broadening of $10^{-3}\ \mathrm{meV}$ for the self-energy~\eqref{eq:self_energy} and $5\!\times\!10^{-2}\ \mathrm{meV}$ for the  Green's functions~\eqref{eq:GkwAkw} and \eqref{eq:Szz} were used. Dashed lines show the bottom of the two-magnon continuum and the arrows indicate the ${\bf Q}$-shifted shadow modes \eqref{eq:Sshadow}.}
\vskip -0.4cm
\label{f:TF_DSF}
\end{figure*}

For comparison with experiments in Ref.~\cite{Robinson2014}, in Fig.~\ref{f:TF_DSF} we illustrate our calculations, where we combined all contributions to the structure factor as given by Eq.~(\ref{eqA:dsf}), and used a momentum transfer in the $bc$ plane  with $k_b/k_c\!=\!0.3$ to have a finite contribution from the shadow mode. We also used artificial broadenings of $10^{-3}\ \mathrm{meV}$ in the  self-energy $\Sigma^{(3)}({\bf k}, \omega)$ (\ref{eq:self_energy}) and $5\!\times\!10^{-2}$~meV in the Green's functions (\ref{eq:GkwAkw}) and \eqref{eq:Szz}.

Figure~\ref{f:TF_DSF} displays our main results. It shows the compilations of the DSF intensity maps by the inelastic neutron scattering,  adapted from Ref.~\cite{fava2020}, with the DSF intensities obtained by the theoretical SCHF$+\Sigma^{(3)}$ approach described above. Each plot consists of the experimental data in the left panel, theoretical results in the middle panel, and the two results overlaid in the right panel, where we have exploited the symmetry of them about the ${\bf \Gamma}$ point. The comparison is presented for the field   $7~\mathrm{T}$ in Figs.~\ref{f:TF_DSF}(a) and \ref{f:TF_DSF}(c) and for $9~\mathrm{T}$ in Figs.~\ref{f:TF_DSF}(b) and \ref{f:TF_DSF}(d), respectively. The upper row, Figs.~\ref{f:TF_DSF}(a) and \ref{f:TF_DSF}(b), shows  theoretical results for the best-fit model of CoNb$_2$O$_6$, see Sec.~\ref{Sec:best_fit}, while the lower row, Figs.~\ref{f:TF_DSF}(c) and \ref{f:TF_DSF}(d), has one parameter from that set modified. The same comparison with the experimental data for $8~\mathrm{T}$ is given in Appendix~\ref{A:DSF}.

One can see that the theoretical results for the best-fit model (upper row of Fig.~\ref{f:TF_DSF}) already yield if not an ideal, but a close quantitative agreement with the experimental data on the gap and magnon bandwidth with no adjustment to the parameters. We note that the mismatch in the energies at the ${\bf Q}$ point might in part be related to the effect of the weak 3D interchain coupling, which affect the experimental data, but are not included in our model. The lower row  of Fig.~\ref{f:TF_DSF} shows that  an even closer agreement can be reached by a modest change of a single parameter in the model, $\lambda_s\!=\!(J_{x_0x_0}\!+J_{y_0y_0})/2J_{z_0z_0}$, used in the parametrization of Ref.~\cite{Woodland2023Parameters}, see also Sec.~\ref{Sec:best_fit}, to which the maximum of the magnon band is most sensitive. Changing it from the best-fit value of $\lambda_s\!=\!0.251$ to $\lambda_s\!=\!0.2$ improves the agreement with our theory. This is not to challenge the  comprehensive multi-dimensional  best-fit strategy of Ref.~\cite{Woodland2023Parameters}, but to highlight, once again, that a self-consistent approach  can turn a theory  plagued with unphysical divergences into a reliable, nearly quantitative tool.

Turning to the other features of the theoretical results shown in Fig.~\ref{f:TF_DSF}, it is clear that the off-shell $\omega$-dependent cubic self-energy successfully regularizes the threshold singularities discussed in Sec.~\ref{Sec:Decay_thresholds}, and, indeed, provides a quantitatively faithful description of the quasiparticle-like and incoherent parts of the single-magnon spectrum. On the inner side of the two-magnon continuum, the magnon spectral lines acquire a substantial broadening in a close agreement with the experimental data, see also the plot  for $8~\mathrm{T}$  in Appendix~\ref{A:DSF}. 

On the outer side of the continuum, a direct intersect of the magnon mode with the continuum  is avoided via a strong renormalization of the magnon energy, creating a gap-like splitting and a characteristic loss of the spectral weight of the magnon line. Although the quantitative agreement with the experimental data on the size of the gap-like feature and its evolution with the field is rather spectacular, the theoretical results contain more details, with the remnant of the magnon mode following the bottom of the two-magnon continuum for an extended range of the momenta. While a recent proposal suggests that in 1D such an edge-mode should survive for all the momenta \cite{Verresen_avoided_19}, this conclusion is an artifact of the one-loop approximation for the magnon self-energy, which is also employed in our study. In reality, it is expected that  the magnon mode should meet the continuum at the so-called termination point~\cite{Zhitomirsky2013,pitaevskii1959,ChZh_triPRB09}, the result that requires  a self-consistent treatment of the higher-order diagrams in the theory, which is not attempted here.  

Because of the constraint provided by the quantitative accord of the experiment and theory on the size of the gap-like splitting in the spectrum, an additional comment can be made on the potential value of   the ``residual" $J_{x_0y_0}$ term in the exchange matrix (\ref{eq:J1stagerredRotated}), discussed in Sec.~\ref{Sec:Perturbative_gamma} and Appendix~\ref{A:RSPT}. This term does not modify the even part of the spin Hamiltonian (\ref{eq:Heven}), but contributes to the cubic coupling from the odd part (\ref{eq:Hodd}).  Because of the staggered nature of this term,  the structure of  the decay vertex is not expected to modify, leading to an enhancement of the decay self-energy according to $\Sigma^{(3)}({\bf k},\omega)\! \propto\! J_{x_0y_0}^2\!+J_{y_0z_0}^2$. Since the best-fit parameters without the $J_{x_0y_0}$-term provide a close quantitative description of the decay-related features described above, see Fig.~\ref{f:TF_DSF}, one can conclude that the  $J_{x_0y_0}$ term must be small compared to $J_{y_0z_0}$. This observation  supports the approach of Refs.~\cite{fava2020, Woodland2023Parameters} and our discussion in Sec.~\ref{Sec:Perturbative_gamma}. 

In Fig.~\ref{f:TF_DSF}, in both theoretical and experimental results, one can also observe  the  ${\bf Q}$-shifted ``shadow'' mode in addition to the main contribution from the single-magnon excitations \cite{Cabrera2014,fava2020,Woodland2023Parameters}.  As is discussed above, it originates from the non-zero component of the transfer momentum along the $b$ axis. In the theory results in Fig.~\ref{f:TF_DSF}, one can also observe a continuum-like contribution from the longitudinal component of the structure factor (\ref{eq:Szz}), with its role being generally minor.

\section{Longitudinal field effects}
\label{Sec:LongFields}

Up to this point, our study of the CoNb$_2$O$_6$ excitation spectrum in the paramagnetic phase concerned  the transverse direction of the field. Now we focus on the effects of an additional longitudinal field component. In this consideration, we use the same spin Hamiltonian  $\hat{\mathcal{H}}_1\!+\!\hat{\mathcal{H}}_2$,  Eqs.~\eqref{eq:spin_hamiltonian}, \eqref{eq:J1}, and \eqref{eq:J2}, with the transverse-field term $\hat{\mathcal{H}}_{\perp}\!=\!g_{y_0}\mu_B B_{\perp} \sum_i S_i^{y_0}$, Eq.~\eqref{eq:H_perp},  now augmented by the longitudinal-field term
\begin{equation}
\hat{\mathcal{H}}_\parallel=g_{z_0} \mu_B B_{\parallel}\sum_i S_i^{z_0}.
\label{eq:H_parallel}
\end{equation} 
with the field component $B_{\parallel}$ along the Ising $z_0$ axis; see Fig.~\ref{f:chain}. We are interested in the regime of the weak longitudinal fields, $B_{\parallel}\!\ll\! B_{\perp}$, with  excitations remaining spin-flip-like and magnon description of the their spectrum still adequate.

Although symmetry-wise the classification of the type of the symmetry-breaking provided by the longitudinal field  in \eqref{eq:H_parallel}  in the case of the zigzag model of  CoNb$_2$O$_6$ is more delicate, relating it to the glide-symmetry breaking~\cite{fava2020}, the main effect is the same as in the paradigmatic Ising model \cite{zamolodchikov89,Lauchli18}. The quantum phase transition of the transverse-field Ising-like model ceases to exist and turns into a crossover, with a finite excitation gap at the former transition point. 

The second effect of the symmetry-breaking longitudinal field is specific to the zigzag chain model, and it is intimately tied to the presence of the staggered bond-dependent $J_{y_0z_0}$  terms,  allowed by the same glide symmetry. Because of the tilt of the spin-quantization axis induced by the longitudinal field  in the paramagnetic state, the two-site unit cell of the zigzag structure becomes explicit in the model of spin flips, doubling the unit cell of the ``simple'' Ising chain that sufficed until now. The description of the excitation spectrum in this more general case requires two distinct branches of spin excitations within the reduced Brillouin zone of the zigzag chain. 

Importantly, these two excitation branches will be split by a band gap. As we argue below, one can  expect strong modifications of the two-magnon DoS as a result of these changes in the single-magnon spectrum,  inducing richer varieties of the Van Hove singularities that are potentially observable. 
Below, we quantify both effects using the LSWT formalism.

\subsection{The excitation gap and the band gap}
\label{Sec:Gaps}

In the tilted field with a small longitudinal component away from the transverse $y_0$ axis toward the Ising $z_0$ axis, the spin-quantization axis will tilt by the angle  $\theta$  in the $y_0z_0$ $(zx)$ plane, see Fig.~\ref{f:chain}(c), found from the minimization of the classical energy, see Appendix~\ref{A:LFields} for details,
\begin{equation}
H_{\perp}\sin\theta - H_c\sin\theta\cos\theta - H_{\parallel} \cos\theta =0,
\label{eq:EclHyHz}
\end{equation}
where $H_{\perp}\!=\!g_{y_0}\mu_B B_{\perp}$ and $H_{\parallel}\!=\!g_{z_0}\mu_B B_{\parallel}$ are the transverse and  longitudinal fields, respectively, in the energy units, and the classical critical field $H_c$ is from Eq.~\eqref{eq:critical_field}.

Here, we focus on the case of the transverse field value equal to the critical field, $H_{\perp}\!=\!H_c$, and study the dependence of the spectrum gap and the band gap on the longitudinal field, as both gaps vanish at  $H_{\parallel}\! =\!0$. For $H_{\parallel}\! \ll \!H_{\perp}\!=\!H_c$, Eq.~\eqref{eq:EclHyHz} can be solved by expanding in the small canting angle, yielding
\begin{equation}
\theta = \left( 2H_{\parallel}/H_c\right)^{1/3}.
\label{eq:CantingAngleLF}
\end{equation}
Note that this fractional power law  is reminiscent of that of the canting angle due to staggered Dzyaloshinskii-Moriya interaction in the isotropic square-lattice antiferromagnet near saturation field \cite{Chernyshev2005}. 

\begin{figure}[t!]
\centering
\includegraphics[width=\linewidth]{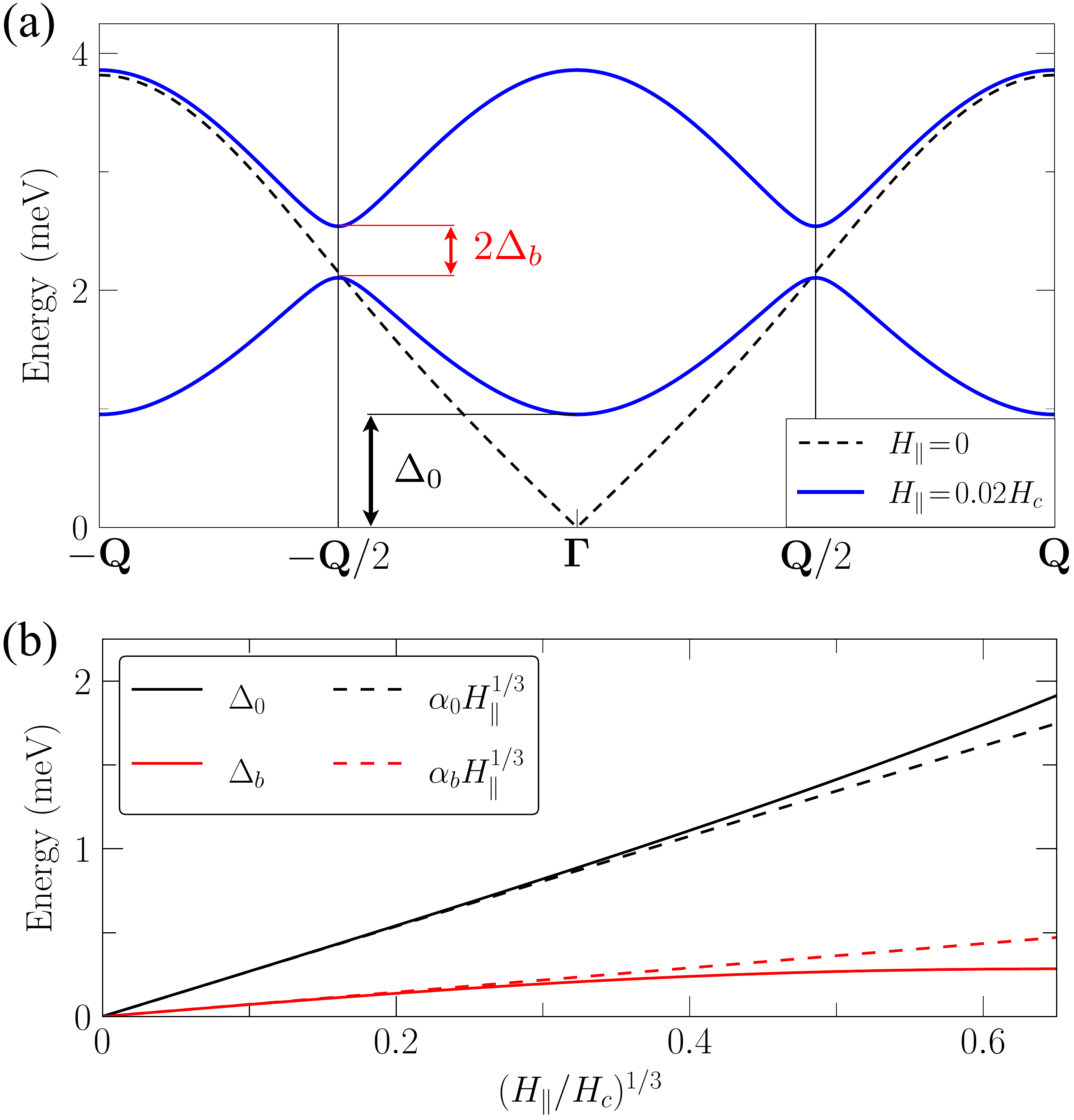} 
\vskip -0.2cm
\caption{LSWT results for the best-fit model of CoNb$_2$O$_6$ and   $H_{\perp}\!=\!H_c$. (a) Magnon spectrum for $H_{\parallel}\! =\!0$ (dashed line) and $H_{\parallel}\! =\!0.02H_c$ (solid line), with gaps $\Delta_{0}$ and $\Delta_{b}$ identified. (b) Excitation gap $\Delta_{0}$ and band gap $\Delta_{b}$ vs $(H_{\parallel}/H_c)^{1/3}$. Lines are the full LSWT results (Appendix~\ref{A:LFields}) and dashed lines are their asymptotics from Eq.~\eqref{eq:GapLF}.}
\vskip -0.5cm
\label{f:GapLF}
\end{figure}

The LSWT consideration of the $1/S$-expansion of the model with the longitudinal field involves a somewhat cumbersome diagonalization of the $4\times 4$ Hamiltonian for the two bosonic species, deferred to  Appendix~\ref{A:LFields}, which gives explicit expressions of the energies of the two magnon branches. In the presence of the longitudinal field, excitation gap $\Delta_{0}$ at the ${\bf \Gamma}$ point and the band gap $\Delta_{b}$ at the ${\bf Q}/2$ point open up, as is discussed above; see Fig.~\ref{f:GapLF}(a) for  a comparison to the $H_{\parallel}\! =\!0$ case.

Using the canting angle \eqref{eq:CantingAngleLF} for small fields, one can obtain asymptotic expressions for the gaps, 
\begin{align}
&\Delta_{0} \approx \alpha_0 H_{\parallel}^{1/3}, \ \
\alpha_0\approx\sqrt{\frac32} H_c \left(\frac{2}{H_c}\right)^{1/3}, \nonumber\\
&\Delta_{b}  \approx \alpha_b H_{\parallel}^{1/3},  \ \ 
\alpha_b\approx 2S\big|J_{y_0z_0}\big| \left(\frac{2}{H_c}\right)^{1/3},
\label{eq:GapLF}
\end{align}
which follow the same  fractional power law vs field, see Appendix~\ref{A:LFields} for the exact proportionality coefficients and Fig.~\ref{f:GapLF}(b), which shows a comparison of the asymptotic  results \eqref{eq:GapLF}  with the full LSWT results for the best-fit model of CoNb$_2$O$_6$.

We note that,  according to  the Ising conformal field theory in $1+1$ dimensions \cite{zamolodchikov89}, the scaling of the spectrum gap with the longitudinal field is known to obey a different  fractional power law with the exponent $8/15$. Still,  the expressions in Eq.~\eqref{eq:GapLF} highlight an important distinction of the two gaps. The spectrum gap $\Delta_{0}$ is essentially the same as it would have been for the ``simple'' Ising-like spin chain, as it is independent of the bond-dependent terms. However, the appearance of the band gap $\Delta_{b}$ is  precisely due to the staggered bond-dependent $J_{y_0z_0}$  terms, rooted in the zigzag nature of the model. 

Quantitatively, because the bond-dependent  terms in  CoNb$_2$O$_6$ are secondary to the main Ising term,  the excitation gap $\Delta_{0}$  grows faster with the longitudinal field than the band gap $\Delta_{b}$.  Comparison of their asymptotics in Eq.~\eqref{eq:GapLF} for the CoNb$_2$O$_6$ model yields
\begin{equation}
\frac{\Delta_{b}}{\Delta_{0}}  \approx 2S\sqrt{\frac23}\, \frac{\big|J_{y_0z_0}\big|}{H_c} \approx 0.27.
\label{eq:GapRatio_LF}
\end{equation} 
As we will see next, this result has a significant impact on the longitudinal field range for which the overlap of the single-magnon branches with the additional Van Hove singularities in the two-magnon spectra is possible.

\subsection{More threshold singularities}
\label{Sec:PhysicalSingLF}

One of the  important consequences of the magnon band splitting is the explicit separation of the two-magnon continuum into three continua,  corresponding to different combinations of the single-magnon species
\begin{equation}
E_{{\bf k},{\bf q}}^{\{\mu,\nu\}}=\varepsilon_{\mu{\bf q}}+\varepsilon_{\nu{\bf k}-{\bf q}}, \label{eq:continuum_LF}
\end{equation}
where $\mu(\nu)\!=\!1,2$. This splitting also necessarily creates richer structure of the Van Hove singularities in the continuum,  which can affect the single-magnon spectrum via the anharmonic coupling. Thus, if allowed by the two-magnon kinematics, the longitudinal field can potentially lead to more singularities in the magnon spectra, in addition to the ones discussed in Secs.~\ref{Sec:Decay_thresholds} and \ref{Sec:DSF}. 

\begin{figure}[t]
\centering
\includegraphics[width=\linewidth]{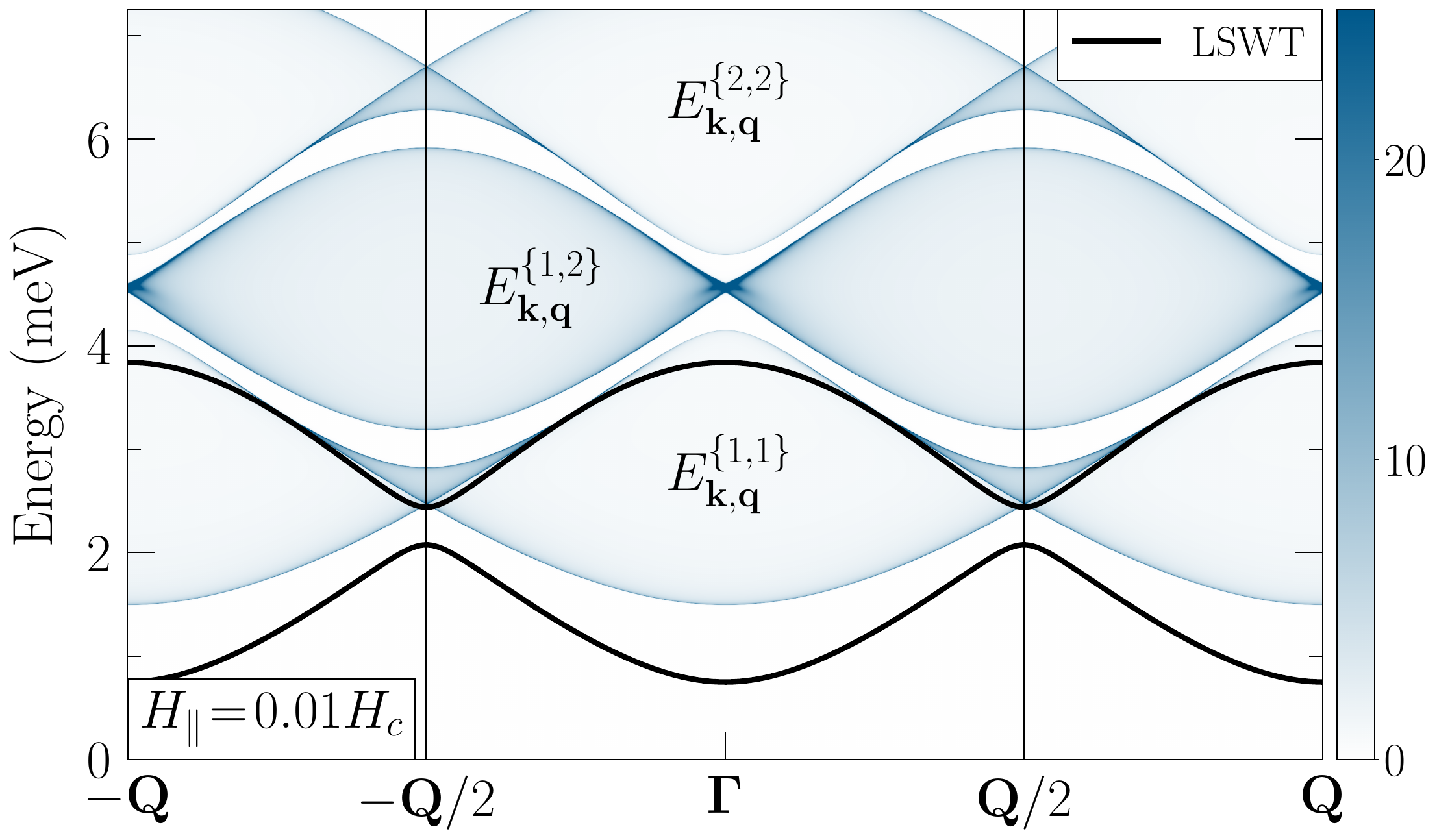} 
\vskip -0.2cm
\caption{Magnon spectrum for $H_{\perp}\!=\!H_c$, $H_{\parallel}\! =\!0.01H_c$ and the best-fit model of CoNb$_2$O$_6$  in the repeated Brillouin zone scheme. The intensity plot is the two-magnon DoS for the continua in (\ref{eq:continuum_LF}).}
\label{f:DoSLF}
\vskip -0.4cm
\end{figure}

In Fig.~\ref{f:DoSLF}, we show magnon spectrum  together with the two-magnon DoS intensity plot for $H_{\perp}\!=\!H_c$ and $H_{\parallel}\! =\!0.01H_c$ ($B_{\parallel}\!\approx\!0.09\ \mathrm{T}$) for the best-fit model of CoNb$_2$O$_6$, from which one can appreciate the more intricate structure of the field-induced Van Hove singularities in the two-magnon continuum. 

However, in practice, because excitation gap $\Delta_0$ grows  faster than the band gap $\Delta_b$  \eqref{eq:GapRatio_LF}, such a trend in the field-induced gaps provides a rather narrow range of the longitudinal fields  for which the kinematics is favorable of the crossing of the additional Van Hove singularities by the single-magnon spectrum.  Thus, already for $H_{\parallel}\! =\!0.01H_c$ shown in Fig.~\ref{f:DoSLF}, the magnon branch barely accesses the extra features in the continuum, prohibiting such a crossing for the larger  fields.

In addition to the kinematics, we would also like to remark on the effect of the small longitudinal field on the structure of the anharmonic cubic term that is responsible for the one-to-two-magnon coupling. There are two parts in it in the presence of the $H_{\parallel}$ field, one that largely retains the same form as in the odd part of the Hamiltonian (\ref{eq:Hodd}), originating from the staggered exchanges $J_{y_0z_0}$, while the other  is due to the tilt angle $\theta \!\propto\! H_{\parallel}^{1/3}$ that allows most other exchanges to contribute to the cubic anharmonicity; see Appendix~\ref{A:LFields} for some more detail. We note that it is the latter term that was previously considered as the main source of the decay singularities in  CoNb$_2$O$_6$ \cite{Robinson2014}. However, not only is it subleading in the  weak longitudinal-field regime, but it is also not staggered, resulting in an unfavorable kinematics  for the decay-related processes.

\section{Conclusions}
\label{Sec:Conclusions}

We conclude by summarizing our  results.
In this study, we have thoroughly reanalyzed the symmetry-based approach to formulating anisotropic-exchange model of the quasi-one-dimensional ferromagnet CoNb$_2$O$_6$. We have proposed a connection of its model to the broader class of such models, studied for a wide variety of materials with complex bond-dependent spin-orbit-induced exchanges. We have also clarified the role of a phenomenological constraint that  has been used to restrict parameter space of CoNb$_2$O$_6$, and have investigated the magnitude and the effects of the residual terms, which  were neglected in the previous studies, using real-space perturbation theory and unbiased DMRG approach.

The main result of the present work is the self-consistent study of the effects of magnon interactions in the excitation spectrum of CoNb$_2$O$_6$ in the quantum paramagnetic phase. We have proposed and applied  a self-consistent Hartree-Fock  regularization of the problematic unphysical divergences in the $1/S$ spin-wave expansion that is common to various anisotropic-exchange models. Not only does this method eliminate such unphysical singularities, but it also preserves the integrity of the threshold phenomena of magnon decay and spectrum renormalization that are present in both theory and experiment of CoNb$_2$O$_6$.

Using the microscopic parameters proposed previously,  we have employed this approach  to study  excitation spectrum in the fluctuating paramagnetic phase of CoNb$_2$O$_6$. For the dynamical structure factor ${\cal S}({\bf k},\omega)$, we have  demonstrated a  close quantitative agreement of our theory with the neutron-scattering data for both the quasiparticle-like and incoherent parts of the single-magnon spectrum,  also in the field regime that is inaccessible by the standard spin-wave theory. Moreover, our results for the spectrum gap are  in a close accord with the complimentary DMRG calculations for the same model parameters. 

These results prove the ability of our approach to provide quantitatively faithful description of the magnetic excitations in the paramagnetic phase of CoNb$_2$O$_6$, despite strong quantum fluctuations, anisotropic exchanges, and low dimensionality of the problem, the factors that make the standard SWT-like approaches fail. Furthermore, it can be expected that our approach should be able to yield further analytical insights into magnon interactions and decay phenomena and shed  light on the important aspects of the excitation spectra in the other anisotropic-exchange magnets in the higher dimensions, where it is free from the remaining minor inconsistencies associated with the 1D nature of the CoNb$_2$O$_6$ model.

Lastly, we have also discussed the effects of additional longitudinal fields in the paramagnetic phase of CoNb$_2$O$_6$. We have demonstrated that due to the zigzag lattice structure and affiliated bond-dependent exchanges in the model,  and due to the symmetry breaking by the longitudinal field, both the excitation gap and the band gap develop in the magnon spectrum. We have  described how the band splitting leads to the additional anomalies in the the two-magnon continuum, potentially resulting in extra threshold singularities in the magnon spectra.

\begin{acknowledgments}

We would like to thank Pavel Maksimov for a prior collaboration on the earlier attempt on this problem and for an important discussion concerning Kitaev-like  bond-dependent terms in 1D that has led us to expand on the general anisotropic-exchange model for the zigzag chain. We are indebted to Leonie Woodland and Radu Coldea for numerous conversations on the phenomenological constrains for CoNb$_2$O$_6$, their implementation, and parameters of the model, as well as for sharing their experimental results, indispensable comments and useful insights, and detailed editorial guidance to ensure coherence of our text and its consistency with the experimental analysis. We would like to thank Jeff Rau and Izabella  Lovas for helpful conversations.  We are grateful to Shengtao Jiang for important guidance regarding DMRG. This entire work, from conception to development, execution, and writing, was supported by the U.S. Department of Energy, Office of Science, Basic Energy Sciences under Award No. DE-SC0021221. We would like to  thank  Aspen Center for Physics (A.~L.~C.) and KITP (A.~L.~C. and C.~A.~G.), where parts of this work were completed. Aspen Center for Physics is supported by National Science Foundation grant PHY-2210452. KITP is supported  by the National Science Foundation under Grants No. NSF PHY-1748958 and PHY-2309135.

\end{acknowledgments}

\appendix

\section{Model}
\label{A:Model}

\subsection{Different parametrizations}
\label{A:Parametrizations}

For the two parametrizations of the exchange matrix in the crystallographic reference frame $\{a,b,c\}$ in Fig.~\ref{f:chain}(a), the original one in Eq.~(\ref{eq:J1stagerred}) and the  one in the ``ice-like" language in Eq.~\eqref{eq:J1ice}, the relations between exchanges are given by
\begin{align}
J=\frac{J_{bb}+J_{cc}}{2}&,\ 
\Delta=\frac{2J_{aa}}{J_{bb}+J_{cc}}, \nonumber\\
J_{z\pm}=\frac{J_{ac}}{\cos\varphi_\alpha}&,\ J_{\pm\pm}=\frac{J_{bb}-J_{cc}}{4\cos\varphi_\alpha}, \nonumber\\ 
\lambda_z=-\frac{(-1)^{\alpha} J_{ab}}{J_{ac}\tan\varphi_\alpha}&,\ \lambda_{\pm}=\frac{2(-1)^{\alpha} J_{bc}}{(J_{bb}-J_{cc})\tan\varphi_\alpha}.
\label{eqA:Jice_Jabc}
\end{align}
The matrix in the laboratory  $\{x_0,y_0,z_0\}$ frame is obtained by rotating the exchange matrix in the crystallographic frame by $\gamma$ about $b$; see Fig.~\ref{f:chain}(b), using the rotation matrix
\begin{align}
\hat{\textbf{R}}_{\gamma} =
\begin{pmatrix}
\cos\gamma  &  0 & -\sin\gamma\\
0 & 1 & 0\\
\sin\gamma & 0 & \cos\gamma\\
\end{pmatrix}.
\label{eqA:rotationMatrix}
\end{align} 
The explicit relation of the matrix elements of the exchange matrix in the laboratory frame to that in the crystallographic frame is 
\begin{align}
J_{x_0x_0}&=J_{aa}\cos^2\!\gamma-J_{ac}\sin{2\gamma}+J_{cc}\sin^2\!\gamma, \nonumber\\
J_{y_0y_0}&=J_{bb}, \nonumber \\
J_{z_0z_0}&=J_{aa}\sin^2\!\gamma+J_{ac} \sin{2\gamma} +J_{cc} \cos^2\!\gamma,\nonumber \\ J_{x_0y_0}&=J_{ab}\cos{\gamma}-J_{bc}\sin{\gamma}, \nonumber\\
J_{x_0z_0}&=J_{ac}\cos{2\gamma}-\frac{1}{2}(J_{cc}-J_{aa})\sin{2\gamma}, \nonumber\\
J_{y_0z_0} & =  J_{bc} \cos{\gamma} + J_{ab} \sin{\gamma}.\label{eqA:Jlab_Jabc}
\end{align}
The best-fit parameters in Ref.~\cite{Woodland2023Parameters} can be translated to the exchanges in the laboratory $\{x_0,y_0,z_0\}$ frame, see Eq.~(\ref{eq:values_JNNlab}) and discussion in Secs.~\ref{Sec:Ising_Axis} and \ref{Sec:best_fit}.

Using Eqs.~\eqref{eqA:Jlab_Jabc} and \eqref{eq:values_JNNlab}, we obtain exchanges in the crystallographic frame
\begin{align}
J_{aa}&=J_{x_0x_0}\cos^2\!\gamma+J_{z_0z_0}\sin^2\!\gamma =-1.05(1)\ \text{meV} ,\nonumber \\
J_{bb}&=J_{y_0y_0}=-0.67(1)\ \text{meV}, \nonumber\\
J_{cc}&=J_{x_0x_0}\sin^2\!\gamma+J_{z_0z_0}\cos^2\!\gamma =-2.00(2)\ \text{meV},\nonumber\\
J_{ab}&=J_{y_0z_0}\sin\gamma=-0.28(1)\ \text{meV}, \nonumber\\
J_{ac}&= \frac{1}{2}(J_{z_0z_0}-J_{x_0x_0})\sin2\gamma=-0.83(1)\ \text{meV}, \nonumber\\
J_{bc}&= J_{y_0z_0}\cos \gamma=-0.49(1)\ \text{meV}.
\label{eqA:Jabc_Jlab}
\end{align}
Finally, combining Eqs.~\eqref{eqA:Jice_Jabc} and \eqref{eqA:Jabc_Jlab}, gives parameters in the ``ice-like" parametrization in Eq.~(\ref{eq:values_Jice}).

\vspace{-.3 cm}
\subsection{Out-of-plane angle}
\vskip -.1cm
\label{A:RSPT}

Real-space perturbation theory (RSPT) \cite{Bergman2007RSPT, Chernyshev2014RSPTKagome, Zhitomirsky2015RSPT} allows to access the effects of quantum fluctuations by expanding around the classical ground state of the ferromagnetic Ising chain in various spin-flip processes. To avoid  unnecessary secondary details, we consider a simplified nearest-neighbor exchange matrix 
\begin{equation}
\hat{\textbf{J}}_{\alpha} \!=\! 
\begin{pmatrix}
0 & (-1)^{\alpha}J_{x_0y_0} & 0\\
(-1)^{\alpha}J_{x_0y_0} & 0 & (-1)^{\alpha}J_{y_0z_0}\\
0 & (-1)^{\alpha}J_{y_0z_0} & {J}_{z_0z_0}\\
\end{pmatrix},\label{eqA:J1minimal}
\end{equation}
where $J_{z_0z_0}$ is the leading ferromagnetic Ising exchange and the two staggered terms, $J_{x_0y_0}$ and $J_{y_0z_0}$, are  perturbations. The Hamiltonian can be written in terms of the spin ladder operators as follows
\begin{align}
\hat{\mathcal{H}}_{0}&=J_{z_0z_0}\sum_{\langle ij\rangle} S_i^{z_0}S_j^{z_0}, \nonumber \\
\hat{V}_{xy}\!&=\!\frac{iJ_{x_0y_0}}{2}\sum_{\langle ij\rangle }(-1)^\alpha \Big\{S_i^-S_{j}^--S_i^+S_{j}^+\Big\},\label{eqA:perturbations} \\
\hat{V}_{yz}\!&=\!\frac{iJ_{y_0z_0}}{2}\!\sum_{\langle ij\rangle}(-1)^\alpha\Big\{\!\left(S_i^-\!-\!S_{i}^+\!\right)\!S^{z_0}_j\!+\!\left(S_j^-\!-\!S_{j}^+\right)\!S^{z_0}_i\!\Big\},\nonumber
\end{align}
where  perturbations $\hat{V}_{xy}$ and $\hat{V}_{yz}$ generate double spin flips and single spin flips, respectively. The ground state of the unperturbed Hamiltonian in Eq.~\eqref{eqA:perturbations} is the ferromagnetic state $|0\rangle$ and its excited states $|n \rangle$ are the  states with $n$  spin flips. 

Since we are interested in the deviations of the ordered moment from the Ising axis, the lowest-order processes that induce a single-spin-flip state $|1 \rangle$ are in question.  Notably, the single-spin-flip term acting on the ground state vanishes identically because of its staggered form, $\hat{V}_{yz}|0\rangle\!=\!0$, providing no spin tilt along the $y_0$ axis. The lowest non-zero contribution that yields the  single-spin-flip state is given by the second-order process involving both single- and double-spin-flip terms in (\ref{eqA:perturbations})
\begin{equation}
|0\rangle \xrightarrow{\hat{V}_{xy}}|2\rangle \xrightarrow{\hat{V}_{yz}}|1\rangle,
\label{eqA:eff_spinflip}
\end{equation}
in which  their mutually-canceling staggered form is important. Then, the fluctuating ground state due to the  process in \eqref{eqA:eff_spinflip} is 
\begin{align}
|\widetilde{0}\rangle = |0\rangle -\frac{J_{x_0y_0}J_{y_0z_0}\sqrt{2S}}{8J_{z_0z_0}^2S\left(1-1/4S\right)} \,  |1\rangle,
\label{eqA:GSfluct}
\end{align}
which yields the angle of the spin tilt  out of the $y_0z_0$ plane along the $x_0$ axis, $\delta\gamma\!\approx\!\langle S^{x_0}_i\rangle/\langle S^{z_0}_i\rangle$, for any site $i$ 
\begin{equation}
\delta \gamma = -\frac{J_{x_0y_0}J_{y_0z_0}}{4J_{z_0z_0}^2S\left(1-1/4S\right)},
\label{eqA:deltagamma}
\end{equation}
where we used  $S^{x_0}_i\!=\!\left(S_i^+\!+\!S_i^-\right)\!/2$ and neglected higher-order corrections to the ground state from the  two consecutive double-spin-flips. The results in (\ref{eqA:GSfluct}) and (\ref{eqA:deltagamma}) are obtained for the model (\ref{eqA:perturbations}) with arbitrary spin $S$, keeping higher-order $1/S$ terms such as $(1-1/4S)$ factors,  originating from the interaction of the nearest-neighbor spin flips. The $S\!=\!1/2$ limit of (\ref{eqA:deltagamma}) is listed in Eq.~(\ref{eq:delta_gamma}).

Importantly, the fluctuating ground state in (\ref{eqA:GSfluct}) continues to respect the glide symmetry and produces no tilt in the $y_0$ direction, $\langle S^{y_0}_i\rangle\!=\!0$. 

\begin{figure}[t]
\centering
\includegraphics[width=\linewidth]{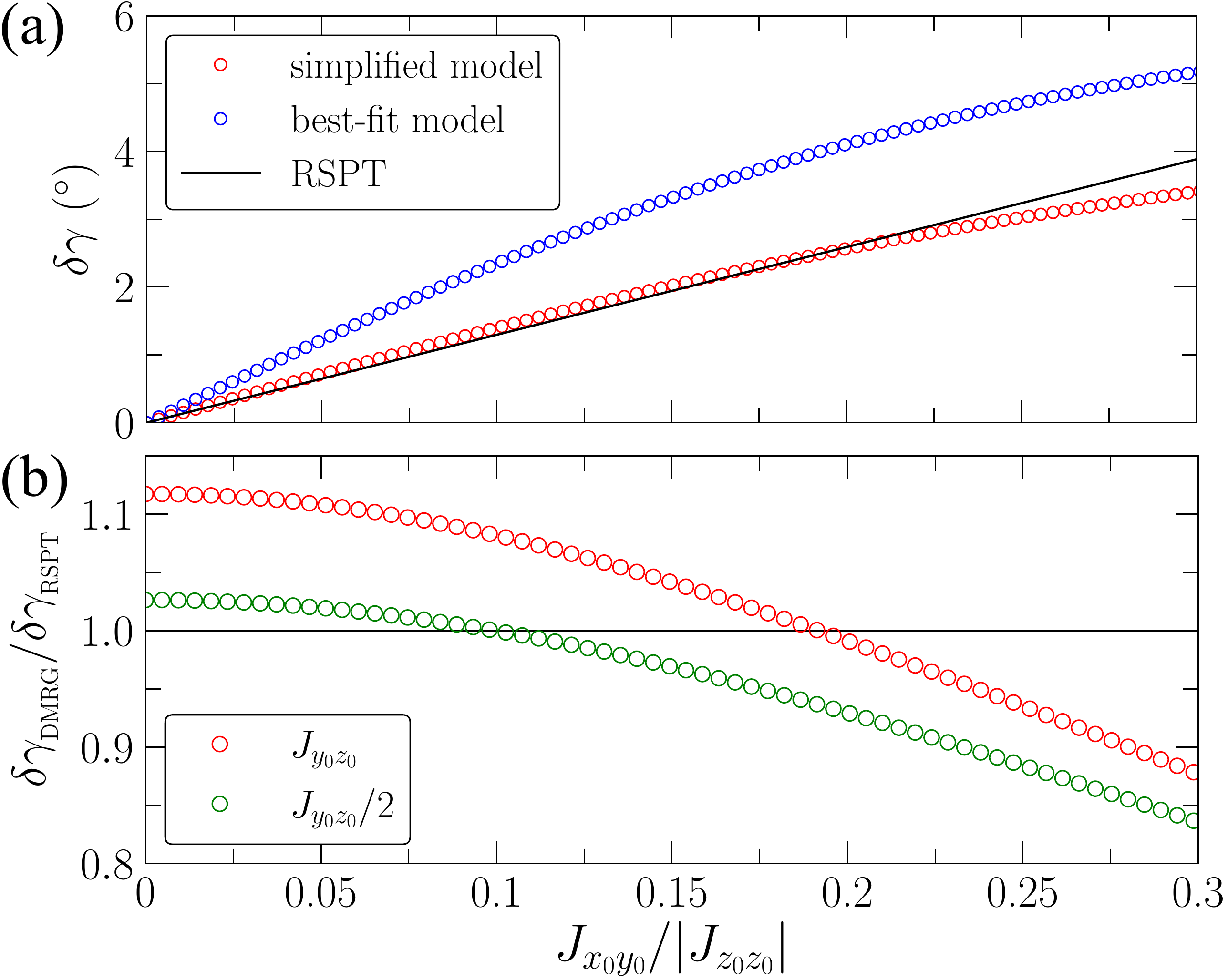} 
\vskip -0.2cm
\caption{(a) The spin tilt angle by RSPT (\ref{eqA:deltagamma}) and DMRG  vs $J_{x_0y_0}/|J_{z_0z_0}|$ for the simplified and full models for the best-fit parameters for CoNb$_2$O$_6$. (b) The ratio of the DMRG and RSPT tilt angles for the simplified model for  two choices of $J_{y_0z_0}$, see the text.}
\vskip -0.4cm
\label{f:DeltaGamma}
\end{figure}

Given the analysis leading to the second-order RSPT result in (\ref{eqA:deltagamma}), one can argue that in order to produce a spin tilt, both $J_{x_0y_0}$ and $J_{y_0z_0}$   terms are necessary   in an {\it arbitrary} order of the theory,  because they have to  cancel their staggered form. The higher-order corrections to (\ref{eqA:deltagamma}) also need to carry {\it odd} powers of each of the staggered terms because they generate different number of the spin flips. These considerations can be expected to remain valid for a  general model in Eq.~(\ref{eq:J1stagerredRotated}), which contains other non-staggered spin-flip terms.

As is discussed in Sec.~\ref{Sec:Ising_Axis} and Sec.~\ref{Sec:Perturbative_gamma} and is clear from Eq.~(\ref{eqA:deltagamma}), the spin tilt angle vanishes for the choice of $J_{x_0y_0}\!=\!0$ made in Refs.~\cite{fava2020, Woodland2023Parameters}, rendering  quantum corrections to the classical Ising spin direction zero. One can verify the accuracy of the second-order perturbative result for the tilt angle in (\ref{eqA:deltagamma}) and elucidate the role of the higher-order fluctuations with the help of the unbiased  DMRG calculations for the ground state. 

The DMRG simulations were performed in chains of up to $5000$ sites using the ITensor library~\cite{ITensor}. We have employed two strategies, the ``scan'' with a slowly varying $J_{x_0y_0}$ along the chain, which provides a real-space variation of the tilt angle, and ``no-scan,'' in which the tilt angle is measured in the middle of the chain away from the edges for each individual    $J_{x_0y_0}$ value.  Both approaches yield numerically indistinguishable results. Because of the Ising nature of the model, a very good convergence is reached with low bond dimensions and small number of the DMRG sweeps~\cite{noteDMRG}.

In Figure~\ref{f:DeltaGamma}(a), we show the RSPT result \eqref{eqA:deltagamma} for $S\!=\!1/2$ and for $J_{z_0z_0}$ and $J_{y_0z_0}$ from Eq.~(\ref{eq:values_JNNlab}), which correspond to the best-fit parameters for CoNb$_2$O$_6$,  as a function of $J_{x_0y_0}/|J_{z_0z_0}|$. It is shown together with the DMRG results for the same $S\!=\!1/2$ simplified model \eqref{eqA:J1minimal}, for which Eq.~\eqref{eqA:deltagamma} was derived. In addition to that,  DMRG results for the full model of CoNb$_2$O$_6$ in Eqs.~(\ref{eq:J1}) and (\ref{eq:J2}) for the best-fit parameters  from  Eqs.~(\ref{eq:values_JNNlab}) and (\ref{eq:values_J2g}) in Secs.~\ref{Sec:best_fit} are shown. 

Clearly, the numerical results for $\delta\gamma$ vanish identically for $J_{x_0y_0}\!=\!0$, in accord with the discussion above. For the simplified model, the agreement of the slope of the DMRG tilt angle with that of the RSPT is close, but not precise, as is also demonstrated in  Fig.~\ref{f:DeltaGamma}(b),  which shows the ratio of the angles. The majority of this difference can be attributed to the next-order correction, corresponding to the fourth-order process
\begin{equation}
|0\rangle \xrightarrow{\hat{V}_{xy}}|2\rangle \xrightarrow{\hat{V}_{yz}}|3\rangle\xrightarrow{\hat{V}_{yz}}|2\rangle\xrightarrow{\hat{V}_{yz}}|1\rangle,
\end{equation}
which is $\mathcal{O}(J_{x_0y_0}J_{y_0z_0}^3)$, conforming to the   rules proposed for the higher-order corrections that are discussed above. 

One can verify the $J_{y_0z_0}^3$-order of the correction to the slope by changing the numerical value of $J_{y_0z_0}$ as is shown in  Fig.~\ref{f:DeltaGamma}(b). Here, the reduction of $J_{y_0z_0}$ by $1/2$ leads to an eightfold decrease of the correction in the DMRG result, all in a close accord with the expectations of the odd powers of each staggered term outlined above. 

The DMRG results for the best-fit parameters in the full model of CoNb$_2$O$_6$  in Fig.~\ref{f:DeltaGamma}(a) show a  qualitative agreement with the perturbative result \eqref{eqA:deltagamma} for the simplified model \eqref{eqA:J1minimal}, but also a substantial quantitative difference. It can be attributed to the fluctuations induced by the other spin-flip terms present in the full model.

Lastly, as is discussed in Sec.~\ref{Sec:Perturbative_gamma}, the phenomenological constraint on the spin direction in the model (\ref{eq:J1stagerredRotated}) allows the ``residual''  $J_{x_0y_0}$ and $J_{x_0z_0}$ terms to be present, but exactly compensating each others' spin tilting, leaving the physical Ising direction intact. The perturbative result in Eq.~\eqref{eqA:deltagamma} together with the classical energy minimization result in Eq.~(\ref{eq:tan2gamma}) immediately suggest an explicit connection between the two terms: $J_{x_0z_0} \!=\! -\delta\gamma J_{z_0z_0}$, see Sec.~\ref{Sec:Perturbative_gamma} for more detail. The DMRG results shown in Fig.~\ref{f:Jzx_vs_Jxy} in that Section, demonstrating a relation between $J_{x_0y_0}$ and $J_{x_0z_0}$, were calculated using the best-fit parameters of the full model of CoNb$_2$O$_6$. For each fixed value of $J_{x_0y_0}$, we performed a DMRG scan vs  $J_{x_0z_0}$ to identify the value of $J_{x_0z_0}$ that corresponds to an exact compensation of the spin tilt away from the Ising axis.

\vspace{-0.2cm}
\section{Spin-wave theory}
\label{A:SWExpansion}
\vskip -.1cm

We consider a standard Holstein-Primakoff (HP) spin representation  with the quantization axis  $z$ 
\begin{equation}
S_i^z = S-n_i, \ \ \ \ S_i^+ = a_i \sqrt{2S - n_i}, 
\label{eqA:holstein_primakoff}
\end{equation}
where $a_i (a_i^\dagger)$ are bosonic operators, $n_i\!=\!a_i^\dagger a_i$, and $i$ is the site index. The expansion of the square roots in \eqref{eqA:holstein_primakoff}  in powers of $n_i/2S$ results in the bosonic Hamiltonian 
\begin{equation}
\mathcal{\hat{H}} = \mathcal{\hat{H}}^{(0)} +\mathcal{\hat{H}}^{(1)}+\mathcal{\hat{H}}^{(2)}+\mathcal{\hat{H}}^{(3)}+\mathcal{\hat{H}}^{(4)} +\mathcal{O}(S^{-1}),
\end{equation}
where the $n$th term $\mathcal{\hat{H}}^{(n)}$ contains the $n$th power of the bosonic operators and  carries an explicit $S^{2-n/2}$ factor, constituting a $1/S$-expansion for a given problem. 

The first term $\hat{\mathcal{H}}^{(0)}$ in such an expansion is the classical energy, and $\hat{\mathcal{H}}^{(1)}$ should vanish upon the  classical energy minimization. The quadratic term $\hat{\mathcal{H}}^{(2)}$ is the harmonic part of the expansion that yields the LSWT and magnon energy spectrum. The lowest $1/S$-order  corrections to the LSWT originate from the two anharmonic  terms in the expansion, $\hat{\mathcal{H}}^{(3)}$ and $\hat{\mathcal{H}}^{(4)}$, describing three- and four-magnon interaction, respectively.

\vspace{-0.2cm}
\subsection{Classical energy}
\label{A:ClassicalEnergy}
\vskip -.15cm

The classical energy of the field-polarized paramagnetic phase considered in Sec.~\ref{Sec:evenodd} is easily obtained from (\ref{eq:Heven}) and is given by $E_{\text{cl}}/N\!=\!- SH +S^2 (J_{y_0y_0}+J_{2})$, with the  linear $\hat{\mathcal{H}}^{(1)}$ from (\ref{eq:Hodd}) vanishing because of the staggered structure of the bond-dependent terms. This is a common situation for collinear states that do not require energy minimization and cannot indicate their phase boundaries from the classical consideration alone. 
 
 One standard approach is to proceed directly with the harmonic term $\hat{\mathcal{H}}^{(2)}$, develop LSWT as in Sec.~\ref{Sec:LSWT}, and obtain the value of the critical field from the condition of stability of the magnon spectrum in (\ref{eq:EkGamma}). 
 
The other approach is to consider the ordered phase of CoNb$_2$O$_6$ in the transverse field in the classical limit and find the critical field of a transition to the fully polarized state from the minimization of its energy. In such a state spins are tilted away from the field toward the Ising axis in the $y_0 z_0$ ($zx$) plane; see Fig.~\ref{f:chain}(c). Denoting this angle as $\theta$,  straightforward algebra in \eqref{eq:Heven} yields
\begin{align}
\frac{E_{\text{cl}}}{N}=&- SH \cos\theta +S^2 (J_{y_0y_0}+J_{2}) \nonumber\\
& -S^2 (J_{y_0y_0}-J_{z_0z_0}+J_{2}-J_{2z_0})\sin^2\theta.
\label{eqA:Eclorderedphase}
\end{align}
Minimizing it with respect to $\theta$ gives $\cos\theta\!=\!H/H_c$ with the critical field given in  Eq.~\eqref{eq:critical_field}. 

\vspace{-0.3cm}
\subsection{Linear spin-wave theory}
\label{A:LSWT}
\vskip -.15cm

The LSWT  is based on the lowest-order expansion in (\ref{eqA:holstein_primakoff})  in the Hamiltonian \eqref{eq:Heven}, leading to
\begin{align}
\hat{\mathcal{H}}^{(2)} & = \frac{S}{2}  \sum_i \Big\{ 2\big(H/S-2(J_{y_0y_0}+J_2)\big) n_i \nonumber \\
&\hspace{-.3cm} +\Big((J_{z_0z_0}+J_{x_0x_0})a_i^\dagger a_{i+1}^{\phantom\dag} + (J_{2z}+J_2) a_i^\dagger a_{i+2}^{\phantom\dag}  \\
&\hspace{-.3cm} +(J_{z_0z_0}\!-\!J_{x_0x_0}) a_i^\dagger a_{i+1}^\dagger +(J_{2z}-J_2) a_i^\dagger a_{i+2}^\dagger + \text{H.c.} \Big)\Big\},\nonumber
\end{align}
Using  Fourier transformation \eqref{eq:fourier} gives the harmonic Hamiltonian in the canonical form \eqref{eq:H2k}. The standard Bogolyubov transformation diagonalizes it with the $u_{\bf k}$ and $v_{\bf k}$ parameters given explicitly as
\begin{align}
u_{\bf k} = \sqrt{\frac{A_{\bf k}+\varepsilon_{\bf k}}{2\varepsilon_{\bf k}}},\ v_{\bf k}  = \text{sgn}(B_{\bf k})\sqrt{\frac{A_{\bf k}-\varepsilon_{\bf k}}{2\varepsilon_{\bf k}}},
\label{eqA:uvk}
\end{align}
with $A_{\bf k}$ and $B_{\bf k}$ from Eq.~(\ref{eq:AkBk}) and $\varepsilon_{\bf k}$ from Eq.~(\ref{eq:omega_onemagn}).
The resultant diagonal form of the LSWT Hamiltonian is
\begin{equation}
\mathcal{\hat{H}}^{(2)} = \sum_{\bf k} \Big\{ \varepsilon_{\bf k} b_{\bf k}^\dagger b_{\bf k}^{\phantom\dag} + \frac{1}{2}\left( \varepsilon_{\bf k} - A_{\bf k} \right) \Big\},
\end{equation}
where the magnon energy in (\ref{eq:omega_onemagn}) can also be written as
\begin{align}
\varepsilon_{\bf k}^2 \! = & \Big[ H\!-\!2S\left(J_{y_0y_0}\!-\!J_{x_0x_0} \gamma^{(1)}_{\bf k}+J_2 \left(1-\gamma^{(2)}_{\bf k}\right) \right)\Big] \nonumber \\ 
\times &\Big[H\!-\!2S\left( J_{y_0y_0} \!+\! J_2 \!- \! J_{z_0z_0} \gamma_{\bf k}^{(1)} \!-\! J_{2z_0} \gamma_{\bf k}^{(2)}\right) \Big].
\label{eqA:omega_onemagn}
\end{align}

\vspace{-0.2cm}
\subsection{Non-linear spin-wave theory}
\label{A:NLSWT}
\vskip -.1cm

\vspace{-0.1cm}
\subsubsection{Cubic terms}
\vskip -.15cm

Using the leading-order HP expansion in the odd part of the Hamiltonian \eqref{eq:Hodd} leads to the cubic term
\begin{equation}
\hspace{-.2cm} \mathcal{\hat{H}}^{(3)}\! =\! -J_{y_0z_0} \sqrt{\frac{S}{2}} \sum_{i}   \! \Big((-1)^i n_i^{\phantomsection} \big(a^\dagger_{i+1}\!- a^\dagger_{i-1}\big) \!+ \text{H.c.} \!\Big).
\label{eqA:H3}
\end{equation}
The Fourier transformation \eqref{eq:fourier} in Eq.~\eqref{eqA:H3} yields
\begin{align}
\mathcal{\hat{H}}^{(3)} = J_{y_0z_0} \sqrt{\frac{2S}{N}}\sum_{{\bf k},{\bf q}}\left(\bar{\gamma}_{\bf p} a^\dagger_{\bf p} a^\dagger_{\bf q}a_{\bf k}^{\phantom\dag} + \text{H.c.} \right),
\end{align}
with ${\bf p}={\bf k}-{\bf q}+{\bf Q}$ and $\bar{\gamma}_{\bf p}=i \sin(p{\sf c}_0)$. The  Bogolyubov transformation  with symmetrization give
\begin{align}
\mathcal{\hat{H}}^{(3)} &= \frac{1}{2!\sqrt{N}} \sum_{{\bf q}+{\bf k}+{\bf p}={\bf Q}}\Big(\Phi_{{\bf qk};{\bf p}}b^\dagger_{\bf q}b^\dagger_{\bf k}b_{-{\bf p}}^{\phantom\dag} 
+ \text{H.c.}  \Big) \nonumber\\ 
& +  \frac{1}{3!\sqrt{N}}\sum_{{\bf q}+{\bf k}+{\bf p}={\bf Q}}\Big( \Xi_{{\bf qkp}}b^\dagger_{\bf q} b^\dagger_{\bf k} b^\dagger_{\bf p}   + \text{H.c.}\Big), \label{eqA:decay_sourceVs} 
\end{align}
with the decay and source vertices,  $\Phi_{{\bf qk};{\bf p}}$ and $\Xi_{{\bf qkp}}$,
\begin{align}
\hspace{-.2cm} \Phi_{{\bf qk};{\bf p}}\!=\! \sqrt{2S}\, J_{y_0z_0} \widetilde{\Phi}_{{\bf qk};{\bf p}},\ \ \Xi_{{\bf qk}{\bf p}}\!=\! \sqrt{2S} \, J_{y_0z_0}\widetilde{\Xi}_{{\bf qk}{\bf p}},
\end{align}
and the dimensionless vertices given by 
\begin{align}
\widetilde{\Phi}_{{\bf qk};{\bf p}} &=\bar{\gamma}_{\bf k}(u_{\bf k}+v_{\bf k})(u_{\bf q}u_{\bf p}+v_{\bf q}v_{\bf p})\nonumber \\ &+\bar{\gamma}_{\bf q}(u_{\bf q}+v_{\bf q})(u_{\bf k}u_{\bf p}+v_{\bf k}v_{\bf p})\nonumber \\ &+\bar{\gamma}_{\bf p}(u_{\bf p}+v_{\bf p})(u_{\bf k}v_{\bf q}+v_{\bf k}u_{\bf q}),\\
\widetilde{\Xi}_{{\bf qk}{\bf p}}&=\bar{\gamma}_{\bf k}(u_{\bf k}+v_{\bf k})(u_{\bf q}v_{\bf p}+v_{\bf q}u_{\bf p})\nonumber \\ &+\bar{\gamma}_{\bf q}(u_{\bf q}+v_{\bf q})(u_{\bf k}v_{\bf p}+v_{\bf k}u_{\bf p})\nonumber \\ &+\bar{\gamma}_{\bf p}(u_{\bf p}+v_{\bf p})(u_{\bf k}v_{\bf q}+v_{\bf k}u_{\bf q}).
\end{align}
The decay and source vertices in (\ref{eqA:decay_sourceVs}) are umklapp-like, with the momentum conserved up to the ${\bf Q}$-vector. 

The resulting lowest-order self-energies are
\begin{align}
\Sigma^{(d)}({\bf k},\omega)\!&=\!\frac{1}{2N}\sum_{\bf q} \frac{|\Phi_{{\bf q},{\bf k}-{\bf q}+{\bf Q};-{\bf k}}|^2}{\omega-\varepsilon_{\bf q}-\varepsilon_{{\bf k}-{\bf q}+{\bf Q}}+i0^+}, \\
\Sigma^{(s)}({\bf k},\omega)\!&=\!-\frac{1}{2N}\sum_{\bf q} \frac{|\Xi_{{\bf q},-{\bf k}-{\bf q}+{\bf Q},{\bf k}}|^2}{\omega+\varepsilon_{\bf q}+\varepsilon_{-{\bf k}-{\bf q}+{\bf Q}}-i0^+}. 
\end{align}

\subsubsection{Quartic terms}

The four-boson terms are obtained from the higher-order expansion in the HP transformation \eqref{eqA:holstein_primakoff} 
\begin{equation}
S_i^+ \approx \sqrt{2S}\left( a_i - \frac{n_ia_i}{4S}\right), \hspace{.5cm} S^z_i = S-n_i.
\label{eqA:HPs-12}
\end{equation}
Using this expansion \eqref{eqA:HPs-12}, the quartic terms come from 
\begin{align}
S_i^{x}S_j^{x} &\rightarrow -\frac{1}{8}\Big( (a_i^\dagger + a_i ) n_ja_j + 
(i\leftrightarrow j) + \text{H.c.}\Big), \nonumber\\
S_i^{y}S_j^{y} &\rightarrow -\frac{1}{8}\Big( (a_i^\dagger - a_i ) n_ja_j + 
(i\leftrightarrow j) + \text{H.c.}\Big), \nonumber\\
S_i^z S_j^z & \rightarrow n_in_j.\label{eqA:HFavrg}
\end{align}
The decoupling of them uses the real-space HF averages
\begin{align}
\hspace{-.2cm} n&=\langle a_i^\dagger a_i^{\phantom\dag} \rangle \!=\! \sum_{\bf k} v_{\bf k}^2, \hspace{.2cm} m_n \!=\! \langle a_i^\dagger a_{j}^{\phantom\dag} \rangle \!= \!\sum_{\bf k} \gamma_{\bf k}^{(n)} v_{\bf k}^2,  \nonumber \\
\delta&=\langle a_i^2 \rangle \!=\! \sum_{\bf k} u_{\bf k} v_{\bf k}, \hspace{.2cm} \Delta_n \!=\! \langle a_ia_{j}\rangle \!=\! \sum_{\bf k} \gamma_{\bf k}^{(n)} u_{\bf k} v_{\bf k},
\label{eqA:HFaverages}
\end{align}
with the index $n\!=\!1(2)$ for the nearest and next-nearest neighbors, yielding
\begin{align}
\hspace{-.2cm} S^{x}_iS^{x}_j \left(S^{y}_iS^{y}_j\right) \rightarrow & -\frac{1}{2}\biggl[ (m_n\pm \Delta_n) (n_i+n_j) \nonumber \\
& +\frac{1}{4}(\Delta_n\pm m_n)\left(a_i^\dagger a_i^\dagger + a_j^\dagger a_j^\dagger + \text{H.c.}\right) \nonumber \\ 
&+\left(n\pm \frac{\delta}{2}\right)\left(a_i^\dagger a_j^{\phantom\dag}+ \text{H.c.}\right)  \nonumber \\
&+\left(\frac{\delta}{2} \pm n\right)\left(a_i^\dagger a_j^\dagger + \text{H.c.}\right) \biggl],
\label{eqA:SxySxydecoupling} \\
&\hspace{-2.4cm} S_i^zS_j^z \!  \rightarrow n (n_i+ n_j)+\!\big(m_n a_i^\dagger a_j + \Delta_n a_i^\dagger a_j^\dagger + \text{H.c.}\big).
\label{eqA:SzSzdecoupling}
\end{align}
Using \eqref{eqA:SxySxydecoupling} and \eqref{eqA:SzSzdecoupling} in the quartic Hamiltonian  from the even part of the model \eqref{eq:Heven}, followed by the Fourier transformation, gives a correction to the LSWT model
\begin{eqnarray}
\delta \hat{\mathcal{H}}^{(4)} = \sum_{\bf k}\Big\{ \delta A_{\bf k} a_{\bf k}^\dagger a_{\bf k}^{\phantom\dag} - \frac{\delta B_{\bf k}}{2} \left(a_{\bf k}^\dagger a_{-{\bf k}}^\dagger +\text{H.c}\right)\Big\},\quad\quad
\label{eqA:dH4k} 
\end{eqnarray} 
with 
\begin{align}
\delta A_{\bf k} &  =  - J_{x_0x_0} \Big( m_1-\Delta_1+\left(n-\delta/2\right)\gamma_{\bf k}^{(1)} \Big) \nonumber\\
& +2J_{y_0y_0} \Big( n+m_1\gamma_{\bf k}^{(1)} \Big) \nonumber\\
&-J_{z_0z_0} \Big(m_1+\Delta_1+\left(n+\delta/2\right) \gamma_{\bf k}^{(1)}\Big) \nonumber\\
&-J_{2}\Big( m_2-\Delta_2-2n+\left(n-\delta/2-2m_2\right)\gamma_{\bf k}^{(2)} \Big) \nonumber \\
& - J_{2z_0}\Big( m_2+\Delta_2 +\left(n+\delta/2\right)\gamma_{\bf k}^{(2)} \Big),
\label{eqA:dAk} 
\end{align}
\begin{align}
\delta B_{\bf k}& = -J_{x_0x_0}\biggl( \frac{1}{2}(m_1-\Delta_1)+(n-\delta/2)\gamma_{\bf k}^{(1)} \biggl) \nonumber \\
&-2J_{y_0y_0} \Delta_1 \gamma_{\bf k}^{(1)} \nonumber\\
& +J_{z_0z_0}\biggl(\frac{1}{2}(m_1 + \Delta_1) +\left(n+\delta/2\right) \gamma_{\bf k}^{(1)} \biggl)\nonumber\\
&-J_{2}\biggl( \frac{1}{2}(m_2-\Delta_2)+(n-\delta/2+2\Delta_2)\gamma_{\bf k}^{(2)} \biggl)\nonumber\\
& + J_{2z_0} \biggl( \frac{1}{2}(m_2+\Delta_2)  + \left(n+\delta/2\right) \gamma_{\bf k}^{(2)}\biggl).
\label{eqA:dBk}
\end{align}

Finally, using  Bogolyubov transformation in (\ref{eqA:dH4k}), yields  the $\omega$-independent $1/S$  energy correction in \eqref{eq:dEk4}. Same results for the quartic terms can be obtained using Dyson-Maleev spin representation~\cite{Dyson1956, Maleev1958}.

\subsection{Self-consistent Hartree-Fock method}
\label{A:SCHF}

Here we provide some further details on the self-consistent  HF method discussed in Sec.~\ref{Sec:SCHF}.

The set of the real-space HF averages in (\ref{eqA:HFaverages}), $\{\mbox{HFs}\}\!=\!\{n,\delta,m_1,m_2,\Delta_1,\Delta_2\}$, are found iteratively for each fixed field value, starting from $H_{0}\!>\!H_c$ and proceeding by decreasing the field with a small step $\Delta H$. 

The initial set of the HF averages, $\{\mbox{HFs}\}_n^0$, for a field $H_n$ in such a sequence of fields  is taken from the final (converged) set of the HF averages from the previous field value, $\{\mbox{HFs}\}_{n-1}^{\rm final}$. The exception is the very first field $H_{0}$, which we choose large enough for the LSWT averages from (\ref{eqA:HFaverages}) to be a good starting point, so we use $\{\mbox{HFs}\}_{0}^{0}\!=\!\{\mbox{HFs}\}_{0}^{\rm LSWT}$ for it. 

For any field $H_n$, the self-consistent iterations follow the cycle shown in (\ref{eq:SCHF_diagram}), with the steps for each subsequent iteration summarized as follows:
\begin{itemize}
\vspace{-0.2cm}
\item[1)] At each $i$th step, the  $\{\mbox{HFs}\}_n^{i-1}$ averages give  the quartic-term contributions to the harmonic theory, $\delta \bar{A}_{\bf k}$ and $\delta \bar{B}_{\bf k}$, according to the expressions in (\ref{eqA:dAk}) and (\ref{eqA:dBk}). The LSWT-like eigenvalue problem of the same form as in Eq.~(\ref{eq:H2k}) with $\bar{A}_{\bf k}\!=\! A_{\bf k}+\delta \bar{A}_{\bf k}$ and $\bar{B}_{\bf k}\!=\! B_{\bf k}+\delta \bar{B}_{\bf k}$ yields the new set of the  Bogolyubov parameters $\bar{u}_{\bf k}$ and $\bar{v}_{\bf k}$. 
\vspace{-0.2cm}
\item[2)] Using $\bar{u}_{\bf k}$ and $\bar{v}_{\bf k}$, a new (temporary) set of HF averages, $\{\mbox{HFs}\}_n^{i_t}$, is calculated using Eq.~\eqref{eqA:HFaverages}.
\vspace{-0.2cm}
\item[3)] The input of the HF averages for the next iteration is updated using \\ $\{\mbox{HFs}\}_n^{i}=\alpha \{\mbox{HFs}\}_n^{i_{t}} +(1-\alpha )\{\mbox{HFs}\}_n^{i-1}$, \\ where $\alpha\!\ll\!1$ ensures a smooth convergence.
\vspace{-0.2cm}
\item[4)] The cycle of the steps from 1) to 3)  is continued until a numerical convergence in the HF averages is reached by meeting a  tolerance $\epsilon$ between the two subsequent iterations. At this step, the final set $\{\mbox{HFs}\}_{n}^{\rm final}$ for the field $H_n$ is defined. Obviously, it also  yields the  SCHF magnon eigenenergies $\bar{\varepsilon}_{\bf k}$ and Bogolyubov parameters $\bar{u}_{\bf k}$ and $\bar{v}_{\bf k}$ used in our results in Sec.~\ref{Sec:SCHF} and Sec.~\ref{Sec:Results}.
\vspace{-0.2cm}
\item[5)] For the next field $H_{n+1}$, the cycle starts at the step 1) with the converged set of $\{\mbox{HFs}\}_{n}^{\rm final}$ used as the initial condition.
\end{itemize}
\vspace{-0.2cm}
In this work, we have used $H_0\!=\!3H_c$, $\Delta H\!=\!3\!\times\!10^{-3}\  \text{meV}$, $\alpha\!=\!0.01$, and $\epsilon \!=\! 10^{-7}$.  The stability of this procedure was verified by varying all of these parameters to ensure independence of the results.

\begin{figure}[t!]
\centering
\includegraphics[width=\linewidth]{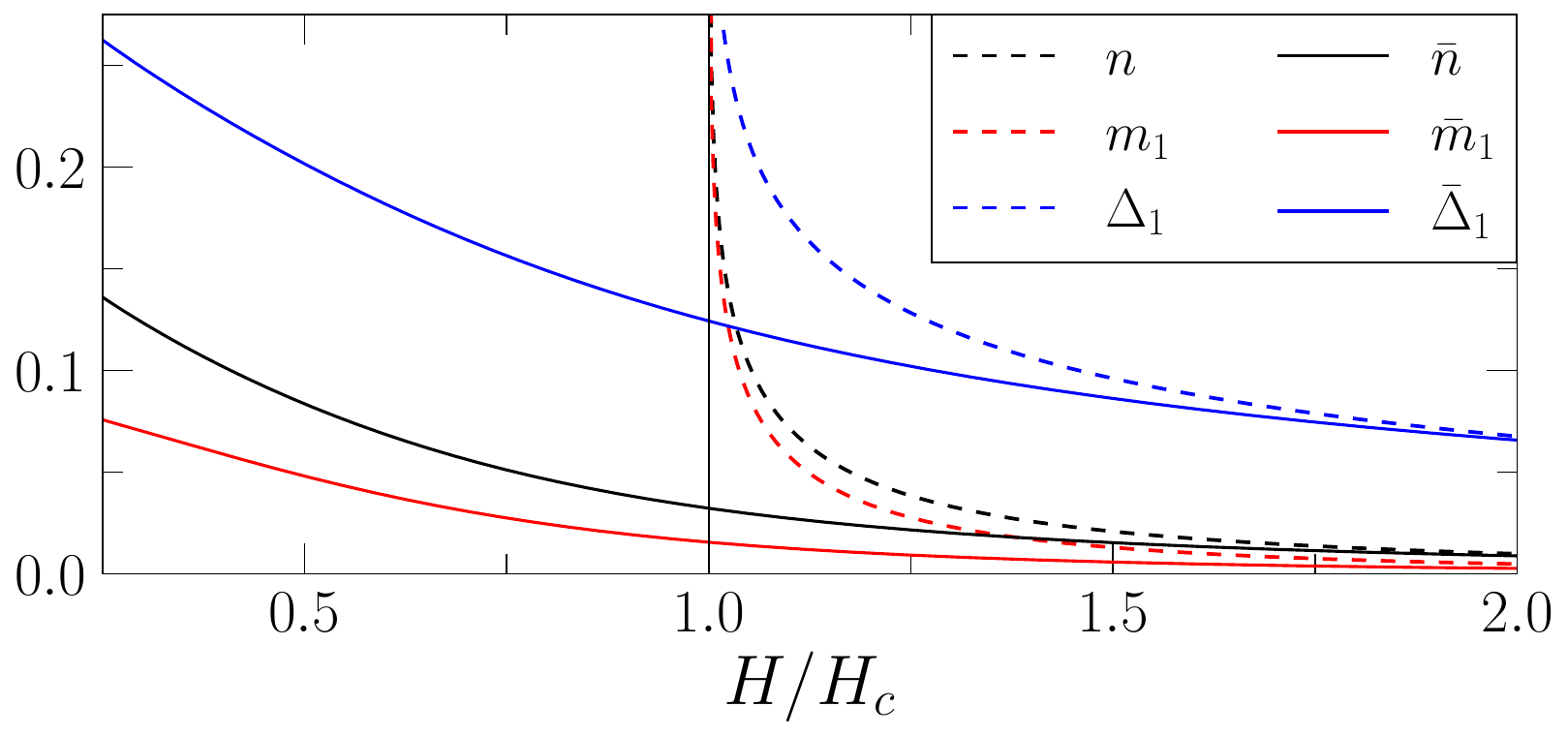} 
\vskip -0.3cm
\caption{Representative Hartree-Fock averages vs  $H$, LSWT (dashed lines) and SCHF (solid lines), respectively.}
\vskip -0.4cm
\label{fA:HFparameters}
\end{figure}

Our Fig.~\ref{fA:HFparameters} illustrates that the logarithmically divergent behavior of the HF averages gets regularized using the SCHF method and that their calculation is successfully extended below the LSWT value of $H_c$.

\section{Results}
\label{A:Results}

\subsection{DMRG calculations of the gap}
\label{A:DMRGHc}

The DMRG simulations for the lowest energy gap in the excitation spectrum of the best-fit model of CoNb$_2$O$_6$ as a function of the field that is shown in Fig.~\ref{f:SC_Gap} in Sec.~\ref{Sec:Results} were performed in the chains of up to $500$ sites. The gap was obtained by calculating the ground state and the first excited state using the ITensor library~\cite{ITensor} for each value of the field. While a good convergence was easily reached for the ground state~\cite{noteDMRG}, a larger number of sweeps (about $75$) were needed to reach the same accuracy when computing the first excited state.

\begin{figure}[t]
\centering
\includegraphics[width=\linewidth]{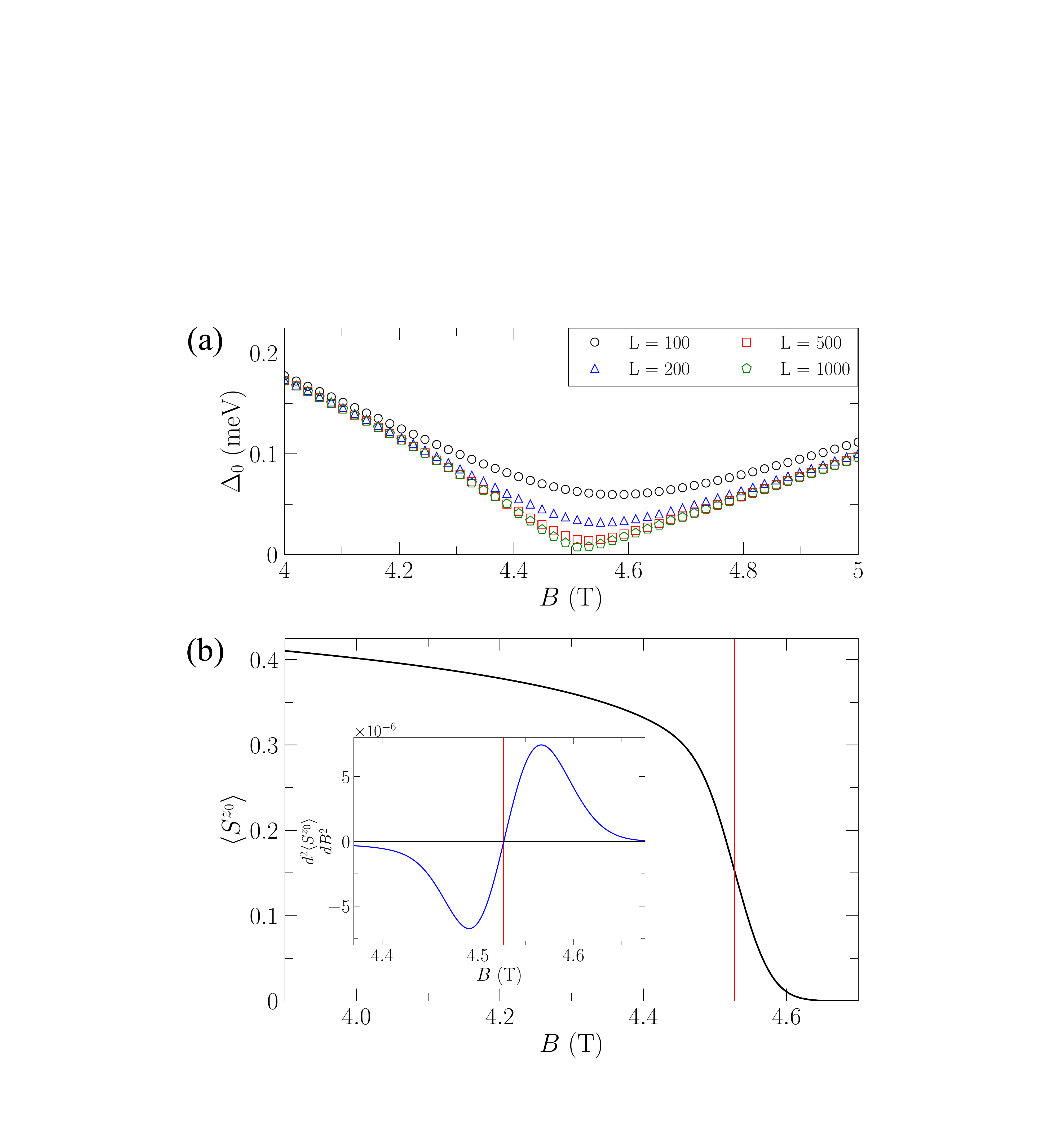} 
\vskip -0.2cm
\caption{(a) Energy gap vs field by DMRG for the chains of different length. (b) The expectation value of the spin component $\langle S_i^{z_0} \rangle$ in the ground state along the long chain in a DMRG  scan vs field. Vertical line indicates the critical field. Inset shows the second derivative of $\langle S_i^{z_0} \rangle$ vs field.}
\vskip -0.4cm
\label{f:BcDMRG}
\end{figure}

Fig.~\ref{f:BcDMRG}(a)  shows the excitation gap in a narrow field region near the critical field for the chains of  lengths up to $L\!=\!1000$. One can see that the finite-size effects are appreciable only near the minimum of the gap. Using the $1/L$ extrapolation of the data in Fig.~\ref{f:BcDMRG}(a) for small gaps, we have verified that the $L\!\rightarrow\infty$ gap vanishes at about $4.52(1)$~T, the value which is also consistent with the linear $|B-B_c|$ extrapolation of the $L\!=\!1000$ data above the critical field. It is worth noting that in the ordered phase, the finite-size effects lift the ground state degeneracy, with the resultant splitting that can be confused with the {\it actual} excitation gap close to the critical field, as both reach $\mathcal{O}(10^{-3})$~meV for $L\!=\!1000$. 

One can also corroborate these results for the critical field using the ground-state DMRG calculations. Our Fig.~\ref{f:BcDMRG}(b) shows the DMRG scan in the ground state with a slowly varying field in a chain of $5000$ sites. The expectation value of the spin component in the Ising direction $\langle S_i^{z_0} \rangle$ suggests a second order phase transition with a critical field at about $4.527(1)\ \mathrm{T}$. This value is determined from the inflection point of the curve, with the change on the sign of the second derivative shown in the inset. This result is consistent with the value suggested by the analysis of the gap from Fig.~\ref{f:BcDMRG}(a).

\subsection{Two-magnon kinematics}
\label{A:Kinematics}

Here we briefly discuss  two aspects of the two-magnon kinematics in the context of the 1D spin model of CoNb$_2$O$_6$: the structure of the two-magnon continuum and  kinematics of the magnon decay.

\subsubsection{The two-magnon continuum}

At any given momentum ${\bf k}$, the ${\bf Q}$-shifted two-magnon continuum is defined within  the energy range
\begin{eqnarray}
E_{{\bf k}, {\bf q}^*}^{\text{min}}\leq E_{{\bf k}, {\bf q}}=\bar{\varepsilon}_{\bf q} + \bar{\varepsilon}_{{\bf k}-{\bf q}+{\bf Q}}\leq E_{{\bf k}, {\bf q}^*}^{\text{max}},
\label{eqA:continuum}
\end{eqnarray}
where the boundaries $E_{{\bf k}, {\bf q}^*}^{\text{min}}$ and $E_{{\bf k}, {\bf q}^*}^{\text{max}}$ should be found from the extremum condition  $\left(\partial E_{{\bf k}, {\bf q}}/\partial{\bf q}\right)|_{{\bf q}^*}\!=\!0$, which translates to the requirement on the  group velocities of the two magnons to be equal for the momentum ${\bf q}^*$
\begin{eqnarray}
 \bar{\bf v}_{{\bf q}^*} = \bar{\bf v}_{{\bf k}-{\bf q}^*+{\bf Q}}.
 \label{eqA:vqvkq}
\end{eqnarray}
Although,  generally, such conditions may require a numerical solution, one class of them, which often describes a majority of the extrema in the two-magnon continua  \cite{Zhitomirsky2013,ChZh_triPRB09}, is easy to find analytically. The equivalence of the magnon velocities in \eqref{eqA:vqvkq} is automatically satisfied when their {\it momenta}  are identical up to a set of the reciprocal lattice vectors. In our case, using that the smallest reciprocal lattice vector is ${\bf G}\!=\!2{\bf Q}$, one obtains two solutions, ${\bf q}^*_\pm\!=\!({\bf k}\pm{\bf Q})/2$, referred to as the ``equivalent magnon'' solutions below. The energies of the two-magnon continua for them are $E_{{\bf k}, {\bf q}^*_\pm}\!=\!2\bar{\varepsilon}_{{\bf q}^*_\pm}$. 

\begin{figure}[t]
\centering
\includegraphics[width=\columnwidth]{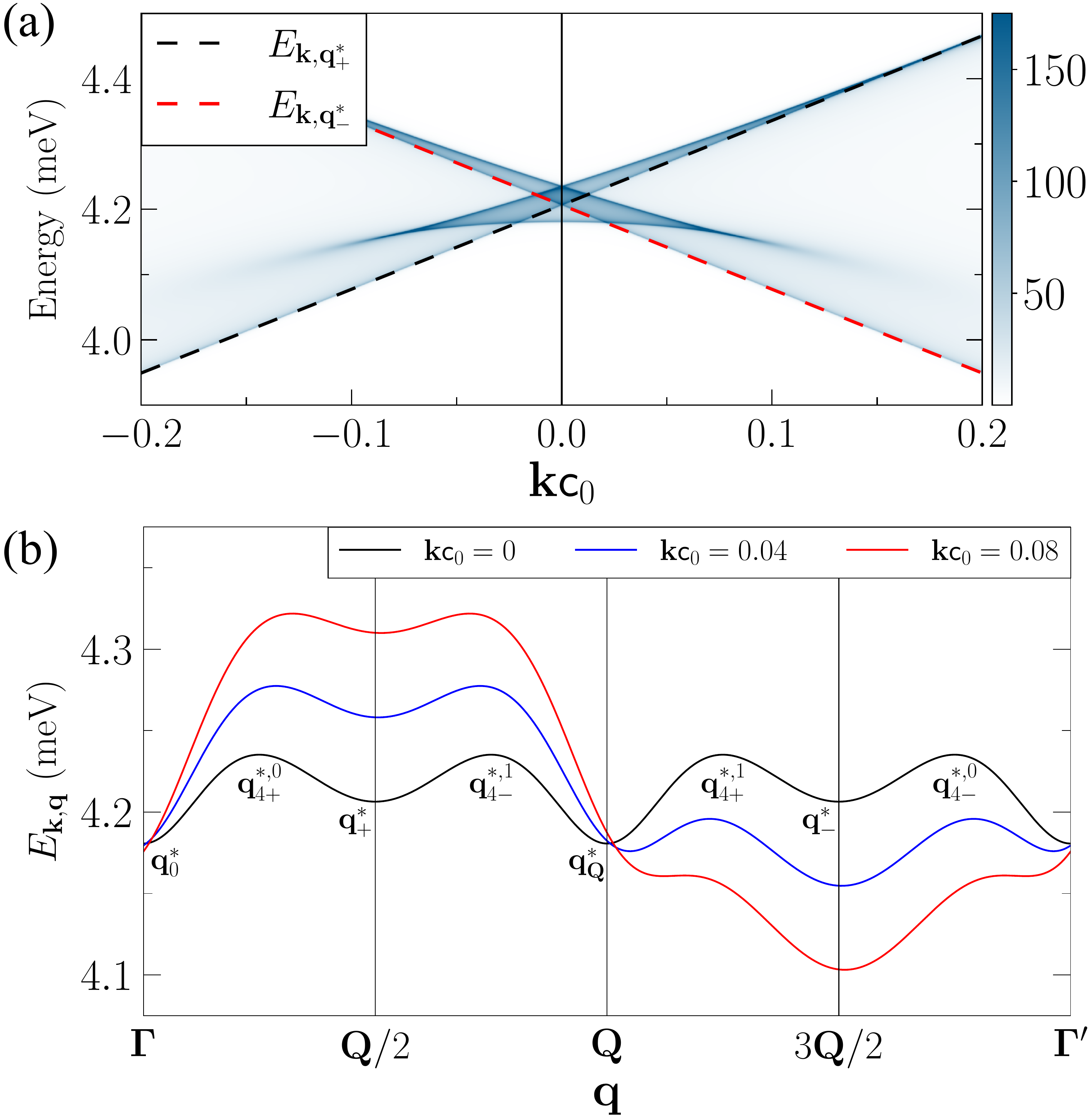}
\vskip -0.2cm
\caption{(a) The two-magnon DoS for the best-fit model of  CoNb$_2$O$_6$ and $B\!=\!7$~T ($H\!=\!0.8H_c$) as in  Fig.~\ref{f:SCHF_08Hc}, focusing on the vicinity of the ${\bf \Gamma}$ point; dashed lines are the equivalent-magnon solutions $E_{{\bf k}, {\bf q}^*_\pm}$. (b) The energy of the two-magnon continuum $E_{{\bf k}, {\bf q}}$ vs ${\bf q}$ for several ${\bf k}$, with the extrema corresponding to different solutions for ${\bf q}^*$ indicated,  see text.}
\vskip -0.4cm
\label{f:TwoMagnon}
\end{figure}

Given the relative simplicity of the magnon dispersion in either the LSWT or SCHF treatment of the model for CoNb$_2$O$_6$, and because of its 1D character, which permits only maxima and minima, these two solutions indeed describe the two edges of the two-magnon continuum and  completely exhaust the number singularities in it for most of the ${\bf k}$ values, as one can see in Fig.~\ref{f:SCHF_08Hc}, Sec.~\ref{Sec:Results}. Our Fig.~\ref{f:TwoMagnon}(a) shows the intensity plot of the two-magnon DoS for the best-fit model of CoNb$_2$O$_6$ for $B\!=\!7$~T from Fig.~\ref{f:SCHF_08Hc}, focusing instead on the vicinity of the ${\bf \Gamma}$ point, which provides a clearer view of the significantly richer structure of the continuum in that region. 

While additional singularities in the continuum in the 2D and 3D cases are common and are usually associated with the saddle points within it \cite{Zhitomirsky2013,ChZh_triPRB09}, the appearance of multiple singularities for the 1D model of CoNb$_2$O$_6$  is somewhat of a surprise. In Fig.~\ref{f:TwoMagnon}(a), one can see  that the equivalent-magnon extrema cease to be the absolute minima and maxima of the continuum and are overtaken by two different ones in the proximity of the $\Gamma$ point. 

Another insight into the structure of the continuum is offered by Fig.~\ref{f:TwoMagnon}(b), which shows the ${\bf q}$-cut of the two-magnon continuum for ${\bf k}\!=\!0$ with the additional minima and maxima,  demonstrating that the equivalent-magnon extremum at  ${\bf q}^*_\pm\!=\!\pm {\bf Q}/2$ is now a local minimum. 

Counterintuitively, this unusual complexity is an outcome of the  simplicity and high symmetry of the magnon spectrum. As is discussed in Sec.~\ref{Sec:Decay_thresholds}, away from the small-gap regime near the critical field, the magnon energy can be well-described by the nearest-neighbor hopping approximation, $\bar{\varepsilon}_{\bf q}\!\approx\!{\sf E_0}+{\sf J_1}\gamma_{\bf q}^{(1)}$, which naturally explains the bow-tie form of the continuum at the ${\bm \Gamma}$ point, as the nearest-neighbor hopping terms of $\bar{\varepsilon}_{\bf q}$ and $\bar{\varepsilon}_{{\bf k}-{\bf q}+{\bf Q}}$ in the continuum (\ref{eqA:continuum})  cancel precisely at ${\bf k}\!=\!0$. 

However, the next-nearest-neighbor exchanges (\ref{eq:Jkpm}) and relativistic form of the magnon dispersion (\ref{eq:omega_onemagn}) produce  small, but essential further-neighbor hoppings. Specifically, the finite width of the continuum at the ${\bm \Gamma}$ point can only be provided  by the effective even-neighbor hoppings. One can verify that the features shown in Fig.~\ref{f:TwoMagnon} can be closely reproduced by the following approximation for the magnon energy
\begin{eqnarray}
\bar{\varepsilon}_{\bf q}\approx {\sf E_0}+{\sf J_1}\gamma_{\bf q}^{(1)}+{\sf J_2}\gamma_{\bf q}^{(2)}+{\sf J_4}\gamma_{\bf q}^{(4)},
 \label{eqA:approxE}
\end{eqnarray}
using ${\sf E_0}\!=\!2.107$, ${\sf J_1}\!=\!-1.4$, ${\sf J_2}\!=\!-0.0065$, and ${\sf J_4}\!=\!-0.01$, all in meV, where $\gamma_{\bf k}^{(n)} \!=\! \cos (n k_c {\sf c}_0)$ as before.

Using the form in (\ref{eqA:approxE}), a simple algebra  yields the maxima of the  ${\bf k}\!=\!0$ continuum in Fig.~\ref{f:TwoMagnon}(b)  at the momenta ${\bf q}^{*,n}_{4\pm}\!=\!\pm\frac12\arccos(-{\sf J_2}/4{\sf J_4})+n\pi$, explicating the essential role of the further-neighbor terms in the additional extrema of the continuum. For small ${\bf k}$, the left pair of these maxima shifts in ${\bf q}$ and up in energy  linearly with ${\bf k}$ and remains the absolute maxima for a range of ${\bf k}$, while the right pair shifts down in energy, becomes the local maxima, and then ceases to be extremal at the larger ${\bf k}$. In Fig.~\ref{f:TwoMagnon}(a), they correspond to the upper singularity, which  merges with the one from the equivalent magnons, and to the one that enters the continuum and annihilates with the nearly flat singularity, respectively.

The last and the most curious is the ``flat'' singularity, which is the absolute minimum of the  continuum in Fig.~\ref{f:TwoMagnon}(a) at ${\bf k}\!=\!0$. The corresponding  minima of $E_{{\bf k}, {\bf q}}$  in Fig.~\ref{f:TwoMagnon}(b) can be found  at ${\bf q}^*_0\!\approx\! -b_0{\bf k}$ and ${\bf q}^*_{\bf Q}\!\approx\! {\bf Q}+b_0{\bf k}$,  where $b_0\!=\!\frac12({\sf J_1}/\widetilde{{\sf J}}_{{\sf 0}}\!-\!1)$ with $\widetilde{{\sf J}}_{{\sf 0}}\!=\!4({\sf J_2}+4{\sf J_4})$. At ${\bf k}\!=\!0$, one magnon in $E_{0, {\bf q}^*}$  is at ${\bf q}^*_0\!=\! 0$ and the other is at ${\bf Q}$,  precisely at the minimum and the maximum of the {\it single-magnon} band. This  arrangement  guarantees that the velocities on both sides of Eq.~(\ref{eqA:vqvkq}) are zero, fulfilling the extremum condition. It also explains the flatness of the singularity in Fig.~\ref{f:TwoMagnon}(a), as the continuum energy is $E_{{\bf k}, {\bf q}^*_0}\!\approx\! \bar{{\sf E}}_{\sf 0}+2\widetilde{{\sf J}}_{{\sf 0}}b_0^2{\bf k}^2$, with $\bar{{\sf E}}_{\sf 0}\!=\!2({\sf E_0}+{\sf J_2}+{\sf J_4})$.

This completes the consideration of the richer set of  the Van Hove singularities in the two-magnon continuum near the ${\bf \Gamma}$ point in our model.

\subsubsection{The kinematics of the two-magnon decay}

\begin{figure}[t]
\centering
\includegraphics[width=\columnwidth]{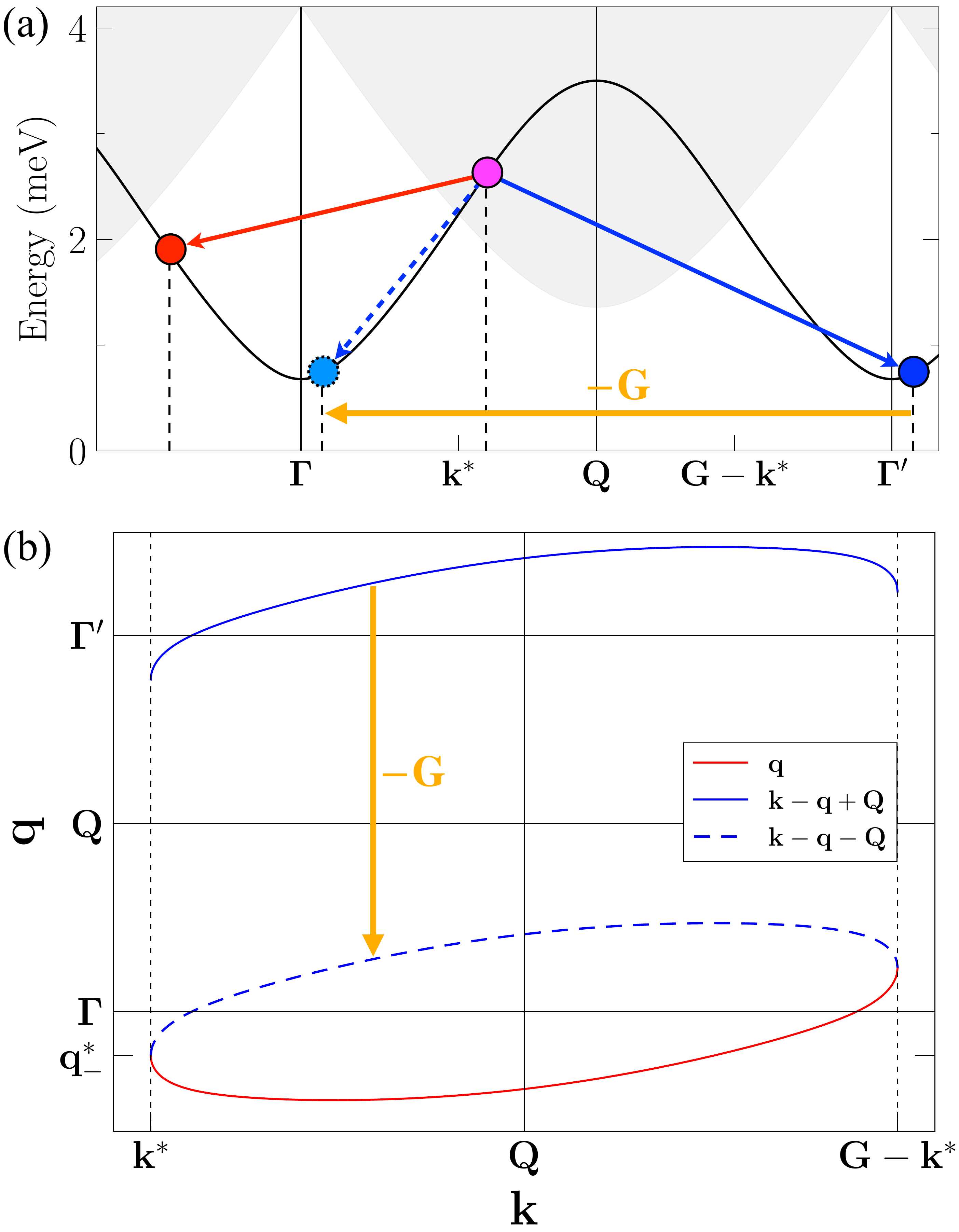}
\vskip -0.2cm
\caption{(a) Schematics of the two-magnon decay of a magnon for ${\bf k}\!>\!{\bf k}^*$; solid line is $\bar{\varepsilon}_{\bf k}$ and shaded area is the continuum. (b) The set of decay ${\bf q}$-points vs  ${\bf k}$; vertical lines are marking decay thresholds ${\bf k}^*$ and ${\bf G}-{\bf k}^*$. Arrows indicate the shift by the reciprocal lattice vector $-{\bf G}$. The best-fit model parameters of CoNb$_2$O$_6$ and $B\!=\!7$~T ($H\!=\!0.8H_c$) are used.}
\label{f:TwoMagnonDecay}
\vskip -0.4cm
\end{figure}

Generally, a magnon with the momentum ${\bf k}$ is kinematically allowed to decay into a pair of magnons if the energy conservation for the single-magnon energy and that of the two-magnon continuum 
\begin{eqnarray}
\bar{\varepsilon}_{\bf k} = E_{{\bf k},{\bf q}},
\label{eq:energy_cons}
\end{eqnarray}
can be satisfied for some momenta ${\bf q}$. This condition naturally separates the ${\bf k}$-space into the stable one, in which (\ref{eq:energy_cons}) cannot be fulfilled, and the decay region, in which it is fulfilled \cite{Zhitomirsky2013}.  They are easy to visualize as having or not having  an overlap of the single-magnon branch with the two-magnon continuum, see Fig.~\ref{f:TwoMagnonDecay}(a), with the decay threshold boundaries ${\bf k}^*$ and ${\bf G}-{\bf k}^*$ separating the stable region from the decay region. 

Such a threshold, or the entry-point of the single-magnon branch into the two-magnon continuum, is necessarily a crossing of the single-magnon branch with the {\it minimum} of the two-magnon continuum at that ${\bf k}^*$, which, in turn, must correspond to a singularity in it as is discussed above, see also Refs.~\cite{ChZh_triPRB09,Zhitomirsky2013}. 

In the case relevant to the 1D model of CoNb$_2$O$_6$ discussed in this work, the singularity associated with the minimum of the two-magnon continuum away from the ${\bf \Gamma}$ point  corresponds to the equivalent-magnon solution of the extremum condition in (\ref{eqA:vqvkq}), yielding an equation on the  decay threshold boundary ${\bf k}^*$
\begin{eqnarray}
\bar{\varepsilon}_{{\bf k}^*} = 2\bar{\varepsilon}_{({\bf k}^*-{\bf Q})/2},
\label{eqA:decay_Ek*}
\end{eqnarray}
for $\Gamma\!\alt\!{\bf k}^*\!\alt\!\Gamma'$. While this equation can be solved numerically for the actual magnon energy $\bar{\varepsilon}_{\bf k}$ of the considered model, a simplified solution for ${\bf k}^*$ can be derived analytically using the nearest-neighbor hopping approximation $\bar{\varepsilon}_{\bf k}\!\approx\!{\sf E_0}+{\sf J_1}\gamma_{\bf k}^{(1)}$ discussed above, which closely  describes $\bar{\varepsilon}_{\bf k}$ when the  gap $\Delta_{0}$ is not too small, yielding 
\begin{eqnarray}
{\bf k}^* = 2\arcsin\left(\frac{\sqrt{5+2\Delta}-1}{2}\right),
\label{eqA:decay_k*}
\end{eqnarray}
where $\Delta\!=\!\Delta_{0}/|{\sf J_1}|\!=\!{\sf E_0}/|{\sf J_1}|-1$ is the dimensionless gap. 

One should also add that the $1/\sqrt{|\Delta{\bf k}|}$ singularities in the on-shell magnon spectrum discussed in Sec.~\ref{Sec:Decay_thresholds} are naturally connected to the $1/\sqrt{|\Delta\omega|}$ one-dimensional Van Hove singularities in the two-magnon DoS, with the latter transferred onto the single-magnon energies via the anharmonic coupling. 

The kinematic consideration of the energy conservation in Eq.~\eqref{eq:energy_cons} also allows to trace the evolution of the decay {\it surfaces}, that is, the sets of the ${\bf q}$ values into which decays are possible, as one traverses along the $\bar{\varepsilon}_{\bf k}$ curve vs ${\bf k}$. In the 1D problem considered here, the decay surfaces are, in fact, the pairs of the discrete ${\bf q}$-points. At the decay threshold ${\bf k}^*$, they correspond to the two equivalent points ${\bf q}^*$ and  ${\bf q}^*+{\bf G}$, as one can see in Fig.~\ref{f:TwoMagnonDecay}(b). As a function of ${\bf k}$, this set of pairs of the ${\bf q}$-points traces a continuous line of the elliptic shape shown in Fig.~\ref{f:TwoMagnonDecay}(b), with the  schematics in  Fig.~\ref{f:TwoMagnonDecay}(a) illustrating the decay process. The best-fit parameters of CoNb$_2$O$_6$ and $B\!=\!7$~T ($H\!=\!0.8H_c$) are used in both figures.

\begin{figure}[t]
\centering
\includegraphics[width=\linewidth]{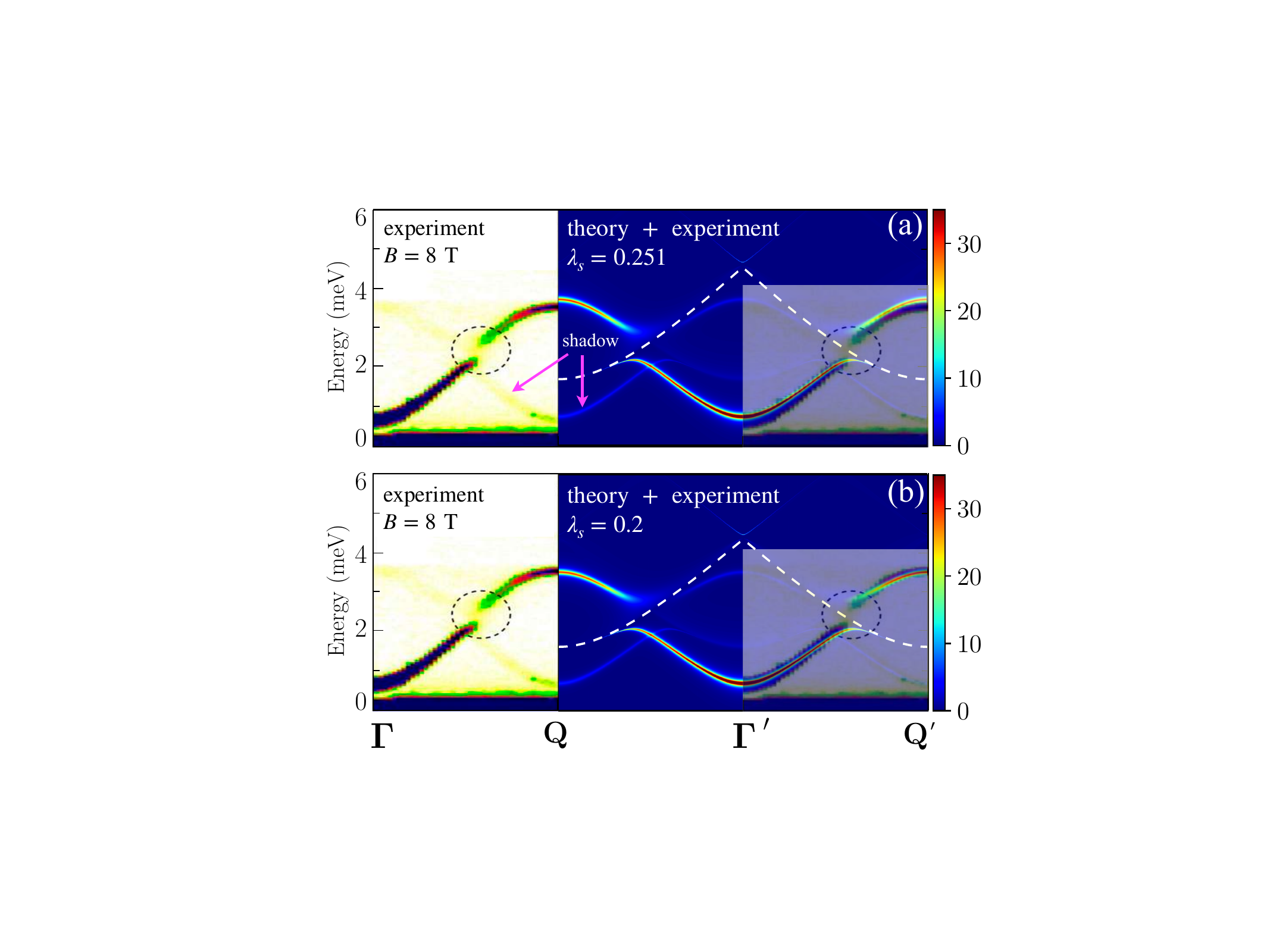} 
\vskip -0.2cm
\caption{Same as Fig.~\ref{f:TF_DSF} in Sec.~\ref{Sec:DSF} for the field $B\!=\!8$~T.}
\label{fA:DSF8T}
\vskip -0.5cm
\end{figure}

\vspace{-.4cm}
\subsection{Dynamical structure factor}
\label{A:DSF}
\vskip -.2cm

Our Fig.~\ref{fA:DSF8T} presents the comparison of the experimental and theoretical DSFs as in Fig.~\ref{f:TF_DSF} in Sec.~\ref{Sec:DSF}, for the field $B\!=\!8$~T.

\section{Longitudinal field effects}
\label{A:LFields}

\subsection{The excitation gap and the band gap}
\label{A:Gaps}

For the  magnetic field tilted in the $y_0z_0$ ($zx$) plane, spin components $\widetilde{\mathbf{S}}_i$ in the local reference frame, which is tilted by the angle $\theta$ away from the $y_0$ axis, are related to the spin components in the laboratory frame $\mathbf{S}_i$ by
\begin{equation}
\widetilde{\mathbf{S}}_i\! =\!\hat{\bf R}_\theta \mathbf{S}_i,    
\end{equation}
where $\hat{\bf R}_\theta$ is the rotation matrix
\begin{align}
\hat{\textbf{R}}_{\theta} =
\begin{pmatrix}
0 & -\sin\theta & \cos\theta\\
1 & 0 & 0\\
0 & \cos\theta & \sin\theta\\
\end{pmatrix}.
\label{eqA:rotationMatrix}
\end{align} 
The Hamiltonian in the tilted local frame is 
\begin{align}
\hat{\cal H}=\sum_{\langle ij\rangle_n} \widetilde{\bf S}_i^{\rm T} 
\widetilde{\bm J}_{ij} \widetilde{\bf S }_j- \sum_i \widetilde{\bf H}^{\rm T} \widetilde{\bf S}_i,
\label{eq:HlocLF}
\end{align}
where $n\!=\!1,2$ for the nearest- and next-nearest-neighbor bonds, respectively. The exchange matrices are
\begin{align}
\widetilde{\bm J}_{ij} =\hat{\bf R}_\theta^{\rm T}\hat{\bf J}_{ij}\hat{\bf R}_\theta
=\left(
\begin{array}{ccc}
\widetilde{J}_{ij}^{xx} &\widetilde{J}_{ij}^{xy} &\widetilde{J}_{ij}^{xz} \\
\widetilde{J}_{ij}^{yx} &\widetilde{J}_{ij}^{yy} &\widetilde{J}_{ij}^{yz} \\
\widetilde{J}_{ij}^{zx} &\widetilde{J}_{ij}^{zy} &\widetilde{J}_{ij}^{zz} 
\end{array} \right),
\label{eqA:JlocLF}
\end{align}
with the exchange matrices $\hat{\bf J}_{ij}$ in the laboratory frame from Eqs.~(\ref{eq:J1}) and (\ref{eq:J2}), and the rotated field given by
\begin{equation}
\widetilde{\bf H}=\mu^{\phantomsection} \hat{\bf R}_\theta^{\phantomsection} \hat{\bf g}^{\phantomsection} {\bf B}^{\phantomsection},
\label{eqA:HlocLF}
\end{equation}
with ${\bf B}\!=\!B_{\perp}\hat{y}_0+B_{\parallel}\hat{z}_0$ and $\hat{\bf g}$ being the diagonal g-tensor. 

Naturally, the Hamiltonian can be split into the even and odd parts, as before. At the LSWT level of approximation, we only consider the even part that reads
\begin{align}
\hat{\cal H}_{\text{even}}=\sum_{\langle ij\rangle_n} &\Big(\widetilde{J}_{ij}^{xx} \widetilde{S}_i^{x} \widetilde{S }_j^x + \widetilde{J}_{ij}^{yy} \widetilde{S}_i^{y} \widetilde{S }_j^y+\widetilde{J}_{ij}^{zz} \widetilde{S}_i^{z} \widetilde{S }_j^z 
\label{eqA:HevenLF} \\
&+\widetilde{J}_{ij}^{xy} \widetilde{S}_i^{x} \widetilde{S }_j^y+\widetilde{J}_{ij}^{yx} \widetilde{S}_i^{y} \widetilde{S }_j^x\Big) - \sum_i \widetilde{H}^{z} \widetilde{S}_i^z. \nonumber
\end{align}
Importantly, compared to the same LSWT consideration in Sec.~\ref{Sec:LSWT}, the diagonal exchanges $\widetilde{J}_{1}^{xx}$ and $\widetilde{J}_{1}^{zz}$ have now acquired bond-dependent contributions from the staggered  $J_{y_0z_0}$ terms because of the local axes tilt. Therefore, the nearest-neighbor bonds in the Hamiltonian (\ref{eqA:HevenLF}) are not equivalent, the unit cell now contains two spins, and two species of bosons need to be introduced with the Holstein-Primakoff transformation that reads
\begin{equation}
\widetilde{S}_{\mu l}^{z}=S-n_{\mu l},\quad \widetilde{S}_{\mu l}^{+}\approx \sqrt{2S} a_{\mu l},
\label{eqA:HPLF}
\end{equation}
where $l$ and $\mu=1,2$ numerate the unit cells of the zigzag chain and the two sublattices, respectively. 

The classical energy, obtained from Eq.~\eqref{eqA:HevenLF}, is (up to a constant $S^2 (J_{y_0y_0}+J_{2})$),
\begin{align}
\frac{E_{\text{cl}}}{N}=-SH_{\perp} \cos\theta-SH_{\parallel} \sin\theta  -\frac{S}{2} H_c \sin^2\theta,
\label{eqA:EclLF}
\end{align}
 with $H_c$ from Eq.~(\ref{eq:critical_field}). Minimizing Eq.~\eqref{eqA:EclLF} with respect to $\theta$ yields Eq.~\eqref{eq:EclHyHz}  in Sec.~\ref{Sec:LongFields}. 

After some algebra, the  LSWT Hamiltonian for the two bosonic species can be written in the matrix form
\begin{equation}
\hat{\mathcal{H}} = \frac{1}{2} \sum_{\bf k} \hat{{\bf x}}_{\bf k}^\dagger \hat{{\bf H}}_{\bf k}\hat{{\bf x}}_{\bf k}^{\phantom\dag},\quad
\hat{{\bf H}}_{\bf k} = \begin{pmatrix}
\hat{{\bf A}}_{\bf k} & \hat{{\bf B}}_{\bf k} \\
\hat{{\bf B}}_{\bf k}^\dagger & \hat{\bf A}^*_{-{\bf k}} \end{pmatrix},
\label{eq:H2kSL}
\end{equation}
where  $\hat{{\bf x}}_{\bf k}^\dagger \!=\! \big( a_{1\bf k}^\dagger, a_{2\bf k}^\dagger, a_{1-{\bf k}}^{\phantom\dag}, a_{2-{\bf k}}^{\phantom\dag} \big)$ are the bosonic vector operators and  the $2\!\times\! 2$ matrices $ \hat{{\bf A}}_{\bf k}$ and $ \hat{{\bf B}}_{\bf k}$ are 
\begin{equation}
\hat{\textbf{A}}_{\bf k} = \begin{pmatrix}
A_{\bf k} & B_{\bf k} \\
B_{\bf k}^* & A_{\bf k}
\end{pmatrix},
\
\hat{\textbf{B}}_{\bf k} = \begin{pmatrix}
D_{\bf k} & C_{\bf k} \\
C_{-{\bf k}} & D_{\bf k}^*
\end{pmatrix},
\end{equation}
with
\begin{align}
A_{\bf k} &= H_{\perp}c_\theta + H_{\parallel}s_\theta -2S(J_{y_0y_0}+J_2)c^2_\theta \nonumber \\
&-2S\big(J_{z_0z_0}+J_{2z_0} \big) s^2_\theta +S\big(J_{2}(1+s^2_\theta)+J_{2z_0}c^2_\theta \big)\gamma_{\bf k}^{(2)}, \nonumber \\
B_{\bf k} &= -SJ_{y_0z_0} s_{2\theta} \bar{\gamma}_{\bf k}+S\big(J_{x_0x_0}+J_{y_0y_0}s^2_\theta +J_{z_0z_0} c^2_\theta\big)\gamma_{\bf k}^{(1)}, \nonumber \\
C_{\bf k} &= -SJ_{y_0z_0} s_{2\theta} \bar{\gamma}_{\bf k}-S\big(J_{x_0x_0}-J_{y_0y_0}s^2_\theta -J_{z_0z_0} c^2_\theta\big)\gamma_{\bf k}^{(1)}, \nonumber \\
D_{\bf k} &=  -S\big(J_{2}-J_{2z_0} \big)c^2_{\theta} \gamma_{\bf k}^{(2)},
\label{eqA:ABCDk}
\end{align}
with the shorthand notations $c_\theta\! =\! \cos \theta$ and $s_\theta \!=\! \sin \theta$. 

The eigenvalue problem for (\ref{eq:H2kSL}) can be solved analytically by diagonalizing $(\hat{\bf g}\hat{ \bf H}_{\bf k})^2$, with the paraunitary matrix $\hat{\bf g}\!=\![1,1,-1,-1]$. Using that $A_{\bf k}\!=\!A_{-{\bf k}}$ and $D_{\bf k}$ are real, $B_{-{\bf k}}\!=\!B_{\bf k}^*$, and $C_{-{\bf k}}\!=\!C_{\bf k}^*$, we get two branches
\begin{align}
&\varepsilon_{1,2{\bf k}}=\sqrt{A_{\bf k}^2 + |B_{\bf k}|^2 - |C_{\bf k}|^2 - D_{\bf k}^2\pm 2\sqrt{{\cal R}}}, \label{eqA:EkLF} \\
&{\cal R}=\!A_{\bf k}^2|B_{\bf k}|^2\! +\!|C_{\bf k}|^2D_{\bf k}^2\!-\!2A_{\bf k}D_{\bf k}{\rm Re}[B_{\bf k}C_{\bf k}^*]\!+\!{\rm Im}[B_{\bf k} C_{\bf k}^*]^2.\nonumber
\end{align}

From the spectrum in Eq.~\eqref{eqA:EkLF}, one can obtain expressions for the excitation gap $\Delta_{0}$ and band gap $\Delta_{b}$ using the canting angle $\theta$ calculated numerically from the nonlinear equation~\eqref{eq:EclHyHz}. For $H_{\parallel}\! \ll\! H_{\perp}\!=\!H_c$, with the approximate solution for the canting angle in Eq.~\eqref{eq:CantingAngleLF}, after some algebra in  Eq.~\eqref{eqA:EkLF}, the excitation gap is 
\begin{align}
\Delta_{0} &\approx \alpha_0 H_{\parallel}^{1/3},\\
\alpha_0 &= \left(\frac{2}{H_c}\right)^{1/3} \sqrt{\frac32}H_c \sqrt{1+\frac{2S(J_{x_0x_0}-J_{y_0y_0})}{H_c}}\nonumber \ ,
\end{align}
and the band gap 
\begin{align}
\Delta_{b} & \approx \alpha_b H_{\parallel}^{1/3},\\ 
\alpha_b & = 2S\big|J_{y_0z_0}\big|\left(\frac{2}{H_c}\right)^{1/3}\sqrt{1+\frac{J_{2}-J_{2z_0}}{J_{z_0z_0}+2J_{2z_0}}}\ . \nonumber
\end{align}
Neglecting small corrections under the square roots leads to the results in Eq.~(\ref{eq:GapLF}).

\subsection{More threshold singularities}
\label{A:PhysicalSingLF}

The anharmonic cubic coupling in the tilted magnetic field is obtained from the  odd part of the Hamiltonian, similarly to the discussion in Sec.~\ref{Sec:evenodd}. Using Eqs.~(\ref{eqA:JlocLF})  and (\ref{eqA:HlocLF}), it can be generally written as 
\begin{align}
\hat{\cal H}_{\text{odd}}\!&=\!\sum_{\langle ij\rangle_n} \widetilde{J}_{ij}^{xz}  \Big(\widetilde{S}_i^{x} \widetilde{S }_j^z \!+\!  \widetilde{S}_i^{z} \widetilde{S }_j^x \Big) - \sum_i \widetilde{H}^{x} \widetilde{S}_i^x.
\label{eqA:HoddLF}
\end{align}
 There are two resulting types of the cubic coupling, one that retains the staggered structure of such a coupling in the pure transverse field, $\propto\!J_{y_0z_0}$, and the other one originating from the tilted component of the spin. In the small longitudinal field regime, $H_{\parallel}\! \ll\! H_{\perp}\!=\!H_c$, the latter is subleading to the former, $\propto\!{\cal O}(J_{z_0z_0}H_{\parallel}^{1/3})$, leaving cubic anharmonicity unaffected by the field. The secondary component  also corresponds to an unfavorable kinematics for the decays, justifying neglecting it in this regime. 

\bibliography{CoNb2O6}

\begin{thebibliography}{86}
\bibitem [{not()}]{noteDMRG}%
  \BibitemOpen
  \href@noop {} {}\bibinfo {note} {{Typically, $5$ sweeps and a maximum bond
  dimension of $m\sim 10$ were needed to ensure good convergence with the
  truncation error of $\mathcal{O}(10^{-10})$}}\BibitemShut {NoStop}%
\end{thebibliography}
\end{document}